\documentclass[aps,twocolumn,superscriptaddress,floatfix,longbibliography]{revtex4-1}

\usepackage{booktabs}
\usepackage{lipsum}
\usepackage{graphicx}
\usepackage{amsmath,amssymb}
\usepackage{braket}
\usepackage{mathtools}
\usepackage[colorlinks,linkcolor=blue,citecolor=blue,urlcolor=blue]{hyperref}
\usepackage{color}
\usepackage[normalem]{ulem}
\usepackage{bm}
\usepackage{multirow}
\usepackage{makecell}
\usepackage{booktabs}
\usepackage{pifont}

\newcommand{\cross}{\ding{55}}

\usepackage{tikz}
\usetikzlibrary{shapes,calc}
\usetikzlibrary{shapes.geometric,arrows.meta,decorations.markings}

\tikzset{
	dot/.style={draw,circle,inner sep=1.5pt,fill=black},
	empty dot/.style={draw,circle,inner sep=1.5pt,fill=white},
	mid arrow/.style={postaction={decorate,decoration={
        markings,
        mark=at position .6 with {\arrow[#1,scale=1.5]{latex}}
    }}},
	spinA/.style={draw=black,thick,circle,inner sep=2.5pt, fill=figBlue},
	spinBC/.style={draw=black,thick,circle,inner sep=2.5pt, fill=figRed},
	faded/.style={opacity=0.2},
}

\begin{document}

\title{Revealing Quantum Geometry in Nonlinear Quantum Materials }

\author{Yiyang Jiang}
\affiliation{Department of Condensed Matter Physics, Weizmann Institute of Science, Rehovot 7610001, Israel}
\affiliation{Department of Physics, The Pennsylvania State University, University Park, Pennsylvania 16802, USA}
\author{Tobias Holder}
\email[]{tobiasholder@tauex.tau.ac.il}
\affiliation{Raymond and Beverly Sackler School of Physics and Astronomy, Tel Aviv University, Tel Aviv, Israel}
\author{Binghai Yan}
\email[]{binghai.yan@weizmann.ac.il}
\affiliation{Department of Condensed Matter Physics, Weizmann Institute of Science, Rehovot 7610001, Israel}
\affiliation{Department of Physics, The Pennsylvania State University, University Park, Pennsylvania 16802, USA}
\date{\today}

\begin{abstract}
Berry curvature-related topological phenomena have been a central topic in condensed matter physics. Yet, until recently other quantum geometric quantities such as the metric and connection received only little attention due to the relatively few effects which have been documented for them.
This review gives a modern perspective how quantum geometric quantities naturally enter the nonlinear responses of quantum materials and demonstrate their deep connection with excitation energy, lifetimes, symmetry, and corresponding physical processes.
The multitude of nonlinear responses can be subdivided into nonlinear optical effects, subgap responses, and nonlinear transport phenomena.
Such a distinction by energy scales facilitates an intuitive understanding of the underlying electronic transitions, giving rise to a unified picture of the electron motion beyond linear order.
The well-known injection and shift currents constitute the main resonances in the optical regime. Exploiting their respective lifetime and symmetry dependencies, this review elucidates how these resonances can be distinguished by a corresponding quantum geometric quantity that shares the same symmetry.
This is followed by a brief exposition of the role of quasiparticle lifetimes for nonlinear subgap responses, which presents a window into the microscopic short-term dynamics as well as the ground state correlation and localization.
We conclude with an account of the anomalous motion due to the Berry curvature dipole and quantum metric dipole in nonlinear transport,
clarifying the correspondence between physical observables and the underlying mechanisms. 
This review highlights the close relationship between quantum geometry and nonlinear response, showing the way  towards promising probes of quantum geometry and enabling novel avenues to characterize complex materials. 
\end{abstract}

\maketitle

\tableofcontents

\section{Introduction}\label{sec1}

\paragraph{Overview}

Quantum geometry has recently emerged as an efficient tool to characterize the properties of quantum materials~\cite{Toermae2023,Yu2025}. On one hand, the quantum geometry of a bulk material 
determines the intrinsic localization properties of the electronic states, making it an economic and physically meaningful language to quantify the charge distribution within the clean bulk solid. 
On the other hand, quantum geometry quantifies the geometric properties such as the distance (quantum metric), area (Berry curvature), and parallelism (quantum connection) of the Bloch wavefunction $|\psi_{\mathbf{k}}\rangle$ in momentum space.

Initially, quantum geometric phenomena have been explored exclusively in relation to Berry phases and curvature effects. Indeed, these quantities are nowadays well-understood as, respectively, measures of the charge polarization~\cite{Resta2007} and self-rotation of Bloch wavepacket~\footnote{The self-rotation can alternatively be expressed as the non-commutativity of the position operator in the ground state manifold~\cite{Haldane2004}.}.
In the early 2000's, the improved understanding of quantum geometry in terms of the Berry curvature played a crucial role in the topological classification of quantum phases beyond Landau's paradigm~\cite{Hasan2010,Qi2011,Hasan2011,Yan2017,Armitage2018,Vanderbiltbook}. Despite these successes, additional geometric quantities such as the quantum metric~\cite{Provost1980} and the quantum connection~\cite{Avdoshkin2023} are needed for a more complete description of the adiabatic evolution of the wavefunction in the ground state. Yet, until very recently these latter quantities have not received the same amount of attention. Thanks to the progress in the engineering and characterization of complex bulk materials as well as layered van-der-Waals heterostructures, the study of multiorbital systems has propelled the arguably subtle effects of quantum geometry to the center of attention: With their many degrees of freedom which can give way to a large number of interband-coherent processes, these so-called \emph{quantum materials} offer an unprecedented and rich phenomenology of correlated phases, light-matter coupling and transport responses.
Importantly, many of these compounds also feature giant nonlinear responses~\cite{Ma2021}.
This review aims to highlight how and why novel phenomena originating from quantum geometry become accessible specifically with the help of such nonlinear quantum materials. 

\begin{figure}%[htbp]
    \begin{center}
        \begin{tabular}{cc}
            \textbf{Momentum Space} & \textbf{Real Space}\\
            \textbf{Band Geometry} & \textbf{Wavepacket Geometry} \\
            \begin{minipage}{0.4\linewidth}
                \centering
                \includegraphics[width=\linewidth]{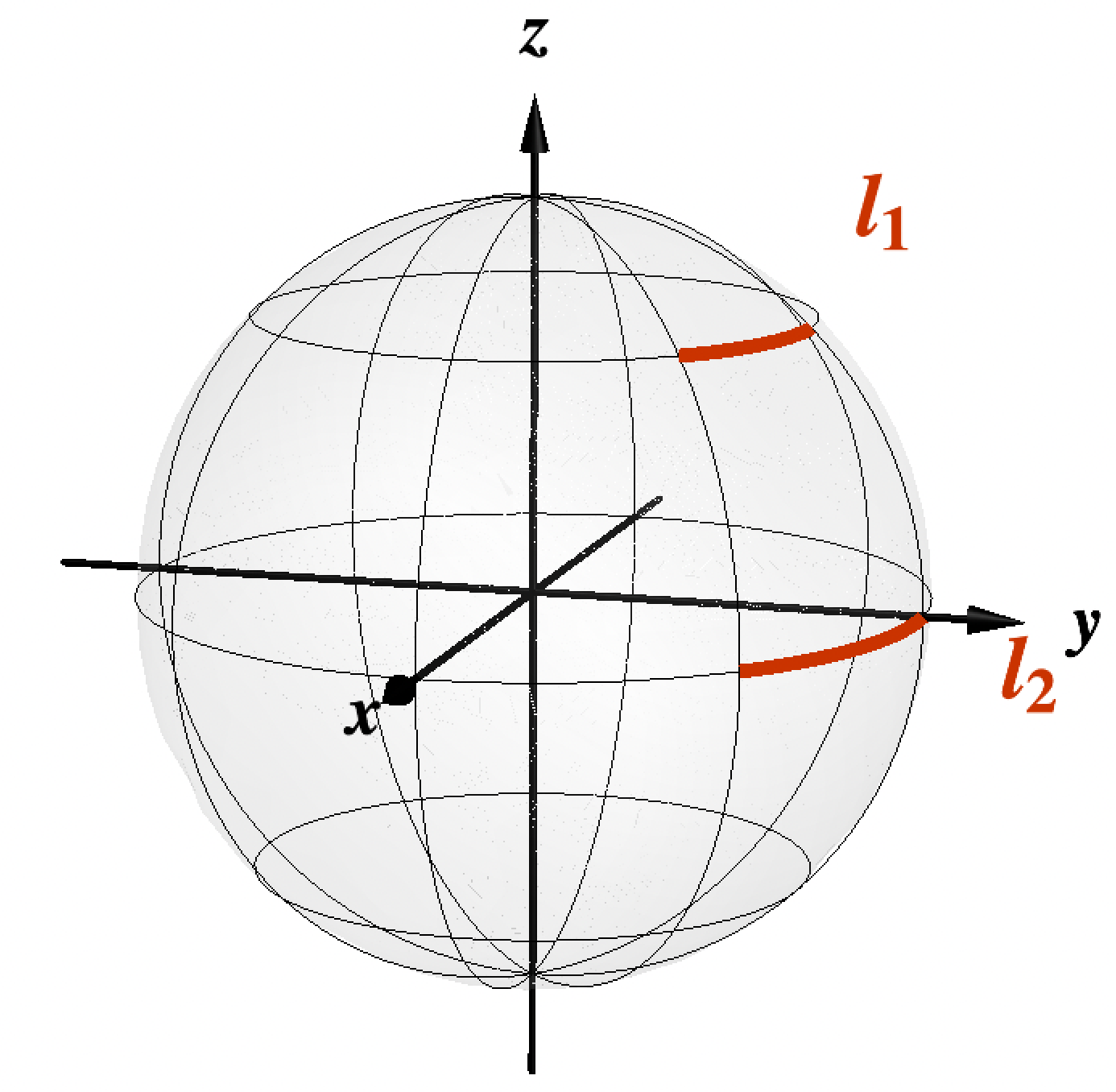}
            \end{minipage}
            & 
            \begin{minipage}{0.45\linewidth}
                \centering
                \includegraphics[width=\linewidth]{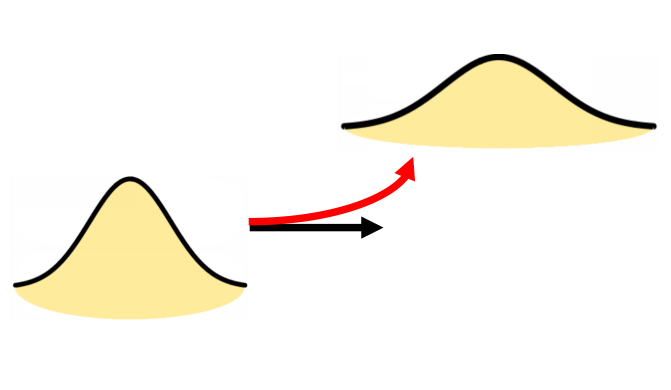}
            \end{minipage}
            \\
            (a) Quantum metric & (d) Deformation \\
            \begin{minipage}{0.4\linewidth}
                \centering
                \includegraphics[width=\linewidth]{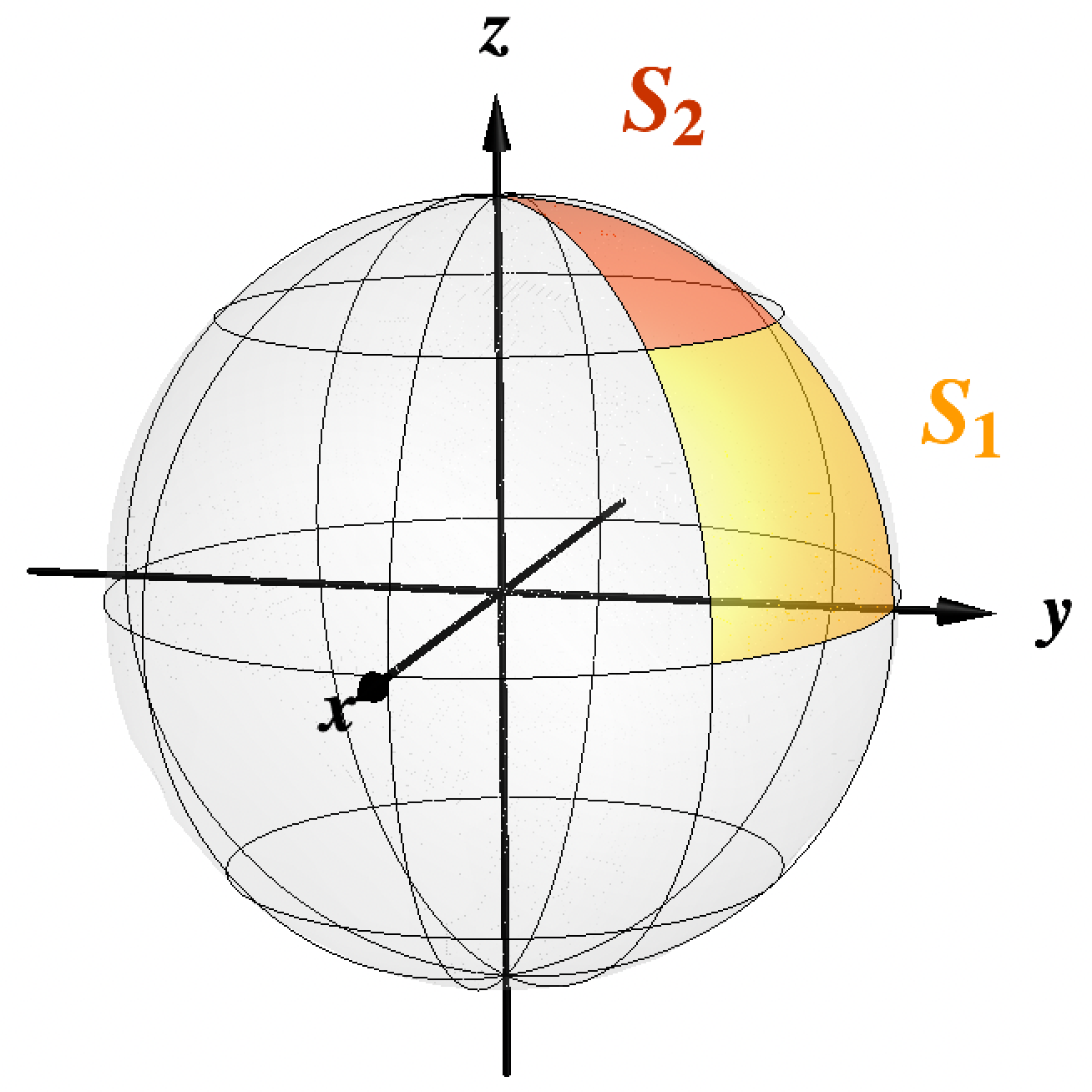}
            \end{minipage}
            & 
            \begin{minipage}{0.45\linewidth}
                \centering
                \includegraphics[width=\linewidth]{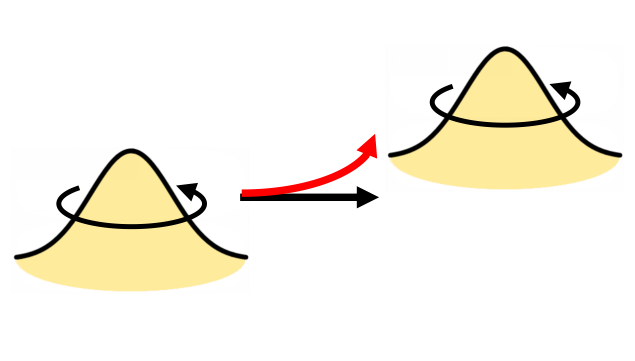}
            \end{minipage}
            \\
            (b) Berry curvature & (e) Rotation \\
            \begin{minipage}{0.4\linewidth}
                \centering
                \includegraphics[width=\linewidth]{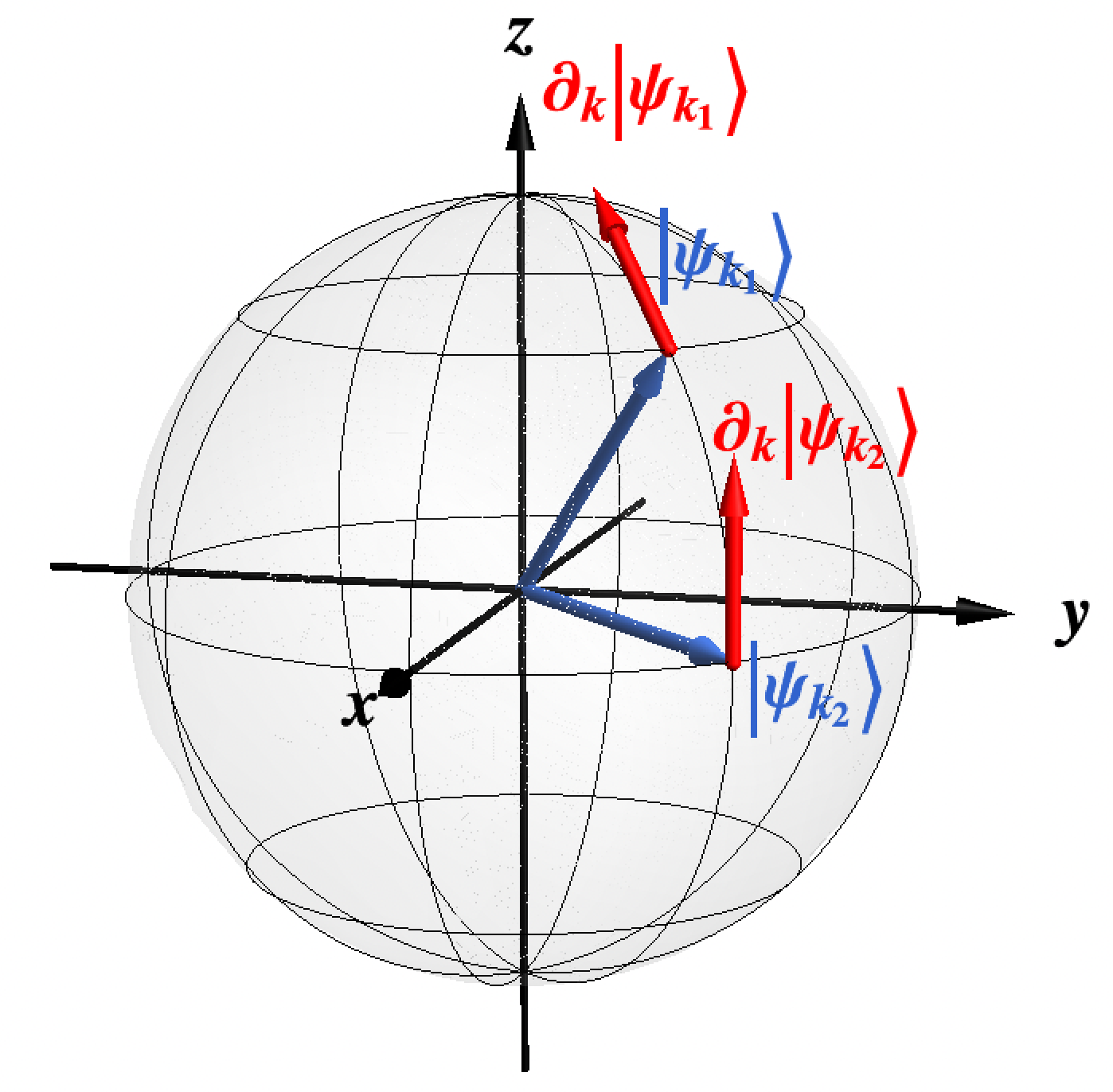}
            \end{minipage}
            & 
            \begin{minipage}{0.45\linewidth}
                \centering
                \includegraphics[width=\linewidth]{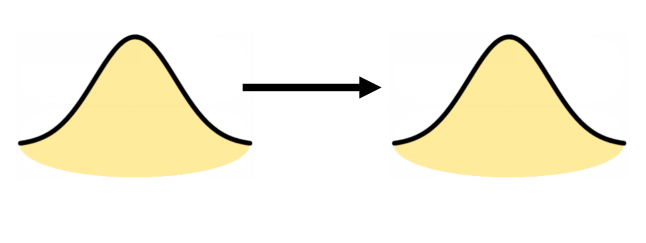}
            \end{minipage}
            \\
            (c) Quantum connection & (f) Positional shift \\[-1em]
        \end{tabular}
    \end{center}
\caption{ Quantum geometry contains both momentum space band geometry and information about the charge distribution in real space. 
In band geometry, quantum metric (a), Berry curvature (b), and quantum connection (c) characterize the distance ($l_{1,2}$), area ($S_{1,2}$), and change of tangent space orientation in Hilbert space (drawn as a Bloch sphere), respectively. 
All three geometric quantities are relevant ingredients for the optical response.
In terms of the quasiparticle motion, quantum metric (a), Berry curvature (b), and quantum connection (c) can be associated with the deformation (d), rotation (e) and nontrivial shift (f) of a traveling wavepacket, respectively.
The corresponding anomalous components of the equations of motion are the quantum metric dipole (d) and Berry curvature dipole (e), as indicated by red curved arrows, while (f) does not appear in the intrinsic nonlinear conductivity.
}\label{fig0}
\end{figure}

\begin{figure*}[htbp]
    \begin{center}
        \begin{tabular}{ccccc}
            \multicolumn{2}{c}{\textbf{Nonlinear Optics}} & \textbf{Subgap Response} & \multicolumn{2}{c}{\textbf{Nonlinear Transport}} \\
            \begin{minipage}{0.192\linewidth}
                \centering
                \includegraphics[width=\linewidth]{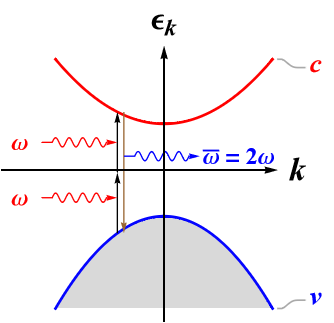}
            \end{minipage}
            & 
            \begin{minipage}{0.192\linewidth}
                \centering
                \includegraphics[width=\linewidth]{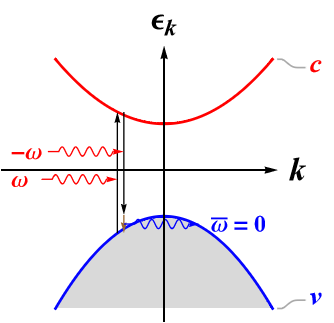}
            \end{minipage}
            & 
            \begin{minipage}{0.192\linewidth}
                \centering
                \includegraphics[width=\linewidth]{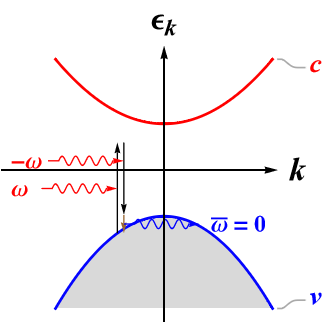}
            \end{minipage}
            & 
            \begin{minipage}{0.192\linewidth}
                \centering
                \includegraphics[width=\linewidth]{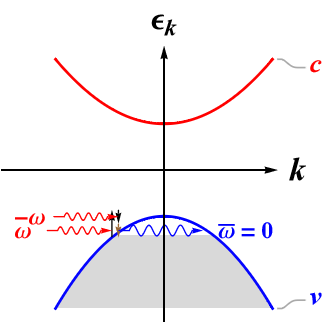}
            \end{minipage}
            & 
            \begin{minipage}{0.192\linewidth}
                \centering
                \includegraphics[width=\linewidth]{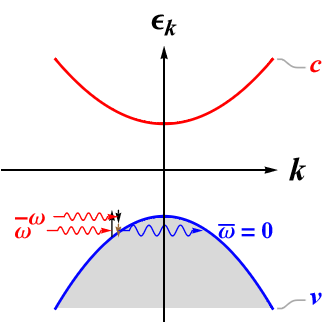}
            \end{minipage}
            \\
            (a) SHG & (b) BPVE & (c) Subgap Response & (d) Metal Transport & (e) Insulator Transport \\[-1em]
        \end{tabular}
    \end{center}
\caption{Nonlinear responses can be subdivided into three main categories: Nonlinear optics consisting of (a) second harmonic generation (SHG) and (b) bulk photovoltaic effect (BPVE), subgap response (c), and nonlinear transport for (d) metal and (e) insulator respectively. In (a-e), red and blue parabolas represent the conduction ($\bm{\mathit{c}}$) and valence ($\bm{\mathit{v}}$) bands, the straight and wavelike arrows represent electronic transitions and corresponding photon absorption/emission (red/blue) processes, while the shaded area represents the filled valence band states. 
}\label{fig1}
\end{figure*}

To set the stage, we can understand the effect of quantum geometry on the charge current response in two ways. Viewed from real space, the response of material to an external electromagnetic field will definitely be affected by the charge distribution, i.e. it depends on the expectation values of various moments of the position operator $\langle \psi | \hat{\mathbf{r}}\dots\hat{\mathbf{r}} | \psi \rangle$. In momentum space,
the perturbation by an electromagnetic field formally couples an external electromagnetic vector potential $\mathbf{A}$ to the non-interacting electronic Hamiltonian: $\hat{H}_0(\mathbf{k}) \rightarrow \hat{H}_0(\mathbf{k} + \frac{e}{\hbar} \mathbf{A})$ by minimal coupling, where $\mathbf{k}$ and $-e$ are the momentum and charge of the electrons.
According to response theory, different orders of charge current response to electromagnetic field $\mathbf{A}$ therefore necessarily involve different orders of momentum derivatives $\partial_{\mathbf{k}}\dots\partial_{\mathbf{k}}\hat{H}_0$, which are related by sum rules to the momentum derivatives of the wavefunction $\langle \psi_{\mathbf{k}} |\partial_{\mathbf{k}}\dots\partial_{\mathbf{k}}| \psi_{\mathbf{k}} \rangle$, thus containing geometric information about the wavefunction in momentum space.
These two arguments, though presented from different perspectives, are fundamentally connected as position and momentum are conjugate variables in quantum mechanics. 
However, the argument above only superficially relates nonlinear response to quantum geometry. To see in detail how the quantum geometry determines the response, a case-by-case classification of the nonlinear response is necessary as different quantum geometric quantities correspond to different nonlinear phenomena. The primary purpose of this review is to elucidate this process and the main outcomes, as summarized in Figs.~\ref{fig0} and~\ref{fig1} as well as  Table~\ref{table1}.

Even though the study of nonlinear responses has a long history dating back to the seventies, the early reviews of the field~\cite{Boyd2003,sturman2021photovoltaic} did not discuss its inherent quantum geometric nature. More recent reviews~\cite{Morimoto2016,Tokura2018,Du2021,Ma2021,Morimoto2023,Nagaosa2022,Dai2023a,liu2024quantum} included these aspects, either focusing solely on  optical responses or the nonlinear Hall effect. 
Here, we aim to highlight how both spatial and temporal (spectral) information form interdependent physical organizing principles, providing the versatile framework required to capture the behavior of intrinsic nonlinear conductivities each for resonant, off-resonant and even semiclassical responses.

\begin{table*}[htbp]
\centering
\renewcommand{\arraystretch}{1.5}
\begin{tabular}{|c|c|c|c|c|c|c|c|c|c|}
\hline
\multicolumn{10}{|c|}{Nonlinear DC Response} \\ \hline
\multirow{2}{*}{\makecell[c]{\textbf{(I)}\\Frequency}} & \multicolumn{4}{c}{$\hbar \omega > E_g$} & \multicolumn{2}{|c}{$\hbar \omega < E_g$} & \multicolumn{3}{|c|}{$\omega \tau \ll 1$} \\ \cline{2-10}
& \multicolumn{4}{c}{Nonlinear Optics} & \multicolumn{2}{|c}{Subgap} & \multicolumn{3}{|c|}{Nonlinear Transport} \\ \hline
\multirow{2}{*}{\makecell[c]{\\ \textbf{(II)}\\Lifetime}} & \multicolumn{2}{c}{$\mathcal{O}(\tau^1)$} & \multicolumn{2}{|c|}{$\mathcal{O}(\tau^0)$} & \makecell[c]{$\frac{\tau_{\mathrm{inter}}}{\tau_{\mathrm{intra}}} \neq 2$\\} & \makecell[c]{$\frac{\tau_{\mathrm{inter}}}{\tau_{\mathrm{intra}}} = 2$\\} & $\mathcal{O}(\tau^2)$ & $\mathcal{O}(\tau^1)$ & $\mathcal{O}(\tau^0)$ \\ \cline{2-10}
& \multicolumn{2}{c}{Injection current} & \multicolumn{2}{|c|}{Shift current} & \makecell[c]{ Weak\\Correlation/\\Localization} & \makecell[c]{ Strong\\Correlation/\\Localization} & \makecell[c]{Nonlinear\\Drude} & \makecell[c]{Berry\\curvature\\dipole} & \makecell[c]{Quantum\\metric\\dipole} \\ \hline
\multirow{2}{*}{\makecell[c]{\\ \textbf{(III)}\\Symmetry}} & $\sigma^{(a,b);c}$ & $\sigma^{[a,b];c}$ & $\sigma^{(a,b);c}$ & $\sigma^{[a,b];c}$ & \multicolumn{2}{c|}{\makecell[c]{\\ \\}} & $\sigma^{(a,b;c)}$ & $\sigma^{(a,b);c}$ &  $\sigma^{(a,b);c}$ \\ \cline{2-5} \cline{8-10}
& $\mathcal{PT}$ & $\mathcal{T}$ & $\mathcal{T}$ & $\mathcal{PT}$ & \multicolumn{2}{c|}{\makecell[c]{$\mathcal{T}$-breaking\\ \\ \\}} & $\mathcal{PT}$ & $\mathcal{T}$ & $\mathcal{PT}$ \\ \hline
\makecell[c]{\textbf{(IV)}\\Phenomenon} & \makecell[c]{Linear\\injection\\
\cite{Sipe2000,Holder2020}} & \makecell[c]{Circular\\injection\\
\cite{Hosur2011,Chang2017,Golub2020,deJuan2017,Avdoshkin2020,Ma2017,Rees2020, Ni2021}} & \makecell[c]{Linear\\shift\\\cite{Wu2017,Osterhoudt2019,Fei2020,Chan2016,Chan2017,Rangel2017,Koenig2017,Zhang2018,Golub2018,Raj2024,Tan2016a,Fregoso2017,Cook2017,Parker2019,Kaplan2022,Chaudhary2022,Watanabe2021b,Watanabe2022,Avdoshkin2024,Alexandradinata2024}} & \makecell[c]{Circular\\shift\\\cite{Zhang2019,Holder2020,Watanabe2021}} & \multicolumn{2}{c|}{\makecell[c]{Subgap\\response\\
\cite{Kaplan2020,Michishita2021,Matsyshyn2023,Kaplan2023}}} & \makecell[c]{Nonreciprocal\\magneto-\\resistance\\(NMR)\\\cite{Sodemann2015,Gao2019,Watanabe2020,Wang2021}} & \makecell[c]{Nonlinear\\Anomalous\\Hall\\(NLAH)\\
\cite{Sodemann2015,Zhang2018a, Matsyshyn2019,Nandy2019,Xu2018a, Ma2019, Kang2019}} & \makecell[c]{Intrinsic\\NMR\\ \& \\NLAH\\
\cite{Gao2014,Liu2021,Kaplan2024,Das2023,jiang2024electrical,zhao2024hybrid,Wang2023,Gao2023}} \\ \hline
\makecell[c]{\\ \textbf{(V)}\\Quantum\\Geometry \\ \\} & \makecell[c]{Quantum\\metric} & \makecell[c]{Berry\\curvature} & \makecell[c]{Symplectic\\Christoffel\\symbol} & \makecell[c]{Christoffel\\symbol of\\1st kind} & \multicolumn{2}{c|}{-} & Translation & Rotation & Distortion \\ \hline
% \multicolumn{10}{|c|}{Quantum Geometry} \\ \hline
\end{tabular}
\caption{ Detailed classifications of nonlinear DC response by perturbation (I) frequency, (II) lifetime, (III) symmetry, together with detailed correspondences between (IV) different nonlinear phenomena and (V) quantum geometry. 
\textbf{(I)} Similar to the overview in Fig.~\ref{fig1}, the possible responses are sorted into three columns according to the excitation frequency .
\textbf{(II)} When a finite quasiparticle lifetime is introduced, injection and shift current can be separated by their lifetime dependence, as can be NLD, BCD and QMD in the transport regime. In the subgap regime between optics and transport, both interband lifetime $\tau_{\mathrm{inter}}$ and intraband lifetime $\tau_{\mathrm{intra}}$ matter. 
\textbf{(III)} The intrinsic symmetry of the response is denoted by round brackets $(\cdot\,,\cdot)$ for symmetric indices, while square brackets $[\cdot\,,\cdot]$ denote antisymmetric spatial components. Time reversal ($\mathcal{T}$) and the combined parity-time reversal symmetry ($\mathcal{PT}$) denote which responses are singled out if present. 
\textbf{(IV)} Different probing methods also help the classification, such as the linear/circular polarization of the perturbation in optics and the longitudinal/Hall response in transport. 
\textbf{(V)}
Recent progress has revealed that each phenomenon can be associated with a geometric property of the underlying Hilbert space.
% \commentYiyang{reference}
}\label{table1}
\end{table*}

\paragraph{Frequency}

The first key factor in distinguishing different types of nonlinear responses is the perturbation frequency $\omega$, which determines the nature of the electronic transitions involved. Here we only focus on monochromatic perturbation and 2nd order response for simplicity. According to the perturbation frequency $\omega$ and its relation with the band gap $E_g$ and electron lifetime $\tau$, nonlinear response can be classified into three main regimes. As depicted in Fig. \ref{fig1} and row \textbf{(I)} of Table \ref{table1}, these include firstly nonlinear optics as shown in Fig. \ref{fig1} (a, b) related to resonant excitations and photocurrent generation, secondly subgap response as shown in Fig. \ref{fig1} (c) where only off-resonant excitations contribute, and finally in nonlinear transport as shown in Fig. \ref{fig1} (d, e), in which case the quasiparticle decoherence rate $\tau^{-1}$ is larger than the frequency $\omega$.

In nonlinear optics, two kinds of responses arise in 2nd order, second harmonic generation (SHG) which generates an alternating current (AC) with frequency $2\omega$ twice the perturbation frequency (Fig.~\ref{fig1}a), and the bulk photovoltaic effect (BPVE, also called bulk photogalvanic effect, BPGE) which describes a direct current (DC) generation (Fig.~\ref{fig1}b). The BPVE can furthermore be separated into the so-called injection current whose magnitude grows with time, and the so-called shift current whose magnitude remains finite.
Quite amazingly, already the original resonant response theory of the intrinsic BPVE~\cite{Belinicher1980,vonBaltz1981,Boyd2003,Kraut1979,vonBaltz1981,Aversa1995,Sipe2000} correctly captured both injection and shift contributions for time-reversal symmetric systems~\cite{Zhang2019}. However, the early interpretation of the shift current as a transient, induced polarization was influenced majorly by the polar semiconductors which were of interest at the time. Therefore, it came as a surprise when topological materials were later found to have both a giant SHG~\cite{Sun2019,Ma2019a} as well as BVPE~\cite{Wu2017,Osterhoudt2019,Fei2020}, leading to a refined understanding of the phenomenology~\cite{Chan2016,Chan2017,Rangel2017,Koenig2017,Zhang2018,Golub2018,Raj2024} using concepts from quantum geometry~\cite{Tan2016a,Fregoso2017,Cook2017,Parker2019,Holder2020,Kaplan2022,Chaudhary2022,Watanabe2021b,Watanabe2022,Avdoshkin2024,Alexandradinata2024}.
Similar conclusions have been reached
for the injection current driven by circular polarized light~\cite{Hosur2011,Chang2017,Golub2020}, most notably leading to the prediction of a 
quantized circular photogalvanic effect for non-interacting two-band systems~\cite{deJuan2017,Avdoshkin2020}. In addition to the quantized circular photogalvanic effect, recent studies have also proposed a quantized shift current in multigap topological phases~\cite{jankowski2024quantized}. Several experiments have been able to resolve the BPVE for circular polarized light~\cite{Ma2017,Rees2020, Ni2021}.
Recent progress has shown that the quantum geometric origin is inherent in all nonlinear optical responses~\cite{Ahn2020,Ahn2022,Kaplan2023,Avdoshkin2024}. 
We note that the magnitude of the nonlinear optical response function depends sensitively on the choice of model and the resulting wave functions~\cite{IbanezAzpiroz2018,Zhu2024}.

Regarding the geometric perspective, the injection current can be related to the quantum metric and Berry curvature of the system, which measure the length and area in the wavefunction space as shown in Fig. \ref{fig0} (a) and (b). Similarly, the shift current has been related to a higher order quantum geometric quantity called quantum connection, which describes the change in tangential vectors of the wavefunction space as shown in Fig. \ref{fig0} (c). 

Quantum geometry also manifest in nonlinear transport when the perturbation frequency $\omega$ approaches zero. 
The theory of semiclassical transport at nonlinear order is well developed~\cite{Sundaram1999,Culcer2005,Chang2008,Gao2019,Bhalla2023}.
By their dependencies on lifetime $\tau$, nonlinear transport can also be categorized into three different parts, including the nonlinear Drude (NLD) term which scales like $\tau^2$~\cite{Sodemann2015,Gao2019,Watanabe2020,Wang2021}, the Berry curvature dipole (BCD) which is linear in $\tau$~\cite{Sodemann2015,Zhang2018a, Matsyshyn2019,Nandy2019}, and the (band-normalized) quantum metric dipole (QMD) term which is independent of $\tau$~\cite{Gao2014,Liu2021,Kaplan2024,Das2023}. The purely dispersion-related NLD term describes the rigid-body-like movement of wavepacket, while QMD and BCD terms, which originate from quantum metric and Berry curvature, describe the deformation and self-rotation of the wavepacket as shown in Fig. \ref{fig0} (d) and (e). These geometric responses have been observed to contribute to the intrinsic nonreciprocal magnetoresistance~\cite{Wang2023,Gao2023} and the nonlinear anomalous Hall effect~\cite{Xu2018a, Ma2019, Kang2019}.
All these three contributions are Fermi surface responses, i.e. they are only nonzero in metals (Fig.~\ref{fig1}d). However, just as the quantum anomalous Hall (QAH) at linear order can arise from a completely filled band, similarly, there exists Fermi sea contribution to the nonlinear Hall effect in insulators~\cite{Kaplan2023a} (Fig. \ref{fig1}e).

Between these two regimes exists the so-called subgap regime (Fig. \ref{fig1}c), where the perturbation frequency $\omega$ is smaller than the band gap $E_g$ so that resonant optical transition can not happen~\cite{Deyo2009,Sturman1992,Sundaram1999}. Although off-resonant, the DC generation does not necessarily vanish entirely inside the gap of a magnetic insulator~\cite{Kaplan2020,Michishita2021,Matsyshyn2023,Kaplan2023}. Instead, it can pertain its magnitude for an limited window of subgap frequencies, where the quasiparticle lifetimes play an important role. 

\begin{figure*}[htbp]
    \begin{center}
        \begin{tabular}{cccc}
            \multicolumn{2}{c}{\textbf{Symmetry of conductivity tensor}} & \multicolumn{2}{c}{\textbf{Symmetry of the material}} \\[1em]
            \begin{minipage}{0.272\linewidth}
                \centering
                \includegraphics[width=\linewidth]{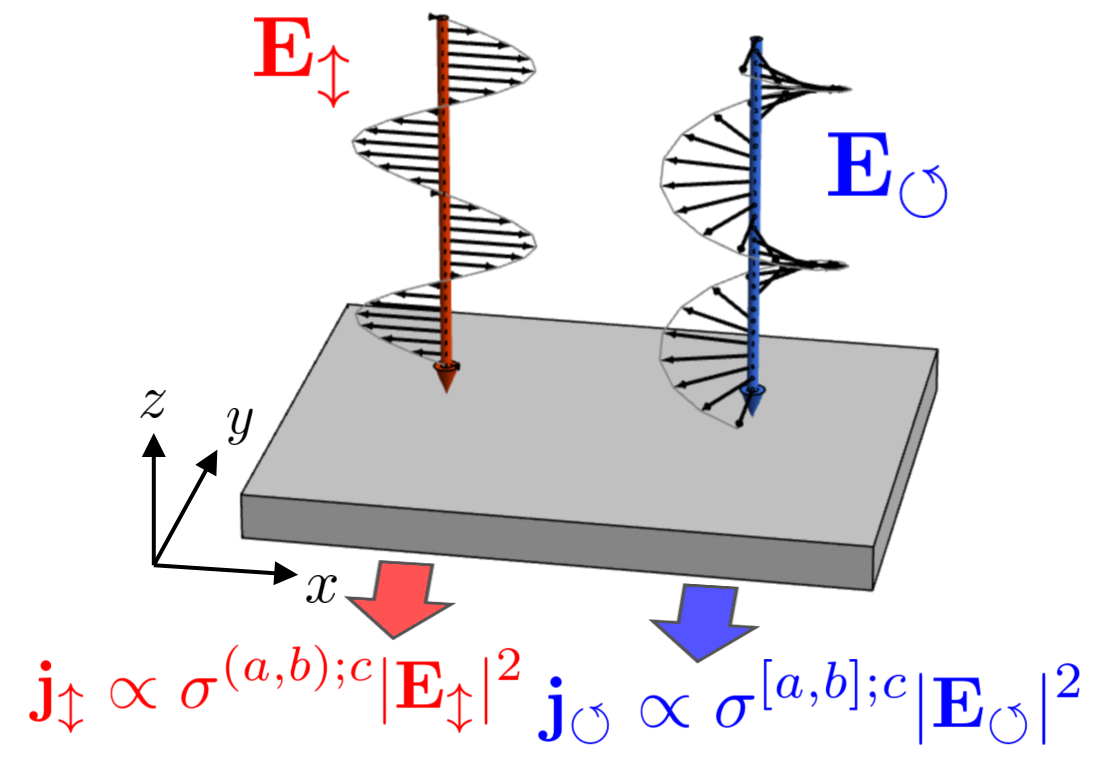}
            \end{minipage}
            & 
            \begin{minipage}{0.212\linewidth}
                \centering
                \includegraphics[width=\linewidth]{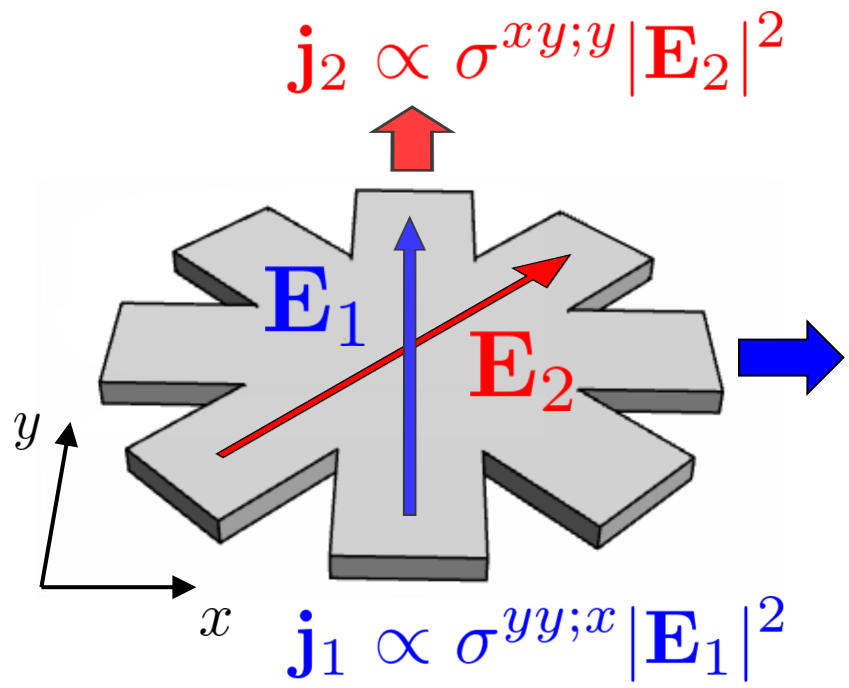}
            \end{minipage}
            &
            \begin{minipage}{0.24\linewidth}
                \centering
                \includegraphics[width=\linewidth]{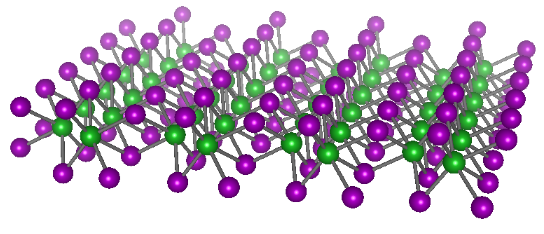}
            \end{minipage}
            & 
            \begin{minipage}{0.24\linewidth}
                \centering
                \includegraphics[width=\linewidth]{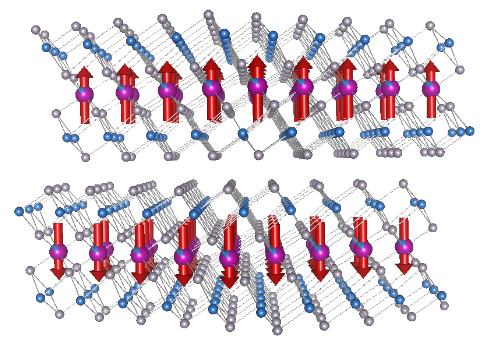}
            \end{minipage}
            \\[5.5em]
            (a) Linear \& Circular BPVE
 & (b) Hall bar Transport & (c) $\mathcal{T}$-symmetric Material & (d) $\mathcal{PT}$-symmetric Material \\[-1em]
        \end{tabular}
    \end{center}
\caption{Overview for the separation of the nonlinear response based on symmetry. (a) In the optical regime, linearly ($\mathbf{E}_{\updownarrow}$) and circularly ($\mathbf{E}_{\circlearrowleft}$) can distinguish responses which are symmetric and antisymmetric with respect to the spatial indices of electric field. (b) In the transport regime, a measurement of different currents ($\mathbf{j}_{1,2}$) with different applied fields ($\mathbf{E}_{1,2}$) allows the investigation of both fully transverse ($\sigma^{yy;x}$) and off-axis directions ($\sigma^{xy;y}$) conductivities, which helps separate the NLD contribution that is fully permutational symmetric ($\sigma^{(a,b;c)}$). It's also possible to distinguish different geometric contributions using materials with different symmetry, e.g. (c) time-reversal ($\mathcal{T}$-) symmetric material WTe${}_2$ and (d) the parity-time inversion ($\mathcal{PT}$-) symmetric material MnBi${}_2$Te${}_4$, the red arrows represents magnetic moments of Mn, showing an anti-ferromagnetic order.
}\label{symmetry separation}
\end{figure*}

\paragraph{Lifetime}

Besides perturbation frequency, the role of finite lifetime is also important in separating different parts of nonlinear responses and unifying three regimes in one unified formalism, as shown in the row \textbf{(II)} of Table~\ref{table1}.
In nonlinear optics, the injection part will not grow indefinitely with time in real materials due to dissipation while the shift part does not explicitly depends on it. The time scale for such dissipation is approximately the quasiparticle lifetime $\tau$.
Based on the lifetime dependence,  injection ($\propto \tau^1$) and shift ($\propto \tau^0$) current can be separated~\cite{Sipe2000,Holder2020}, as noted in the ``Nonlinear Optics'' column of Table~\ref{table1}. 
However, this seemingly simple classification  does not fully account for the fact that nonlinear responses require the introduction of at least two quasiparticle lifetimes, the interband lifetime $\tau_{\mathrm{inter}}$ and the intraband lifetime $\tau_{\mathrm{intra}}$, that play a significant role in regulating both the optically resonant as well as the transport response~\cite{Mikhailov2014,Mikhailov2016,Cheng2014,Cheng2015,Cheng2017,Ventura2017,Passos2018,Kaplan2023}. It is therefore no surprise that these lifetimes compete in the subgap regime, where their ratio $\frac{\tau_{\mathrm{inter}}}{\tau_{\mathrm{intra}}}$ can indicate the difference between strong and weak correlation effects, as indicated in the ``Subgap'' column of Table~\ref{table1}.

The introduction of two lifetimes helps to establish a connection across optical, subgap and transport regime under a unified Feynman diagrammatic formalism.
Historically, the above three categories have been discussed under very different perspectives, even though they belong to the same order nonlinear processes. For instance, optical responses are usually derived in the reduced density matrix formalism~\cite{Sipe2000,Ventura2017}, while transport responses are typically discussed in a semiclassical approach~\cite{Gao2019}. In these latter approaches, only a single phenomenological lifetime $\tau$ is introduced, which has a vague physical origin and does not distinguish between interband and intraband processes. Yet, as shown in Fig. \ref{fig3}, using a diagrammatic approach, interband and intraband transitions are clearly separated and the relaxation times can be connected to a physical mechanism via the self-energies $\Sigma=i/(2\tau)$ of the quasiparticle Green's functions~\cite{Holder2020}. 
In such a diagrammatic formalism, the subgap response likewise arises naturally as an incomplete cancellation between injection and shift current. 
This incompletely canceled conductivity can therefore serve as a probe of the quasiparticle self-energies in the system,
and shed light on the 
microscopic scattering mechanisms. 

Finally, from the same diagrams the transport regime is accessible by a combined limit $\omega\rightarrow 0$  and $\frac{\tau_{\mathrm{inter}}}{\tau_{\mathrm{intra}}}\rightarrow2$. Here, the lifetime ratio being 2 might seem somewhat special. However, it is merely implementing the requirement that the quasiparticle cannot decay (and thus escape) while being in the excited state, which enforces charge conservation in the adiabatic case. 
In terms of the underlying electronic processes, it is not surprising that both intraband and interband transitions are important in all nonlinear response regimes: DC current generation (like in the bulk photovoltaic effect) always arises from intraband transitions as shown in Fig. \ref{fig3} (a) and (b), while on the flip side even in the transport limit virtual (i.e. off-resonant) excitations continue to contribute. This further highlights the importance of keeping track of the microscopic quasiparticle properties.

\paragraph{Symmetry}

Contributions of these quantum geometric quantities can be further separated or singled out by their symmetry dependencies since different geometric quantities and their corresponding nonlinear effects possess dissimilar symmetries as shown from the row \textbf{(III)} to \textbf{(V)} of Table~\ref{table1}.
As we point out in the row \textbf{(III)} of Table~\ref{table1} and Fig.~\ref{symmetry separation}, two kinds of symmetry properties dictate the behavior of nonlinear phenomena and their corresponding quantum geometric quantities, firstly the symmetry of conductivity tensor and secondly the symmetry of the material.

The symmetry of conductivity tensor originates from the permutation symmetry between spatial indices. In Table.~\ref{table1} and throughout this review, we use $\sigma^{(a,b);c}$ and $\sigma^{[a,b];c}$ to represent the conductivity tensors that are symmetric and antisymmetric after permutation of electric field spatial indices $a$ and $b$. Likewise, the symbol $\sigma^{(a,b;c)}$ will indicate a fully symmetric conductivity tensor under any permutation among both electric field spatial indices $a$ and $b$ as well as the current spatial index $c$ (cf. NLD column of Table.~\ref{table1}).

The symmetry of the conductivity tensor can be exploited by the spatial configuration of perturbation to separate different quantum geometric responses. In the optical regime, this is achieved by perturbing the system with linearly or circularly polarized light, which detects the symmetric and anti-symmetric part of the optical conductivity tensor, respectively [cf. Fig.~\ref{symmetry separation} (a)]. This allows, for instance, separating a quantum metric contribution from the Berry curvature in the injection current, where the former is spatial indices symmetric while the latter is antisymmetric. In the transport regime, this is achieved by measuring current at different Hall bar contacts when applying distinct electric field [cf. Fig.~\ref{symmetry separation} (b)], which allows separating NLD from BCD and QMD terms that are not symmetric under full spatial indices permutation~\cite{Zhang2023a}. It is worth mentioning that the difference of permutation symmetry is actually a manifestation of the axial-gravitational anomaly~\cite{Holder2021a}, a close relative of the chiral anomaly. This may provide an promising avenue to explore a quantum anomaly which is otherwise rather difficult to access.
Finally, also at nonlinear order surface contributions can exist which often enhance the nonlinear photocurrents~\cite{Kim2017,Chang2020,Steiner2022}.

Apart from the symmetry of conductivity tensor, it is also possible to utilize the symmetry of materials to single out certain quantum geometric contribution (cf. Fig.~\ref{symmetry separation} c,d). 
For the second order charge response, the most important symmetries are the magnetic point group symmetry and the space-time symmetry.
This encompasses, in particular, inversion ($\mathcal{P}$) and time-reversal ($\mathcal{T}$) as well as the combined parity-time reversal symmetry ($\mathcal{PT}$).

Generally, all 2nd-order current responses require breaking of $\mathcal{P}$. This is because the current $\mathbf{j}$ changes sign while the electric field, which enters quadratically, does not.
In a $\mathcal{P}$-breaking system, we can further separate different geometric contributions in $\mathcal{T}$- or $\mathcal{PT}$-symmetric materials.
In nonlinear optics, the injection current growing with time and the shift current with constant amplitude possess distinct $\mathcal{T}$- and $\mathcal{PT}$- symmetries, and so do the linear and circular polarizations of the electromagnetic field. $\mathcal{T}$- and $\mathcal{PT}$- symmetry transformation can also be used to distinguish different geometric quantities~\cite{Holder2020,Watanabe2021}; for instance, the Berry curvature vanishes in a $\mathcal{PT}$-symmetric system.
Similar symmetry constraints appear in nonlinear transport, where $\mathcal{PT}$ enforces a vanishing BCD contribution, such that only the NLD and QMD terms survive. On the other hand, $\mathcal{T}$ enforces vanishing NLD and QMD terms, leaving only the nonlinear anomalous Hall (NLAH) effect nonzero. It is this NLAH which famously can lead to an anomalous Hall signal in a time-reversal symmetric system~\cite{Ma2019}, which helped ignite much of the recent developments regarding nonlinear responses.

Additionally, one can consider more intricate space-time symmetries such as rotational symmetries ($\mathcal{C}_n$) and their combination with $\mathcal{T}$. Although all 2nd-order conductivity tensors share the same rotational transformation properties, it is still possible to exclude a pure Hall contribution based on the presence of spatial rotational symmetries such as $\mathcal{C}_{3z}$. 
It is also possible to separate different responses by their combined space-time symmetries. For example, the combination $\mathcal{C}_n\mathcal{T}$  usually results in a separation of the response in different directions. Namely, $\mathcal{C}_{2z}\mathcal{T}$ acts as $\mathcal{T}$ in the $z$-direction while also acting like $\mathcal{PT}$ in the $x$-$y$ plane, therefore responses that respect $\mathcal{PT}$ survive in the $x$-$y$ plane while those respecting $\mathcal{T}$ survive in the $z$-direction. We also point out here that for the most general system with no specific symmetry, it's possible to contain contributions from all geometric quantities, then it's still possible to utilize the permutational symmetry of the corresponding conductivity tensor to conduct relevant measurement to distinguish different contributions, as shown in Fig.~\ref{symmetry separation} (a) and (b).

\paragraph{Experiments}

The nonlinear Hall effect has been studied experimentally in transition metal dichalcogenides (TMDs) like WTe$_2$ and MoTe$_2$~\cite{Du2018a,Ma2019,Kang2019,Tiwari2021}, and in the semimetals
BaMnSb$_2$~\cite{Min2023} and TaIrTe$_4$~\cite{Kumar2021a}. These and other Weyl semimetals like TaAs, CoSi and RhSi can likewise exhibit a large nonlinear optical response~\cite{Wu2017,Ma2017,Rees2020, Ni2021,Osterhoudt2019,Ma2019a}. 
Another promising platform are even-numbered layers of CrI$_3$, which constitute a PT-symmetric band structure~\cite{Sun2019,Zhang2019}.
Enhanced nonlinear responses have also been reported van-der-Waals (vdWs) heterostructures stacked at a finite twist angle like 
twisted multilayer graphene~\cite{Duan2022,He2022}, 
twisted WSe$_2$~\cite{Huang2022}
as well as twisted WTe$_2$~\cite{He2021a}.
Bi$_2$Se$_3$ is a well-studied topological insulator for which second harmonic generation has been mapped out~\cite{He2021}.
Finally, experiments in the weakly doped semiconductor MnBi$_2$Te$_4$ have seen signatures of the transverse and longitudinal contributions of the quantum metric dipole~\cite{Wang2023,Gao2023}, and a BPVE has also been reported~\cite{Fei2020}.

\paragraph{Extrinsic scattering}

Despite the fact that the emerging physical picture regarding nonlinear responses is rather well-understood, we emphasize that this phenomenology does not capture the entire range of nonlinear current generation. The mechanisms covered by the diagrammatic approach mentioned above are usually referred to as \emph{intrinsic} nonlinear response formalism, which includes the effect of disorder, phonons and other relaxation mechanisms only indirectly in the form of the self-energy of the one-particle Green's functions.
However, the underlying physical processes can lead to additional contributions to the current, which are usually termed \emph{extrinsic} mechanisms. 
Extrinsic AC-effects are related to the current carrying steady-state that develops in the presence of additional relaxation channels, for example via skew scattering~\cite{Isobe2020} or phonon creation~\cite{Sturman1992,Sturman2019,Zhu2024}. 
In the dc-limit, the major sources of extrinsic current generation are resonant impurity scattering and related momentum-relaxing processes, which enter as a renormalization of velocity vertices or even higher-order contributions in the diagrammatic expansion. Such effects beyond lifetime corrections are known to contribute prominently to the anomalous Hall effect at linear order~\cite{Nagaosa2010}. Similar extrinsic mechanisms are to be expected and have been discussed in the nonlinear regime~\cite{Xiao2019,Papaj2019,Du2019,Nandy2019,Koenig2021,Du2021a,Atencia2022}. Most importantly, in some cases they might be comparable or even larger than  the intrinsic contributions~\cite{Du2021}.
Although extrinsic and intrinsic mechanisms generally coexist in most systems, it is often possible to separate, suppress, or even exclude extrinsic contributions through certain symmetries~\cite{Kaplan2024} or measurement methods~\cite{Wang2023,Gao2023}.

\paragraph{Structure of the review}

The relationship between quantum geometry and nonlinear responses extends beyond formal associations, revealing a rich collection of physical phenomena where quantum geometric quantities, especially those other than Berry curvature, directly play a role. The recent establishment of this perspective thus makes nonlinear responses promising probes of quantum geometric properties. Additionally, the insights from the quantum geometry can be leveraged to better control for example photosensing in nonlinear optics or rectification in nonlinear transport devices.

\begin{table}[tbp]
    \centering
    \renewcommand{\arraystretch}{1.2} % Adjust row spacing
    \begin{tabular}{cc}
        \hline\\[-1.2em]
        \hline\\[-1.2em]
        \textbf{Abbr.} & \textbf{Meaning} \\ 
        \hline
        SHG  & Second Harmonic Generation \\ 
        BPVE & Bulk Photovoltaic Effect \\
        BPGE & Bulk Photogalvanic Effect \\
        AC(ac)   & Alternating Current \\
        DC(dc)   & Direct Current \\
        NLD  & Nonlinear Drude \\ 
        BCD  & Berry Curvature Dipole \\ 
        QMD  & Quantum Metric Dipole \\ 
        QAH  & Quantum Anomalous Hall \\ 
        NLAH  & Nonlinear Anomalous Hall \\
        NMR & Nonreciprocal Magneto-Resistance\\
        NQAHE & Nonlinear Quantum Anomalous Hall Effect \\
        $\mathcal{T}$  & Time-reversal \\ 
        $\mathcal{P}$   & Parity (spatial inversion) \\ 
        $\mathcal{PT}$  & Parity-Time inversion \\ 
        $\mathcal{C}_{nz}$ & $n$-fold rotational symmetry with respect to $z$-axis \\
        $\mathcal{M}_{z}$ & Mirror symmetry with respect to $z$-axis \\
        TMDs & Transition Metal Dichalcogenides \\
        vdWs & van-der-Waals \\ 
        QGT & Quantum Geometric Tensor \\
        QC & Quantum Connection \\
        BZ & Brillouin Zone \\
        \hline\\[-1.2em]
        \hline
    \end{tabular}
    \caption{List of abbreviations used in the text.}
    \label{tab:abbreviations}
\end{table}

% Outline
This review is structured as follows.
In Sec. \ref{sec2}, we briefly review the perturbation theory underlying nonlinear responses, elucidating three major regimes: nonlinear optics, subgap response and nonlinear transport and the corresponding physical process using Feynman diagrammatics.
In Sec. \ref{sec3}, we discuss the BPVE in nonlinear optics, illuminating the direct correspondence between quantum geometry and nonlinear responses. In particular, we show how the physical observables including the injection and shift current follow from the quantum geometry. The symmetry property shared by quantum geometry and the corresponding phenomena is also discussed.
In Sec. \ref{sec4}, we give an exposition of the subgap regime, a much less studied parameter range, focusing on lifetime effects and the 2-lifetime prescription. We clarify how the subgap response arises from the incomplete cancellation between injection and shift current in off-resonant optics, which highlights the subtle but important interplay of intra- and interband processes in all nonlinear charge responses.
Sec. \ref{sec5} is dedicated to the nonlinear transport regime, illustrating the geometric interpretation of NLD, BCD, and QMD as translation, self-rotation, and distortion of semiclassical wavepackets. 
We further discuss the strong constraints on lifetimes and symmetry properties, in particular how the transport limit arises when $\tau_{\mathrm{inter}}=2\tau_{\mathrm{intra}}$,  a consequence of charge conservation.
The resulting intrinsic nonlinear response formalism not only unifies different regimes in nonlinear responses, but also predicts new effects such as a nonlinear Hall effect in magnetic insulators.
Finally, in Sec. \ref{sec7}, we provide a summary of the key points discussed and offer an outlook on future research directions in the field of nonlinear responses and quantum geometry.
A number of abbreviations are used for frequently recurring terms, which are summarized in Table~\ref{tab:abbreviations} for convenience.

\section{Basic Concepts}\label{sec2}

\subsection{Quantum geometric quantities}\label{sec2A}
For a set of Bloch-periodic eigenfunctions  $| u_n(\mathbf{k}) \rangle$ with band index $n$, it is possible to study the geometric properties of the complex projective space spanned by the momentum $\mathbf{k}$ in the Brillouin zone. To this end, one can define the \emph{quantum geometric tensor} (QGT)~\cite{Provost1980,Kolodrubetz2017,bouhon2023quantum},
\begin{align}
    Q^{ab}_n(\mathbf{k})&=\langle \partial_{k_a} u_n(\mathbf{k})|\bigl[1-|u_n(\mathbf{k})\rangle\langle u_n(\mathbf{k})|\bigr]
    |\partial_{k_b} u_n(\mathbf{k}) \rangle\notag\\
    &=g^{ab}_n(\mathbf{k})-\tfrac{i}{2}
    \Omega^{ab}_n(\mathbf{k}),
    % &= \langle PxP PyP \rangle = \langle \{PxP,PyP\} \rangle + \langle [PxP,PyP] \rangle
    \label{eq:def_Q}
\end{align}
whose imaginary part is the Berry curvature, whereas the real part is called \emph{quantum metric}.
In the complex manifold spanned by a given band $n$, $g^{ab}_n$ measures distances, while $\Omega^{ab}_n$ quantifies local sources of the Berry phase.
It's also common to encounter vector form of Berry curvature $\mathbf{\Omega}_n$ which is related to the antisymmetric tensor form of $\Omega^{ab}_n$ by $(\mathbf{\Omega}_n)^a \equiv \frac{1}{2} \epsilon^{abc} \Omega_n^{bc}$ where $\epsilon^{abc}$ is the Levi-Civita symbol.

One can also define the $Q^{ab}_{M}(\mathbf{k})$ for manifolds $M$ containing multiple bands, which is in general not the sum of Eq.~\eqref{eq:def_Q}~\cite{Mera2022}. Of particular interest is the QGT of the ground state, a fundamental property of an bulk solid. 
For the ground state, the QGT has an intuitive interpretation as the second moment of the position operator, $\langle(\hat X^a-\langle \hat X^a\rangle)(\hat X^b-\langle \hat X^b\rangle)\rangle$.
Denoting the projection to the target manifold by $\hat P$ and momentum derivatives as $\partial_{k_a}\hat P=\partial_a\hat P$, the QGT tensor can be written very compactly as
\begin{align}
    Q^{ab}&=\mathrm{tr}[\hat P(\partial_{a}\hat P)(\partial_{b}\hat P)].
\end{align}
Based on this form, it is possible to generalize the notion of quasi-local geometric objects, with the canonical next order term being the quantum connection (QC)~\cite{Avdoshkin2023}
\begin{align}
    Q^{a;bc}&=\mathrm{tr}[\hat P(\partial_{a}\hat P)(\partial_{b}\partial_{c}\hat P)].
\end{align}
The imaginary part of the quantum connection is related to the third cumulant $\langle \hat X^3\rangle$.
In optical responses, it further becomes necessary to define multistate (i.~e. multiband) variants of the QGT and QC, which read respectively~\cite{Ahn2022,Mitscherling2024},
\begin{align}
    Q^{ab}_{mn}&=\mathrm{tr}[\hat P_n(\partial_{a}\hat P_m)(\partial_{b}\hat P_n)],\\
    Q^{a;bc}_{mn}&=\mathrm{tr}\bigl[[\hat P_n(\partial_{a}\hat P_m)[(\partial_{b}\partial_{c}\hat P_n)+(\partial_{b}\hat P_m)(\partial_{c}\hat P_n)]\bigr].
\end{align}

Similar to QGT, QC can also be separated into an symmetric $Q_{mn}^{a;(b,c)}$ part and anti-symmetric part $Q_{mn}^{a;[b,c]}$, which are termed ``Symplectic Christoffel symbol'' and ``Christoffel symbol of 1st kind'' respectively, as shown in Table~\ref{table1}.

A big driving force of the recent developments regarding nonlinear responses is the fact that sometimes the matrix elements of response functions simplify to the QGT or the QC. Explaining the physical origin behind this observation is the main purpose of the remainder of this review.

\subsection{General nonlinear response theory}\label{sec2B}

The basic concepts in nonlinear response theory have been reviewed before, see for example~\cite{Boyd2003}. This section mostly introduces the associated notation, focusing on the nonlinear response induced by a monochromatic perturbation and identify three primary regimes: nonlinear optics, subgap responses, and nonlinear transports.

If system and perturbation both exhibit time translation symmetry, the respective response functions also adhere to it. Taking the electrical conductivity $\sigma$ as an example, the response function is given by
\begin{equation}
\sigma(t;t_1,\dots,t_n) = \sigma(t + \Delta t;t_1 + \Delta t,\dots,t_n + \Delta t)
\end{equation}
where $\{t_i\}_{i=1}^{n}$ are the times at which perturbations occur, which contribute to the induced response at time $t$, while $\Delta t$ represents an arbitrary time translation. 

Therefore, after a Fourier transform to the frequency domain, the response functions is proportional to a delta funtion, i.e.

\begin{equation}
\sigma(\bar\omega;\omega_1,\omega_2,\dots,\omega_n)  \propto \delta ( \bar\omega - \sum_{j=1}^{n} \omega_j )
\end{equation}
where $\{\omega_i\}_{i=1}^{n}$ are the frequencies contained in the perturbation that contribute to the response frequency $\bar{\omega}$. This is known as the frequency sum rule of $n$-th order nonlinear responses. We point out that the limit $\bar\omega\rightarrow 0$ is subtle when the quasiparticle lifetime $\tau$ is finite~\cite{Holder2020,deJuan2020}, resulting non-commuting limits between 
$\bar\omega\rightarrow 0$, $\tau\rightarrow\infty$ and $\Delta t\rightarrow\infty$.

\begin{figure*}[htbp]
\includegraphics[width=0.92\linewidth]{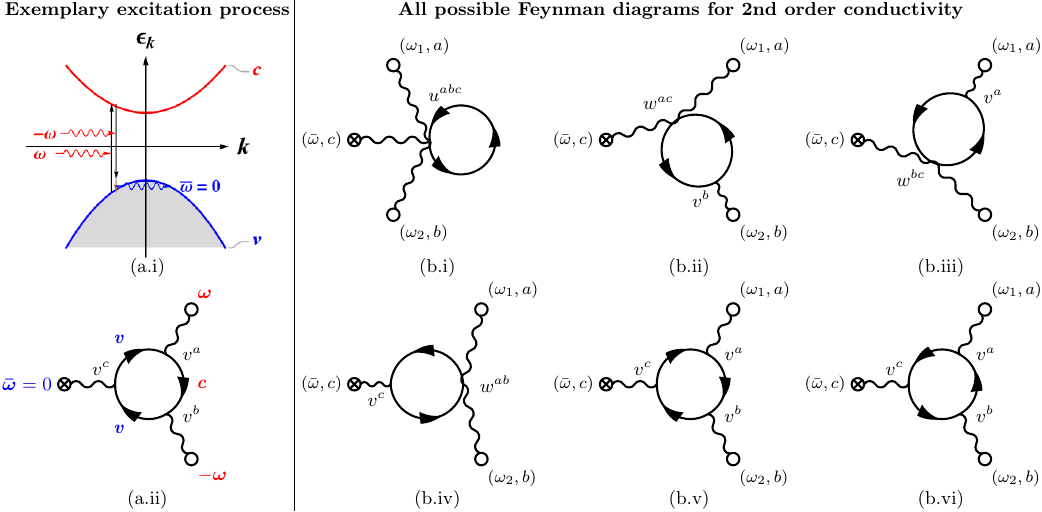}
\vspace*{-.5\baselineskip}
\caption{ (a) Typical 2nd order DC current generation process in band systems (a.i) with corresponding Feynman diagram (a.ii). An electron in the valence ($\bm{\mathit{v}}$) band absorbs a photon with frequency $\omega$ from vertex $v^a$ together with an interband transition to conduction ($\bm{\mathit{c}}$) band, then it absorbs a photon with frequency $-\omega$ from vertex $v^b$ together with an interband transition to $\bm{\mathit{v}}$ band, then it generates DC current with frequency $\bar{\omega}=0$ from vertex $v^c$ with an intraband transition this time. (b) all Feynman diagrams contributes to nonlinear conductivity $\sigma^{ab;c}(\bar{\omega};\omega_1,\omega_2)$. 
}\label{fig3}
\end{figure*}

\subsection{Perturbation theory and Feynman diagrams}\label{sec2C}
The microscopic photon absorption and emission processes occurring during a nonlinear perturbation can be effectively represented using Feynman diagrams~\cite{Parker2019,Holder2020}. 

When electromagnetic field is introduced to a crystalline system with unperturbed Bloch Hamiltonian $\hat{H}_0(\mathbf{k})$, the perturbed Bloch Hamiltonian by minimal coupling becomes $\hat{H}_0(\mathbf{k} + \frac{e}{\hbar}\mathbf{A})$, where $\mathbf{A}$ is the vector potential of the electromagnetic field. 
For a spatially homogeneous electric field, we can adopt a time-dependent yet spatially homogeneous vector potential $\mathbf{A}(t)$ such that $\mathbf{E} = -\partial \mathbf{A}/\partial t$. This perturbation method is known as the ``velocity gauge'' because it introduces the electric field through the vector potential. 

It is worth noting that another approach, known as the ``length gauge''~\cite{Sipe2000}, introduces the external electric field via a spatially inhomogeneous perturbation $\hat{H}' = -e\mathbf{E} \cdot \hat{\mathbf{r}}$. Although the ``length gauge'' and ``velocity gauge'' are equivalent up to a gauge transformation of the single-particle basis~\cite{Ventura2017,Passos2018}, the ``length gauge'' breaks the lattice translation symmetry of the system. This symmetry breaking necessitates the use of a covariant derivative in momentum space to represent the effect of the coordinate operator $\hat{\mathbf{r}}$.

In the velocity gauge
the perturbed Bloch Hamiltonian yields,
\begin{equation}
\hat{H}_0(\mathbf{k} + \tfrac{e}{\hbar} \mathbf{A}) = \hat{H}_0(\mathbf{k}) + \tfrac{e}{\hbar}\hat{v}^a A^a + \left( \frac{e}{\hbar}\right)^2 \frac{\hat{w}^{ab}}{2!} A^a A^b + \dots
\end{equation}

Here, $\hat{v}^a \equiv \frac{\partial \hat{H}_0}{\partial k_a}$, $\hat{w}^{ab} \equiv \frac{\partial^2 \hat{H}_0}{\partial k_a \partial k_b}$, etc. serve as coefficients in perturbative expansions, where the Roman subscripts ($a,b,c$) represent the spatial indices ($x,y,z$) and the Einstein summation convention is assumed. These coefficients can also be regarded as electron-photon scattering vertices in Feynman diagrams, representing the physical processes occurring during a nonlinear response. 
These vertices are intrinsic properties of a crystalline system and encapsulate quantum geometric information, which will be reflected in the nonlinear responses.

To understand how these vertices take part in the electromagnetic transitions, we consider the time evolution of an unperturbed quasiparticle eigenstate $|\psi_n(\mathbf{k})\rangle$, where $n$ is the band index. The time evolution satisfies the time-dependent Schrodinger equation:
\begin{equation}
i \hbar \partial_t |\psi_n \rangle = \left[ \hat{H}_0 + \frac{e}{\hbar} \hat{v}^a A^a + \left( \frac{e}{\hbar}\right)^2 \frac{\hat{w}^{ab}}{2!} A^a A^b  + \dots \right] |\psi_n \rangle
\end{equation}
The first term, $\hat{H}_0 |\psi_n \rangle = \hbar\varepsilon_n |\psi_n \rangle$, results in a dynamical phase $e^{-i \varepsilon_n t}$ of the normal propagating quasiparticle eigenstate and will not contribute a transition to other eigenstates. 

On the other hand, terms involving $\mathbf{A}$ can be interpreted as
a one-photon absorption (emission) processes, accompanied by the transition of an occupied quasiparticle state into an unoccupied state.
Such a process is easy to identify in Feynman diagrams. For instance, a one-photon absorption process (photon with frequency $\omega$), which induces a transition from valence band to conduction band in Fig. \ref{fig3} (a.i) can be represented by a scattering across a one-photon vertex in Feynman diagram as shown by the $\hat{v}^b$ vertex in Fig. \ref{fig3} (a.ii), where the subscript indicates the contribution from the $b$-direction component of the electric field $\mathbf{E}$ (which shares the same direction with the vector potential $\mathbf{A}$).

\begin{table}[tbp]
\includegraphics[width=0.91\columnwidth]{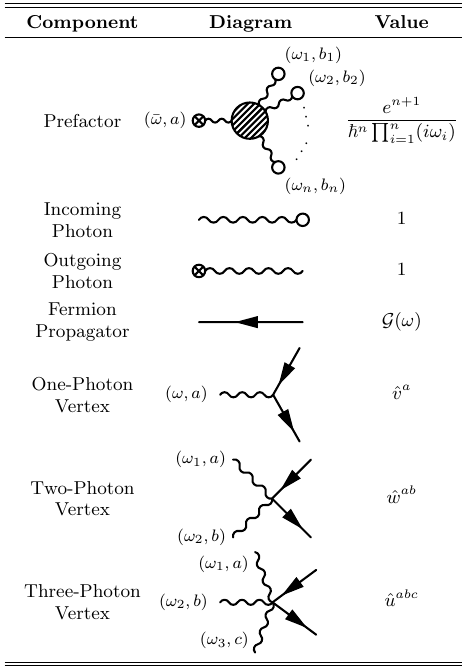}
\caption{Feynman rules and definition of symbols.}\label{tab:feynman_rules}
\end{table}

Similarly, we can interpret the higher-order couplings as multi-photon absorption/emission processes accompanied by a quasiparticle transition. For example, the multi-photon processes $\hat{u}^{abc}$ and $\hat{w}^{ab}$ are depicted in Fig. \ref{fig3} (b.i) and (b.iv). It is worth noting that multi-photon absorption/emission happens simultaneously, in contrast to the sequential one-photon processes depicted for example in Fig. \ref{fig3} (b.v).

Combining these processes, we can describe the behavior of a Bloch state in a perturbed system as follows: The electromagnetic perturbation can induce one-photon and multi-photon absorptions/emissions, causing electronic transitions among Bloch bands. After each such a transition, the perturbed state continues to propagate in a new band until another transition occurs. Finally, the current vertex is by definition $\frac{1}{\mathcal{V}}\frac{\delta H}{\delta \bm{A}}$ where $\mathcal{V}$ is the volume of the full system. Thus the current vertex enters the diagrammatic expansion in exactly the same fashion as electron-photon vertices.
All processes at second order in the electric field are shown in Fig.~\ref{fig3} (b.i - b.vi). Combined, they provide a comprehensive description of the intrinsic nonlinear response. 

\subsection{Feynman rules}\label{sec2D}

Feynman diagrams not only helps differentiate different physical processes contributing to each order of nonlinear responses, but also provide a simple way to formulate the nonlinear conductivity. 
In principle, the concepts elucidated below can be employed for the nonlinear response at any order~\cite{Fregoso2019,Liu2022a,Cheng2014,Cheng2015,xiang2023third,Fang2024,jankowski2025optical}. 
However, for the sake of clarity, we restrict the discussion to the 2nd-order conductivity. We define the 2nd-order current generation to be described by,
\begin{equation}
    j^c(\bar{\omega})=\sigma^{ab;c}(\bar{\omega};\omega_1,\omega_2)E^a(\omega_1)E^b(\omega_2)
\end{equation}
where $j(\bar{\omega})$ is the the magnitude of 2nd-order current response with frequency $\bar{\omega}$, $E(\omega_i)$ is the magnitude of electric field with frequency $\omega_i$. 
Since the frequency summation law is always satisfied, we can neglect the delta function part for notational simplicity, more details about this notation is discussed in App.~\ref{app00}.

To formulate the contribution of different physical processes depicted in Fig. \ref{fig3} (b.i-b.vi), one collects all vertices ($\hat{v},\hat{w},\hat{u}$) and fermionic Matsubara Green's functions $\mathcal{G}(i\omega)=\frac{1}{i\hbar \omega - H_0}$ along the fermionic loop in the corresponding Feynman diagram and calculates its trace over degree of freedom of band indices. The internal momentum is integrated over the Brillouin zone (BZ) and the internal frequency is unrestricted. At each vertex we enforce frequency and momentum conservation. 
The Matsubara summation is done by residues, which is followed by an analytical continuation to the real axis.
The full expression of the 2nd-order conductivity is just the sum of all 2nd-order Feynman diagrams Fig. \ref{fig3} (b.i-b.vi) in a symmetrized form under internal permutation symmetry $(\omega_1, a \leftrightarrow \omega_2, b)$ as shown in Eq. \eqref{tot}, whose derivation can be found in App.~\ref{app1}.
\begin{widetext}
\begin{equation}
\begin{aligned}
    \sigma^{ab;c}(\bar{\omega};\omega_1,\omega_2) = -\frac{e^3}{\hbar^2 \omega_1 \omega_2} \int_\mathbf{k} & \sum_m f_m \left( \frac{\hat{u}^{abc}}{2} + \frac{1}{2} \left[ \hat{w}^{ab}, \frac{\hat{v}^c}{\tilde{\bar{\omega}}-\hat{\varepsilon}} \right] + \left[ \frac{\hat{v}^a}{\tilde{\omega}_1 + \hat{\varepsilon}}, \hat{w}^{bc} \right] + \left[ \frac{\hat{v}^a}{\tilde{\omega}_1 + \hat{\varepsilon}}, \left[ \hat{v}^b , \frac{\hat{v}^c}{\tilde{\bar{\omega}} - \hat{\varepsilon}} \right] \right]\right)_{mm} \\ & + (\omega_1, a \leftrightarrow \omega_2, b) \label{tot}
\end{aligned}
\end{equation}
\end{widetext}

Here, $f_m \equiv \frac{1}{e^{\beta\hbar(\varepsilon_m-\mu)}+1}$ represents the Fermi-Dirac distribution of $m$-th band where $\mu$ is the chemical potential, $\beta^{-1}=k_BT$ for temperature $T$, and $\int_{\mathbf{k}} \equiv \int \frac{d^D \mathbf{k}}{(2\pi)^D}$, where $D$ is the spatial dimension of the system. Most of the discussion will be restricted to temperature $T=0$. The fractions of operators in Eq. \eqref{tot} are defined element-wise. For example, $\left(\frac{\hat{v}^a}{\omega_i + \hat{\varepsilon}}\right)_{nm} \equiv \frac{\hat{v}^a_{nm}}{\omega_i + (\hat{\varepsilon})_{nm}}$, where $(\hat{\varepsilon})_{nm} = \varepsilon_n - \varepsilon_m$. The tilde over the frequency is defined via analytical continuation $\tilde{\omega} = \omega + i 0^+$, which becomes important for finite quasiparticle particle lifetimes and will be addressed more in details in Sec.~\ref{sec4}.

\subsection{Three regimes in 2nd-order responses}\label{sec2E}

In practical scenarios, the most common and simplest perturbation is a monochromatic perturbation which has the form of $\cos \omega t \sim e^{i \omega t} + e^{-i \omega t}$, which contains both $\omega$ and $-\omega$ components. This type of perturbation serves as a fundamental basis for understanding more complex nonlinear processes.

For a linear order response ($n=1$), the absolute value of response frequency would always be $|\bar\omega|=|\omega|$, meaning the linear response shares the same frequency as the perturbation. For a second order response, we can have both $|\bar\omega| = |\omega + \omega| = 2|\omega|$ or $|\bar\omega| = |\omega + (- \omega)| = 0$. The response with $|\bar{\omega}|=|2\omega|$ frequency corresponds to the SHG while the response with $|\bar{\omega}| = 0$ correspond to the BPVE.

% Nonlinear optics
When the perturbation frequency $\omega$ is comparable with the band gap $E_g$, the corresponding nonlinear response is termed ``nonlinear optics'', where resonant excitation can happen between valence and conduction bands. By frequency sum rule, the perturbation frequency is $\hbar \omega > E_g/2$ for SHG or $\hbar \omega > E_g$ for BPVE as shown in Fig. \ref{fig1} (a) and (b) to achieve this regime.

% Subgap response
When the frequency $\omega$ does not satisfy the previous conditions, yet it is still comparable to the band gap, off-resonant optical excitation can still occur. Specifically, for BPVE with $\hbar \omega < E_g$, the DC current will not necessarily vanish since off-resonant excitation also contributes to the nonlinear response. We define this regime where only off-resonant processes contributes to the DC generation in an insulator as the so-called ``subgap response''.

% Nonlinear transports
When the perturbation frequency is further reduced to the full DC limit, where also $\omega$ approaches zero, the quasiparticle lifetime is much shorter than the perturbation period, i.e. $\omega \ll \tau^{-1}$. This defines the so-called ``nonlinear transport'' regime where a semiclassical (Fermi-surface dominated) phenomenology is expected. Since $\omega \rightarrow 0$, the formerly distinct processes SHG and BPVE merge into the same DC response.

\section{Bulk Photovoltaic Effect}\label{sec3}

\begin{figure*}[htbp]
\includegraphics[width=\linewidth]{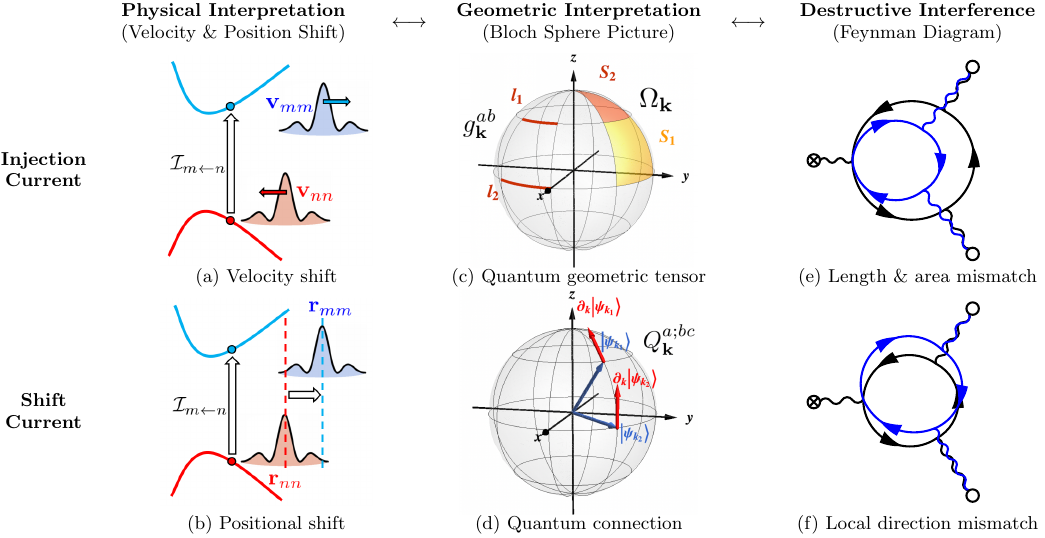}
\caption{Overview regarding the photocurrent creation in inversion-symmetry broken bulk materials. The linear response formulation at second order in the electric field can either be approached as a collection of interband corrections to the semiclassical expectations values (a-b), in terms of the contained quantum geometry (c-d), or using diagrammatics (e-f). 
Here, the Feynman diagrams illustrate the fact that the time-reversed Green's functions (propagators) of the clockwise and anticlockwise triangle diagrams will not be identical even if all other parameters a kept fixed. Importantly, there is more than one way in which the time-reversal properties can be broken, as we elucidate in detail in the text.
}\label{fig4}
\end{figure*}

The creation of a photocurrent via rectification has a long history, but it was not until the works of Kraut and von Baltz~\cite{Kraut1979,vonBaltz1981} that it was realized that the bulk band structure of an inversion-symmetry breaking material can give rise to such. 
Phenomenologically, the bulk photovoltaic effect (BPVE) can be separated into two parts according to their time dependence, namely the \emph{injection current} whose magnitude grows linearly with time and the \emph{shift current} whose magnitude is fixed.

In a real material, no current can grow indefinitely. Instead, injection and shift current are differentiated by their dependence on quasiparticle lifetime because the injection current grows until the quasiparticle decoheres. Thus the quasiparticle lifetime cuts off the injection of current into the system and making the steady state current proportional to the quasiparticle lifetime $\tau$. On the other hand, the shift current is not created by this kind of mechanism and thus does not depend on $\tau$ altogether.

In the following, we present and compare between the semiclassical and quantum perturbative approaches, which have been developed in order to rationalize the quantitative properties of shift and injection currents (cf. Fig.~\ref{fig4}).

\subsection{Injection and shift current}\label{sec3B}

The BPVE is a resonant phenomenon. As such, it can be interpreted as a sequence of excited intermediate states~\cite{Ahn2020} (Fig.~\ref{fig4}a,b). For injection current, let us focus on the change in velocity  which the transition between valence and conduction bands causes. The transition rate of electrons from $n$-th band to $m$-th band is determined by Fermi's golden rule,
\begin{equation}
\mathcal{I}_{m \leftarrow n} = \frac{2 \pi}{\hbar^2} \left| \langle m | - e \hat{\mathbf{r}} \cdot \mathbf{E} | n \rangle \right|^2 \delta (\omega - \varepsilon_{mn})
\end{equation}
where $\varepsilon_{mn}\equiv (\varepsilon_m - \varepsilon_n)$ and the momentum dependencies have been suppressed. Further considering the velocity change during the transition $\Delta_{nm}^c \equiv v_{nn}^c - v_{mm}^c = \hbar\partial_{k_c} \varepsilon_{nm}$, the rate of increase in current can then be expressed as,
\begin{align}
\partial_t j_{\mathrm{inj}}^c = \int_{\mathbf{k}} \sum_{\substack{n \in \text{occ} \\ m \in \text{unocc}}} (-e \Delta_{mn}^c) \mathcal{I}_{m\leftarrow n} f_{nm}
\label{partial_t}
\end{align}
where $f_{nm} \equiv f_n - f_{m}$ represents the difference in equilibrium Fermi-Dirac distribution function. Since there is a squared electric field in the transition rate, we can rewrite it as $\left| \langle m | \hat{\mathbf{r}} \cdot \mathbf{E} | n \rangle \right|^2 = r_{mn}^a r_{nm}^b E^a E^b$ and read the injection conductivity from it,
\begin{align}
\sigma_{\mathrm{inj}}^{ab;c} = - \tau \frac{2 \pi e^3}{\hbar^2}  
\sum_{nm} 
\int_{\mathbf{k}} f_{nm} \Delta_{mn}^c r_{nm}^a r_{mn}^b \delta (\omega - \varepsilon_{mn})
\label{eq:injphysical}
\end{align}

Because of the $\partial_t$ in Eq. \ref{partial_t}, we first get a frequency denominator $(- i \bar{\omega})$ in the injection conductivity, which nominally diverges for the BPVE ($\bar{\omega}\rightarrow0$). However,  in a real system it is regularized by a finite quasiparticle lifetime which shifts $\bar{\omega} \rightarrow \bar{\omega} + i/\tau$ and makes the injection current proportional to $\tau$.

Similarly, the shift current can be understood as the positional shift during the transition. To this end, we first recall that the position of a wavepacket center composed of $n$-th Bloch band can be represented by the momentum space integral over the Berry connection $\langle \hat{\mathbf{r}} \rangle_n \equiv \langle u_n | i \nabla_{\mathbf{k}} | u_n \rangle$ up to a lattice constant according to the modern theory of polarization~\cite{Xiao2010}, where $| u_n \rangle$ is the periodic part of the Bloch wavefunction. Therefore, the positional shift during an interband transition should contain the difference $\langle \hat{\mathbf{r}} \rangle_m - \langle \hat{\mathbf{r}} \rangle_n$, which is, however, not invariant under the gauge transformation $|u_n\rangle \rightarrow e^{i \theta_n} | u_n \rangle$. To remove this ambiguity, the positional shift has to be corrected for the interband phase difference, yielding,
\begin{align}
R_{mn,a}^{c} = r_{mm}^c - r_{nn}^c + i \partial_{k_c} \operatorname{log} r_{mn}^a,
\end{align}
which is a gauge invariant observable describing  the positional shift during the interband transition caused by the transition dipole $r^a_{nm}$.

The longitudinal shift current can then be expressed as the positional shift multiplied by the transition rate $\mathcal{I}_{m \leftarrow n}$ and a occupancy factor $f_{nm}$, i.e.
\begin{align}
j_{\mathrm{shift}}^c = \int_{\mathbf{k}} \sum_{\substack{n \in \text{occ} \\ m \in \text{unocc}}} (-e R_{mn,c}^{c}) \mathcal{I}_{m\leftarrow n} f_{nm}
\end{align}
which leads to the corresponding shift conductivity for arbitrary spatial directions in the resonant optical regime, i.e.

\begin{equation}
\begin{aligned}
\sigma_{\mathrm{shift}}^{ab;c} = - \frac{2 \pi e^3}{\hbar^2} 
\sum_{nm} 
\int_{\mathbf{k}} & f_{nm} r_{nm}^a r_{mn}^b R_{mn,a}^c \delta (\omega - \varepsilon_{mn}) \\
& + (a,\omega \leftrightarrow b,-\omega)
\end{aligned}
\label{eq:shiftphysical}
\end{equation}

Since there is one time derivative difference between position and velocity, the factor $(-i\omega)$ in the denominator of $\sigma_{\mathrm{inj}}$ is absent in the expression of $\sigma_{\mathrm{shift}}$, which explains the different lifetime scaling of both resonances.

The argument presented above employs the length gauge where the perturbation has the form $H' = -e \hat{\mathbf{r}} \cdot \mathbf{E}$. It is equivalent to the velocity-gauged perturbation $H_0(\mathbf{k}) \rightarrow H_0(\mathbf{k} + \frac{e}{\hbar}\mathbf{A})$ up to a time-dependent unitary transformation~\cite{Ventura2017}, where the latter one can be formulated easily using Feynman diagrams. 
However, we point out that the expression of the shift current based on this ``transition rate''$\times$``positional shift'' only gives the dominant part when resonant transitions happen. For broad resonances, the complete expressions based on the Kubo formula for shift conductivity is needed~\cite{Sipe2000, Holder2020},
\begin{align}
\sigma_{\mathrm{shift}}^{ab;c} &= \frac{e^3}{\hbar^2} \int_{\mathbf{k}} \sum_{nm} \frac{f_{nm}r^{a}_{nm}r^{b}_{mn}R_{mn,a}^c}{\omega-\varepsilon_{mn}+i0^+} 
+ (a,\omega \leftrightarrow b,-\omega)
\end{align}

Using the formula $\frac{1}{x \pm i0^+} = \mathcal{P}\frac{1}{x} \mp i\pi\delta(x)$, it is easy to find that the physically motivated formulae Eq.~\eqref{eq:injphysical} and Eq.~\eqref{eq:shiftphysical} are the resonant (delta-function) part of the full response. 
On the other hand, the principal parts become important far from resonance, where they may provide useful information about the localization and interaction of system's ground state~\cite{Kaplan2020,Kaplan2023}.

\subsection{Geometric interpretation}\label{sec3C}

As we have seen in the previous section, both injection and shift conductivity contain matrix elements of the positional operator $r_{mn}^c$ and shift vector $R_{mn,a}^{c}$, which offer the attractive opportunity to draw conclusions about the encoded quantum geometry.

Yet such a geometric interpretation of nonlinear optical response is subtle compared to more direct geometric nature of the linear order anomalous Hall effect, where the anomalous velocity and Hall current is directly the integral of a single geometric quantity, the Berry curvature. In contrast, the injection and shift conductivities are related to several geometric properties, and multiplied by occupation and dispersion-related quantities. Also, unlike Hall current where the geometric response is completely determined by the ground state property, the optical response necessarily involves information of both ground state and excited states, making the discussion of geometry quite intricate.

Not surprisingly, the geometric nature of nonlinear optical responses has only recently been appreciated, starting with the observation that the injection current induced by circular polarized light is related to the Berry curvature~\cite{deJuan2017}. More recently,  injection and shift current have been associated with the quantum geometric tensor and quantum connection, respectively~\cite{Ahn2020}.

The geometric nature of injection and shift current can be elucidated through a mathematical formula equivalence (cf. Fig.~\ref{fig4}). It can also be related with the geometric properties of general projected Hilbert space, which becomes a simple Bloch sphere in systems with only two bands (Fig.~\ref{fig4}c,d). Alternatively, both resonant responses can be interpreted as an interference effect of coherently counterpropagating quasiparticles (Fig.~\ref{fig4}e,f). 

\paragraph{Formula equivalence}
Since the optical response indispensably involves information from both occupied and unoccupied states, let us focus on a pair of occupied and unoccupied bands with a momentum subspace where the transition between unoccupied $m$-th band and occupied $n$-th band matches the frequency of incident light, i.e. $ \omega = \varepsilon_{mn}(\mathbf{k})$. One can then define a manifold $S_{\mathbf{k}}$ of optically allowed transitions as $S_{\mathbf{k}}\equiv\{\mathbf{k}\in \text{BZ}|\omega=\varepsilon_{mn}(\mathbf{k})\}$.
Similar to the linear order anomalous Hall conductivity, which is proportional to the integral of Berry curvature over all occupied states, the injection and shift current are then  proportional to the integral of their corresponding quantum geometric quantities over $S_{\mathbf{k}}$.

Specifically, the injection current is proportional to the momentum space integral of quantum geometric tensor $Q^{ab} = \sum_{\substack{n \in \text{occ} \\ m \in \text{unocc}}} r^{a}_{nm} r^{b}_{mn}$ over $S_{\mathbf{k}}$, i.e.
\begin{align}
\sigma_{\mathrm{inj}}^{ab;c} & 
= - \tau \frac{2 \pi e^3}{\hbar^2} \hat{c} \cdot \int_{\omega_{mn}=\omega} d\vec{S}_k Q^{ab}
\label{eq:injgeometric}
\end{align}
where the $Q^{ab}$ is composed of the quantum metric $g^{ab}$ (real/symmetric part) and Berry curvature $\Omega^{ab}$ (imaginary/antisymmetric part), and can be roughly expressed as the expectation value of two momentum derivative over Bloch states, i.e. $\langle \partial_k \partial_k \rangle$. 

Similarly, 
Refs.~\cite{Ahn2020, Ahn2022,Avdoshkin2024} found that the shift current is proportional to the momentum space integral of the quantum connection over $S_{\mathbf{k}}$, i.e.
\begin{align}
\sigma_{\mathrm{shift}}^{ab;c} & 
= - \frac{2 \pi e^3}{\hbar^2} \int_{\mathbf{k}} \sum_{\substack{n \in \text{occ} \\ m \in \text{unocc}}}\mathrm{Im}[Q^{c;ab}_{mn}] \delta (\omega - \epsilon_{mn}).
\end{align}
This connection describes how the tangential vector changes as it moves along a curved manifold, which in this case is the projected Hilbert space. 
However, a much more in-depth analysis reveals that the (interband) quantum connection actually contains two distinct geometric pieces, one related to the third moment of position, while the second piece constitutes torsion tensor, a true multi-state quantity which has no ground-state equivalent~\cite{Avdoshkin2024,Mitscherling2024}.

\paragraph{Bloch sphere picture}
For illustration let us repeat the previous analysis for a 2-band system. Then, the projected Hilbert space is just a Bloch sphere whose geometry is easy to draw (Fig.~\ref{fig4}c,d). Moreover, there is only one pair of occupied and unoccupied bands that contributes to the optical response while in general systems all combinations of occupied and unoccupied bands may contribute as long as $\omega_{mn}=\omega$. Most importantly, the unoccupied states are completely determined by the occupied states since the remaining state vector is orthogonal. 
The Hamiltonian of a general 2-band system can be described as,
\begin{equation}
\hat{H}_0(\mathbf{k}) = \mathbf{d}(\mathbf{k}) \cdot \vec\sigma + d_0 (\mathbf{k}) \sigma_0
\end{equation}
where $\sigma_i,\,i\in \{1,2,3\}$ are Pauli matrices of a generic pseudo-spin degree of freedom, and $\sigma_0$ is the 2 by 2 identity matrix. 
The wavefunction information is completely stored in the vector $\mathbf{d}(\mathbf{k})$ or equivalently $\hat{\mathbf{d}}(\mathbf{k}) \equiv \frac{\mathbf{d}(\mathbf{k})}{|\mathbf{d}(\mathbf{k})|}$, where the eigenstates can be defined as poloarized states, $| u_{\pm,\mathbf{k}}\rangle \equiv | \hat{\mathbf{d}}({\mathbf{k}}), \pm \rangle$, satisfying
$
\hat{H}_0 | \hat{\mathbf{d}}, \pm \rangle = ( d_0  \pm |\mathbf{d}| ) | \hat{\mathbf{d}}, \pm \rangle
$.
The manifold composed of $\hat{\mathbf{d}}$ forms a unit sphere in $\mathbb{R}^3$ space, i.e. the Bloch sphere. 
The wavefunction maps a momentum vector $\mathbf{k}$ to a point on the Bloch sphere $\hat{\mathbf{d}}(\mathbf{k})$, or equivalently its corresponding state vector $| u_{n,\mathbf{k}}\rangle$, assuming the band index $n$ is determined. 
Such a map is analogous to the map from spherical coordinate $\mathbf{x} \equiv (x^1,x^2)=(\theta, \phi)$ to the sphere $S^2$. In this spherical coordinate map, the metric $g_{\mu\nu}$ is defined by the squared distance $ds^2 = g^{ab} dx^a dx^b$ between two close points $x^a$ and $x^a + dx^a$, the curvature $\Omega^{ab}$ is defined by the oriented area of small parallelogram $dA = \Omega^{ab}dx_1^a \wedge dx_2^b$ formed by $dx_1^a$ and $dx_2^b$ near point $\mathbf{x}$, and the connection $Q^{a;bc}$ is defined by the shift in the orientation of a basis tangent vectors when parallel-transported along certain directions.

Therefore, in order to discuss the distance between two state vectors mapped from $\mathbf{k}$ and $\mathbf{k}+d\mathbf{k}$ in the projected Hilbert space, or the area of small parallelogram formed by state vector mapped from $\mathbf{k}$ and $\mathbf{k}+d\mathbf{k}_1$ to $\mathbf{k}$ and $\mathbf{k}+d\mathbf{k}_2$, or a of the local coordinate frame upon translation from $\mathbf{k}$ to $\mathbf{k}+d\mathbf{k}$, this information is contained in the metric, curvature, and connection induced by the map between the BZ and the projected Hilbert space.

Returning to the nonlinear optical resonances, the quantum metric and Berry curvature depict the infinitesimal length and area which a state travels in the projected Hilbert space, which is measured by the injection current.
For shift current, the quantum connection can be regarded as the change in tangent vector when a given state moves adiabatically along the Bloch sphere. Of course, the same quantity also represents the shift of the center point of a Bloch wave packet in the unit cell. Since the geometric understanding of the connection is the shift of the tangent vector during parallel transport, the physical understanding of the connection in the projected Hilbert space should be as a shift of the wavepacket position along certain k-path, thereby recovering the previously outlined interpretation.

Nonetheless, we caution against taking the Bloch sphere trajectory too literally.
First, the Bloch sphere here should be considered as an "interband" manifold made up of interband transition since nonlinear optics inevitably involves one unoccupied and one occupied band. Only for two bands such an ``interband'' Bloch sphere actually isomorphic to the usual Bloch sphere.
Secondly, even when an $N$-band system is considered, the $N$ single-band geometric quantities should be generalized to $N(N-1)$ interband geometric quantities where quantum metric, Berry curvature and quantum connection are all redefined for pairs of bands made up of $m$-th and $n$-th index, which means that a suitable generalized Bloch sphere grows rapidly in dimension~\cite{Avdoshkin2024,Ahn2020,Mitscherling2024}.

\subsection{Destructive interference}\label{sec3D}
In weakly disordered system, the weak localization mechanism features a constructive interference between two time-reversed paths, which maximizes the probability amplitude at the origin where the wavepacket starts to move. Therefore, a wavepacket has an enhanced probability to stay at its original position,  indicating the tendency towards localization. 
However, breaking time-reversal symmetry, for example by magnetic impurities, destroys the constructive interference at the origin, thus increasing the mobility of charge carriers in an effect termed \emph{weak anti-localization}.

A similar argument can be applied to explain the physical origins of the injection and shift current: If the response breaks time-reversal symmetry, two time-reversed diagrams cannot interfere constructively, thus the electron can become delocalized and contribute a nonzero conductivity. 
The destructive interference results from in-equivalent time-reversed closed trajectories, which can be caused by two mechanisms. One comes from the inequivalence in distance and area, the other comes from the inequivalence caused by local direction (cf. Fig.~\ref{fig4}e,f). 

Taking injection current as an example, the appearance of injection current can be regarded as the inequivalence of time-reversal counterpart Feynman diagram because of nontrivial quantum metric and Berry curvature.
From diagrammatic language, the 2 closed trajectories in the triangular diagrams are related by time-reversal (denoted by $\circlearrowleft$ and $\circlearrowright$, respectively) and contribute,
\begin{widetext}
\begin{equation}
\sigma_{\triangle;\circlearrowleft}^{ab;c} + \sigma_{\triangle;\circlearrowright}^{ab;c} = \frac{e^3}{\hbar^2 \omega^2} \sum_{n,m,l} \int_{\mathbf{k}} \frac{f_{nm}}{\omega+\varepsilon_{nm}+i0^+} \left[\frac{v_{nm}^a v_{ml}^b v_{ln}^c}{\bar{\omega}+\varepsilon_{nl}+i0^+} - \frac{v_{nm}^a v_{ml}^c v_{ln}^b}{\bar{\omega}+\varepsilon_{lm}+i0^+}\right] + (a,\omega \leftrightarrow b,-\omega)
\end{equation}

\end{widetext}

Using that $\bar{\omega} = 0$, the summation of time-reversal counterpart diagrams gives,
\begin{equation}
\frac{1}{\omega+\varepsilon_{nm}-i0^+} - \frac{1}{\omega+\varepsilon_{nm}+i0^+} = 2\pi i\delta(\omega + \varepsilon_{nm})
\end{equation}
which corresponds to the energy conserving delta function in the ``transition rate''$\times$``velocity/position shift'' -based physical interpretation. Because of the delta function, one can substitute $\omega$ with $\varepsilon_{mn}^a$ which changes $v_{nm}^a$ into $r_{mn}^a$ by $v_{mn}^a = i \varepsilon_{nm} r_{nm}^a$. This yields the same expression as Eq.~\eqref{eq:injphysical} as we derived in Sec.~\ref{sec3B}.

The two transition dipoles indicate a proportionality to quantum geometric tensor. Such a incomplete cancellation is caused by the nontrivial distance and area property on the projected Hilbert space.
Similarly, the shift current can also be regarded as caused by destructive interference between two time-reversal counterparts via the nontrivial nonparallelism of the projected Hilbert space, which manifests the quantum connection of the system.
The latter mechanism is much weaker because time-reversal is broken only in terms of the rectified current, i.e. an inverted trajectory is not guaranteed to trace out the same positions, while velocity and distances are equal for both directions.
Of course, we hasten to note that the analogy with weak anti-localization works in some features, but fails for others. For example, nonlinear responses do not entail any resummation over repeated perturbations up to infinite order. 

Altogether, the physical interpretation by a ``transition rate''$\times$``velocity/positional shift'' argument, the geometric interpretation via a Bloch sphere picture, as well as the interpretation via quantum interference provide a complete and versatile understanding of the nonlinear optical response in both physical and quantum geometric aspects. This shows the power of quantum geometry in re-imagining convoluted perturbation expressions, which can help us develop new avenues towards measuring quantum geometry or, conversely, sharpen the predictions regarding nonlinear response functions.

\subsection{Symmetry separation}\label{optical symmetry}

We have shown in nonlinear optical regime that the total current response can be separated into the injection and shift current, which are further linked with different quantum geometric quantities. However, the current response always comes as a whole, which makes the investigation of a specific kind of quantum geometric quantity difficult. As we explain henceforth, such difficulties can be bypassed by exploiting different symmetry constraints, which include the intrinsic permutation symmetry of the nonlinear conductivity tensor and the symmetry of the material's microscopic Hamiltonian. 

Regarding the permutation symmetry of nonlinear conductivity tensor, it essentially boils down to inducing bulk photovoltaic currents with different spatial configurations of the perturbing electromagnetic field, i.e. linearly polarized light vs. circularly polarized light. As detailed in App.~\ref{LPLCPL}, the BPVE induced by linearly polarized light is proportional to the symmetric part of the conductivity tensor $\sigma^{(a,b);c}$ while the difference in the current induced by left and right handed circularly polarized light is proportional to the antisymmetric part of the conductivity tensor $\sigma^{[a,b];c}$. Since the conductivity tensor and their corresponding quantum geometric quantities share the same symmetry, we can expect linearly polarized light to single out the contribution of quantum metric $g^{ab}$ in the injection current as well as the contribution of symplectic quantum connection $Q^{a;(b,c)}$ in the shift current, while the circular BPVE extracts those of Berry curvature $\Omega^{ab}$ and Christoffel symbol of 1st kind $Q^{a;[b,c]}$, as shown in Table~\ref{table1}.

Regarding the symmetry of the bulk material, the most important two symmetries are spatial inversion ($\mathcal{P}$, parity) and time-reversal ($\mathcal{T}$). 
As mentioned, $\mathcal{P}$ symmetry renders the second order current $j^{c} (\bar{\omega}) = \sigma^{ab;c}(\bar{\omega};\omega_1,\omega_2) E^a(\omega_1) E^b(\omega_2)$ zero since $\mathcal{P}$ changes the sign of current $j^{c}(\bar{\omega})$ yet leaves $E^{a}E^{b}$ invariant.
With respect to $\mathcal{T}$, 
the shift current $\mathbf{j}_{\mathrm{shift}}$ and the injection current $\partial_t \mathbf{j}_{\mathrm{inj}}$ possess opposing symmetry properties, similar to the behavior of velocity and acceleration under $\mathcal{T}$. Such distinct $\mathcal{T}$-symmetry properties of different nonlinear responses result from different symmetry properties of their corresponding quantum geometric quantities under $\mathcal{T}$ transformation. Therefore, Berry-curvature-, quantum-metric-, and quantum-connection-related conductivities are also expected to be separable by $\mathcal{T}$ and other magnetic point group symmetries.

To be more specific, $\mathcal{T}$ transformation of injection conductivity (Eq.~\ref{eq:injphysical}) and shift conductivity (Eq.~\ref{eq:shiftphysical}) require time-reversal transforming all corresponding matrix elements. As further explained in the App.~\ref{optical spacetime symmetry}, we have ${f}_{\mathcal{T},nm} (\mathbf{k})=- f_{mn}(-\mathbf{k})$, ${v}_{\mathcal{T},nm}^a (\mathbf{k}) = - v_{mn}^a (-\mathbf{k})$, ${\Delta}_{\mathcal{T},mn}^c (\mathbf{k}) = \Delta_{nm}^c (-\mathbf{k})$, $R^c_{\mathcal{T},mn,a}(\mathbf{k}) = - R^c_{nm,a} (-\mathbf{k})$, ${\varepsilon}_{\mathcal{T},nm}(\mathbf{k}) = \varepsilon_{nm}(-\mathbf{k})$. Substituting these results into the expression of injection conductivity yields $\sigma_{\mathcal{T},\mathrm{inj}}^{ab;c} (\omega) = - \sigma_{\mathrm{inj}}^{ab;c} (-\omega)$ and $\sigma_{\mathcal{T},\mathrm{shift}}^{ab;c} (\omega) = \sigma_{\mathrm{shift}}^{ab;c} (-\omega)$. Also as detailed in App.~\ref{intrinsic symmetry}, a real-valued physical response further imposes an intrinsic permutation symmetry on conductivity tensor, i.e. $\sigma_{\mathrm{inj/shift}}^{ab;c} (-\omega) = \sigma_{\mathrm{inj/shift}}^{ba;c} (\omega)$. Furthermore, time reversal symmetry enforces the constraint $\sigma_{\mathrm{inj/shift}}^{ab;c} (\omega) = \sigma_{\mathcal{T},\mathrm{inj/shift}}^{ab;c} (\omega)$, which implies,
\begin{align}
\sigma_{\mathrm{inj}}^{ab;c} (\omega) \stackrel{\mathcal{T}}{=} -\sigma_{\mathrm{inj}}^{ba;c} (\omega), \quad \sigma_{\mathrm{shift}}^{ab;c} (\omega) \stackrel{\mathcal{T}}{=} \sigma_{\mathrm{shift}}^{ba;c} (\omega)
\end{align}

The real-valued dc current requires the second order BPVE conductivity to be symmetric in the real part, which therefore contributes to the response for linear polarization, while the imaginary part contributes to the circular polarized light response, giving rise to circular dichroism, i.e. the difference in currents induced by left and right handed circularly polarized light (cf. App.~\ref{LPLCPL}). Based on this analysis, it becomes obvious why in $\mathcal{T}$-symmetric system the injection current can only be induced by circular polarized light. 
Along the same lines, one can show that in $\mathcal{T}$-symmetric systems the shift current is induced by linear polarized light.

With respect to the combined symmetry transformation of space-time inversion ($\mathcal{P}\mathcal{T}$), which induces another $(-1)^3$ factor because of the $\mathcal{P}$ transformation on 3 spatial indices, the situation is reversed, i.e.
\begin{align}
\sigma_{\mathrm{inj}}^{ab;c} (\omega) \stackrel{\mathcal{PT}}{=} \sigma_{\mathrm{inj}}^{ba;c} (\omega) ,\quad \sigma_{\mathrm{shift}}^{ab;c} (\omega) \stackrel{\mathcal{PT}}{=} -\sigma_{\mathrm{shift}}^{ba;c} (\omega)
\end{align}
Therefore, in $\mathcal{PT}$-symmetric system the resonant responses exchange roles, i.e. current injection is induced by linear polarized light while the shift current forms in response to circular polarized light~\cite{Zhang2019,Watanabe2021}.

More general magnetic point group symmetries lead to more subtle constraints~\cite{Zhang2023a}. For example, $\mathcal{C}_{2z}\mathcal{T}$ symmetry acts like $\mathcal{T}$ symmetry on the $z$ axis while acting like $\mathcal{PT}$ symmetry on the $x$-$y$ plane. This will cause a separation of the response in the material based on spatial direction. For instance, linear polarized light can only induce shift currents along $z$-direction while inducing injection current in the $x$-$y$ plane. Similar response separation effects based on the effective symmetry in a certain sub-direction or sub-plane, are summarized in Table.~\ref{effective symmetry}.

\begin{table}[htbp]
    \centering
    \renewcommand{\arraystretch}{1.2} % Adjust row spacing
    \begin{tabular}{ccccccccccc}
    \hline\\[-1.2em]
    \hline
        & $\mathcal{C}_{2z}$ & $\mathcal{C}_{3z}$ & $\mathcal{C}_{4z}$ & $\mathcal{C}_{6z}$ & $\mathcal{M}_{z}$ & $\mathcal{C}_{2z} \mathcal{T}$ & $\mathcal{C}_{3z} \mathcal{T}$ & $\mathcal{C}_{4z} \mathcal{T}$ & $\mathcal{C}_{6z} \mathcal{T}$ & $\mathcal{M}_{z} \mathcal{T}$ \\ \hline
        $x$-$y$ & $\mathcal{P}_{xy}$ & $\mathcal{C}_{3z}$ & $\mathcal{P}_{xy}$ & $\mathcal{P}_{xy}$ & $\mathbb{I}$ & $\mathcal{P}_{xy}\mathcal{T}$ & $\mathcal{T}$ & $\mathcal{P}_{xy}$ & $\mathcal{P}_{xy}\mathcal{T}$ & $\mathcal{T}$ \\ 
        $z$ & $\mathbb{I}$ & $\mathbb{I}$ & $\mathbb{I}$ & $\mathbb{I}$ & $\mathcal{P}_{z}$ & $\mathcal{T}$ & $\mathcal{T}$ & $\mathcal{T}$ & $\mathcal{T}$ & $\mathcal{P}_{z} \mathcal{T}$ \\ 
        \hline\\[-1.2em]
        \hline
    \end{tabular}
    \caption{Effective symmetry of (magnetic) point groups along different spatial directions. $\mathbb{I}$ represents identity transformation, while $\mathcal{P}_{xy}$ and $\mathcal{P}_z$ represents spatial inversion only in $x$-$y$ plane and $z$-axis, which renders all 2nd order charge current response in that sub-direction or sub-plane zero.}
    \label{effective symmetry}
\end{table}

As a corollary of these symmetry considerations, we point out that magnetic systems with low internal symmetry generically exhibit both injection and shift current.

\subsection{Extrinsic photocurrents}
Besides the response created by the bulk band structure, a number of extrinsic effects of nonlinear photocurrent generation have been investigated. 
Historically, these are usually known as ballistic currents~\cite{sturman2021photovoltaic}, which arise in the driven steady state alongside the shift current, mostly due to phonon-assisted relaxation
\cite{Gong2018,Dai2021,Xu2022,Dai2023b,Zhu2024}, and which have been observed experimentally~\cite{Burger2019}. Another important factor outside the purview of the non-interacting band structure are excitonic in-gap states~\cite{Fei2020a,Chang2024,Nakamura2024}.
Scattering can also lead to additional AC-DC effects~\cite{Bhalla2020}, and even facilitate a
reverse BVPE which converts DC current into an AC signal, known as the gyrotropic Hall effect~\cite{Koenig2019}.
Compared to intrinsic contributions, less is known about the underlying quantum geometry of extrinsic photocurrent generation, but recent progress~\cite{Zhu2024} suggests that generalized shift vectors can be constructed which account for the changes in the Bloch wave function in  diagonal transitions which can involve different energies or momenta due to extrinsic scattering and are related to both the diagonal and off-diagonal components of the Berry curvature. 

\section{Finite Lifetime and Subgap Response}\label{sec4}

When the frequency is smaller than the bandgap, in an insulator one might expect the DC response to vanish, as resonant interband transitions are no longer possible~\cite{Belinicher1986,Onishi2022a,Golub2022}. However, as it turns out in magnetic materials the second order response is not entirely extinguished~\cite{Kaplan2020,Michishita2021,Matsyshyn2023,Shi2023}.
To appreciate this aspect, let us point out that the derivation of the nonlinear optical resonances as detailed in the previous sections implicitly assumes a clean limit, where the only lifetime effect is to broaden the resonance peak (i.e. the response consists entirely of regularized delta-functions). Not contained in such an approach are broad features in the response which contribute an off-resonant background that is unavoidably present in second order responses.
Indeed, in the most general case the subgap response remains nonzero and offers a valuable avenue for probing the effects of finite lifetimes, as well as the interaction and localization mechanisms underlying the system.

\subsection{Introduction of two lifetimes}\label{sec4A}

As alluded to in the introduction, from the nonlinear Kubo expressions it is expected that both interband and intraband lifetimes impact the conductivity~\cite{Kaplan2020}. 
Phenomenologically, on one hand nonlinear optical responses are entirely derived from resonant interband transitions. On the other hand, nonlinear transport probes the adiabatic regime where the frequency is so small that only intraband transitions matter. It is then logical that an intermediate crossover regime exists where the frequency is neither resonant to interband nor intraband transitions and both two lifetimes matters.

On a formal level, compared to implementing a simple lifetime broadening in all Green's functions $G(\omega) \rightarrow G(\omega + i/(2\tau))$, in a multiband setting it is worthwhile to consider several, band-specific broadening factors. For example, when comparing conduction band and valence band electrons, one can immediately deduce from their respective density of states that a more dispersive band will exhibit weaker disorder scattering, thus giving rise to a dissimilar lifetimes. This holds true even before taking into account orbitally selective interaction or disorder effects. 
If the quasiparticle excitations are relatively long-lived, one can assume that each decay process is statistically independent. In that case, staying within the two-band picture, the interband and intraband lifetime can be related to individual lifetimes from conduction and valence band through Matthiessen’s Rule, i.e.
\begin{align}
\tau_{\mathrm{inter}}^{-1} &= (2\tau_{c})^{-1} + (2\tau_{v})^{-1}\\
\tau_{\mathrm{intra}}^{-1} &= (2\tau_{v})^{-1} + (2\tau_{v})^{-1}.
\end{align}

Two physically important cases can be distinguished. 
In the optical regime (\(\omega \tau \gg 1\)), the quasiparticles in the conduction and valence band are both on-shell, carrying their respective microscopic lifetimes $\tau_{c}$ and $\tau_{v}$, which can lead to a nonlinear subgap response with a substantial amplitude and distinct frequency-dependence.

On the other hand, in the transport regime (\(\omega \tau \ll 1\)), the transport problem becomes entirely restricted to the low-energy regime close to the Fermi level.
In the latter case, it becomes important to enforce charge conservation in the low-energy sector. Mainly, this means that electrons which are (virtually) excited to the conduction band should not decay within the excited band but instead return to the valence band unperturbed, because otherwise the state is lost for the low-energy description. This requirement is straightforwardly implemented in the Feynman diagrammatic formalism by choosing $\tau_{v}/\tau_{c}\rightarrow 0$, which provides a natural transition from the optical regime to the transport regime. As outlined below, this approach can yield all key results of the intrinsic nonlinear transport such as the nonlinear Drude response, Berry curvature dipole, and quantum metric dipole - findings that align with those derived from a semiclassical Boltzmann formalism. Moreover, the quantum-coherent treatment extends beyond the capabilities of a semiclassical approach. On the other hand, a real material might violate the condition $\tau_{v}/\tau_{c}\rightarrow 0$, in which case a more elaborate quantum kinetic ansatz must be pursued.

\subsection{Subgap responses}\label{sec4B}
In subgap regime, the separation of the injection and shift current is no longer guaranteed because the response is completely off-resonant. Yet it still makes sense to discuss both off-resonant parts separately and see how they merge upon lower the drive frequency.

To make progress, the perturbation theory needs to be modified to include finite lifetimes such that they account for interband and intraband processes. To this end, the poles in the response functions are adjusted such that  $\omega - \epsilon_{mn}$ is replaced either by by $\omega - \epsilon_{mn} + i \tau_{\mathrm{inter}}^{-1}$ for the case $n \neq m$, which represents the interband transition. Conversely, in the case $n = m$ representing intraband transitions, it is replaced by $\omega - \epsilon_{nn} + i \tau_{\mathrm{intra}}^{-1}$.
For sake of clarity, let us focus on the longitudinal response, i.e. $a=b=c$. 
The leading-order, off-resonant conductivity in a time-reversal breaking compound is then~\cite{Kaplan2023},
\begin{align}
\sigma^{aa;a}_{\mathrm{inj}} + \sigma^{aa;a}_{\mathrm{shift}} 
&\approx -\frac{e^3}{\hbar^2 \omega^2} \left(\frac{2}{\alpha} - 1 \right) \sum_{nm} \int_{\mathbf{k}} f_{nm}  |r^a_{nm}|^2 \Delta_{nm}^a 
\end{align}
where a lifetime dependence enters through the parameter $\alpha=\tau_{\mathrm{inter}}/\tau_{\mathrm{intra}}$.
For large enough frequencies, typically near the band edge, the subgap signal is still sufficiently strong. Since the value of $\alpha$ depends on the microscopic lifetimes in the system, this subgap signal thus offers a window into the short-time dynamics across the band gap. On the other hand, as $\omega$ decreases, charge conservation enforces that $\tau_{\mathrm{inter}}\rightarrow 2\tau_{\mathrm{intra}}$, hence $\alpha\rightarrow 2$ and the off-resonant ac conductivity thus tends to zero very quickly inside the gap.

\section{Nonlinear Transport}\label{sec5}
The charge conductivity at linear order famously carries an anomalous Hall effect which is due to a modification of the semiclassical equation of motion by the Berry curvature. The Berry curvature is a ground state property which encodes the self-rotation of the corresponding Bloch state. Crucially, while the Berry curvature is a ground state property, it can only assume nonzero values if the band structure entails more than a single band. 
Therefore, the \emph{anomalous motion} which is induced by the Berry curvature is clearly a multi-band effect, seemingly depending on the entire band structure, despite the fact that the wavepacket propagation itself exclusively involves states close to the Fermi surface. This apparent contradiction is resolved by noting that quantum-mechanical operators can acquire renormalized expectation values due to virtual (interband) fluctuations.

\begin{figure}[htbp]
    \begin{center}
        \begin{tabular}{cccc}
            \multirow{3}{*}{\makecell[c]{Nonlinear\\Drude\\term}} & 
            \begin{minipage}{0.2\linewidth}
                $$\propto \int_{\mathbf{k}} f^{(2)}_{\mathbf{k}} \mathbf{v}^{(0)}_{\mathbf{k}}$$
            \end{minipage}
            & &
            \begin{minipage}{0.4\linewidth}
                \centering
                \includegraphics[width=\linewidth]{positional_shift.png}
            \end{minipage}
            \\
            & & & \\
            & (a) & & (d) \\
            & & & \\
            \multirow{3}{*}{\makecell[c]{Berry\\Curvature\\Dipole}} & 
            \begin{minipage}{0.2\linewidth}
                $$\propto \int_{\mathbf{k}} f^{(1)}_{\mathbf{k}} \mathbf{v}^{(1)}_{\mathbf{k}}$$
            \end{minipage}
            & &
            \begin{minipage}{0.5\linewidth}
                \centering
                \includegraphics[width=\linewidth]{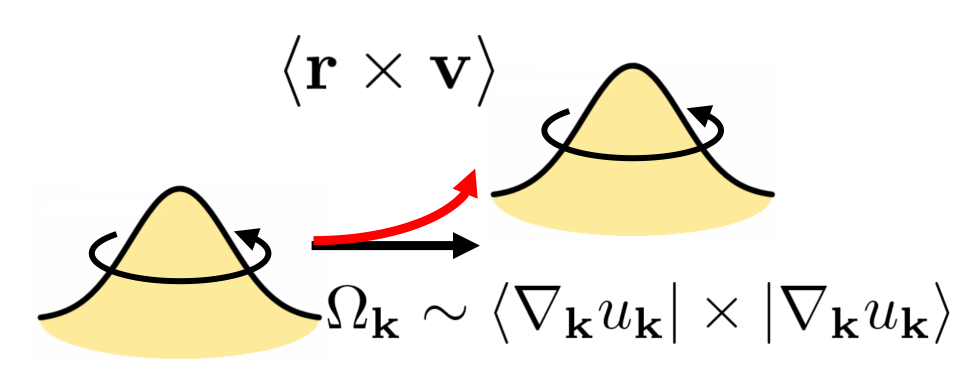}
            \end{minipage}
            \\
            & & & \\
            & (b) & & (e) \\
            & & & \\
            \multirow{3}{*}{\makecell[c]{Quantum\\Metric\\Dipole}} & 
            \begin{minipage}{0.2\linewidth}
                $$\propto \int_{\mathbf{k}} f^{(0)}_{\mathbf{k}} \mathbf{v}^{(2)}_{\mathbf{k}}$$
            \end{minipage}
            & &
            \begin{minipage}{0.5\linewidth}
                \centering
                \includegraphics[width=\linewidth]{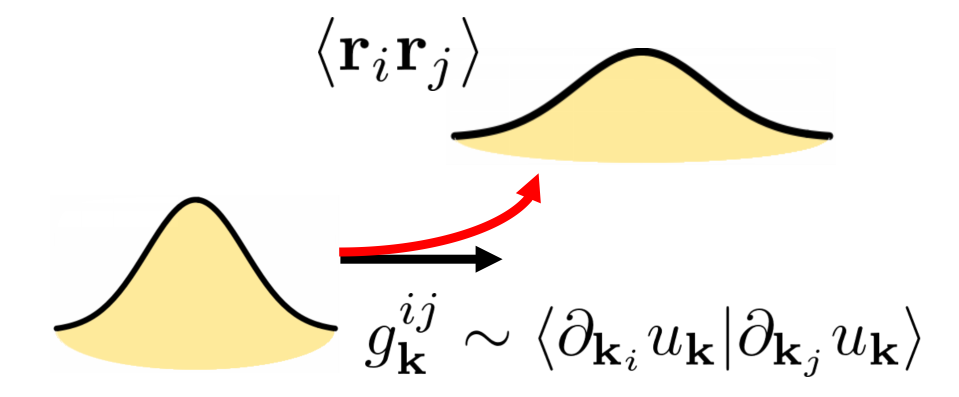}
            \end{minipage}
            \\
            & & & \\
            & (c) & & (f) \\
            & & & \\
            \multicolumn{2}{c}{\makecell[c]{\textbf{Response Theory}\\(Renormalized $\varepsilon_{\mathbf{k}}$ \& $\Omega_{\mathbf{k}}$)}} & $\leftrightarrow$ & \makecell[c]{\textbf{Geometric Interpretation}\\(Deformable Wavepacket)} \\[-1em]
            % \multicolumn{2}{c}{(Renomalized $\varepsilon_{\mathbf{k}}$ \& $\mathbf{\Omega}_{\mathbf{k}}$)} &  & (Deformable Wavepacket) \\
        \end{tabular}
    \end{center}
\caption{ Physical features of nonlinear charge transport, which couples to the average acceleration of the wavepacket motion. The kinetic and anomalous parts of the motion can be understood either in terms of semiclassical renormalizations to the Fermi velocity and Berry curvature, or as deformations of the semiclassical wavepacket. Finally, the triangle diagram gives rise to a triangle anomaly which is of the mixed axial-gravitational type and results in a longitudinal nonreciprocity in the current on the level of the quantum metric dipole.
}\label{fig5}
\end{figure}

The same guiding principles regarding anomalous components of the motion can be applied in the nonlinear regime.
Nevertheless, semiclassically it is tough to capture all pieces of the anomalous motion at second order in the electric field, depending on which operators are modified in the semiclassics~\cite{Chang2008,Gao2014,Sodemann2015,Gao2019,Kaplan2024,Jia2024}.
Starting from the finite-frequency expressions and performing a
low-frequency expansion is likewise a subtle endeavor~\cite{Parker2019,deJuan2020,Kaplan2023} because the limit $\omega\tau\rightarrow0$ has to be taken while keeping $\tau<\infty$.
Even though all recent works agree that the nonlinear dc-conductivity has three major components, i.e. \emph{nonlinear Drude}, \emph{Berry curvature dipole}, and \emph{quantum metric dipole} (cf. Fig.~\ref{fig5}), it is no surprise that varying predictions have been made regarding the coefficients and signs of the respective dc-conductivity terms~\cite{Watanabe2020,Watanabe2021,Oiwa2022,Das2023,Jia2024,Kaplan2024}.
In the following, we present a particularly simple perspective which emerges based on the renormalized kinetic approach as well as Kubo formalism.

\subsection{Berry curvature and quantum metric dipole}\label{sec5A}

We first elucidate within the renormalized kinetic approach how nonlinear order anomalous transport including Berry curvature dipole and quantum metric dipole arise.
Semiclassically, the current density in a material can be expressed as:
\begin{equation}
\mathbf{j} = -e \sum_{n} \int_\mathbf{k} f_n \mathbf{v}_n
\label{eq:smcurrent}
\end{equation}
where \(f_n\) is the (non-equilibrium) distribution function of the \(n\)-th band, \(v_n^a\) is the velocity of the \(n\)-th band in $a$-direction. Momentum arguments will be suppressed throughout in this section. 

The external electric field \(\mathbf{E}\) is considered as the perturbation to the system, under which the evolution of the distribution function \(f_n(\mathbf{r},\mathbf{k}, t)\) follows the Boltzmann transport equation, i.e.
\begin{equation}
\partial_t f_n - e \frac{\mathbf{E}}{\hbar} \cdot \nabla_{\mathbf{k}} f_n + \mathbf{v} \cdot \nabla_{\mathbf{r}} f_n = - \frac{f_n - f_n^{(0)}}{\tau},\label{boltzmann equation}
\end{equation}

where \(\tau\) is the transport relaxation time and \(f_n^{(0)}\) is the equilibrium distribution function of $n$-th band, i.e. Fermi-Dirac distribution function. Under the assumption of a steady-state and spatially uniform system, time and spatial derivative terms can be excluded in Eq.~\eqref{boltzmann equation} and the distribution function can be solved perturbatively by expanding with respect to the amplitude of electric field: \(f_n = f_n^{(0)} + f_n^{(1)} + f_n^{(2)} + \dots\), whose solution yields $f_n^{(l)} = \left( \frac{e \tau}{\hbar} \mathbf{E} \cdot \nabla_{\mathbf{k}} \right)^l f_n^{(0)}$.

Crucially, a determination of the current [Eq.~\eqref{eq:smcurrent}] also requires an expansion of \(\mathbf{v}_n\) with respect to the perturbation. The velocity of an electron in $n$-th band is given by the semiclassical equation of motion~\cite{Xiao2010}, which also holds at nonlinear order~\cite{Gao2014,Gao2019}:
\begin{equation}
\mathbf{v}_n = \frac{1}{\hbar} \nabla_{\mathbf{k}} \varepsilon_n - \frac{e}{\hbar} \mathbf{E} \times \mathbf{\Omega}_n.
\end{equation}

The first term $\frac{1}{\hbar} \nabla_{\mathbf{k}} \varepsilon_n$ represents the normal group velocity of semiclassical wavepacket, which only depends on the energetics of band. The second term captures the anomalous contribution due to Berry curvature $\mathbf{\Omega}_n$. Under the perturbation of electric field $\hat{H}' = -e \hat{\mathbf{r}} \cdot \mathbf{E}$, the band dispersion $\varepsilon_n$ and the wavefunction $|u_n\rangle$ as well as the Berry curvature $\mathbf{\Omega}_n$ will be renormalized. They can also be formally expanded with respect to $\mathbf{E}$, denoted as $\varepsilon_n = \varepsilon_n^{(0)} + \varepsilon_n^{(1)} + \varepsilon_n^{(2)} + \dots$ and $\mathbf{\Omega}_n = \mathbf{\Omega}_n^{(0)} + \mathbf{\Omega}_n^{(1)} + \mathbf{\Omega}_n^{(2)} + \dots$, which further results in a renormalized velocity expanded as $\mathbf{v}_n = \mathbf{v}_n^{(0)} + \mathbf{v}_n^{(1)} + \mathbf{v}_n^{(2)} + \dots$, where $\mathbf{v}_n^{(l)} = \frac{1}{\hbar} \frac{\partial \varepsilon_n^{(l)}}{\partial \mathbf{k}} - \frac{e}{\hbar} \mathbf{E} \times \boldsymbol{\Omega}_n^{(l-1)}$. The renormalized dispersion and Berry curvature can be determined in perturbation theory by a Schrieffer–Wolff transformation as detailed in App.~\ref{appF1}. 

After inserting the renormalized quantities, the charge current and the corresponding conductivity can be evaluated at arbitrary order. Specifically, linear order and second order current densities have the form,
\begin{align}
& \mathbf{j}^{(1)}=-e \int_\mathbf{k} \sum_n f_n^{(1)} \mathbf{v}_n^{(0)}+f_n^{(0)} \mathbf{v}_n^{(1)} \\
& \mathbf{j}^{(2)}=-e \int_\mathbf{k} \sum_n f_n^{(2)} \mathbf{v}_n^{(0)}+f_n^{(1)} \mathbf{v}_n^{(1)}+f_n^{(0)} \mathbf{v}_n^{(2)}.
\end{align}

For linear order current, these are the well-known Drude term ($\propto f^{(1)}\mathbf{v}^{(0)}$) and the anomalous velocity term ($\propto \mathbf{v}^{(1)}$), where the former is proportional to relaxation time $\tau$ because of linear order change in distribution function $f^{(1)}$ while the anomalous velocity term does not carry an explicit dependence on the relaxation time $\tau$. 

Similarly in the second order response, nonlinear Drude term (NLD), Berry curvature dipole (BCD), and quantum metric dipole (QMD) can be differentiated by their relaxation time dependences, which originate from different ordered perturbation on the distribution function as shown in Fig.~\ref{fig5},a-c. After substituting the expression of renormalized velocity, the 2nd order transport conductivity then becomes $\sigma^{ab;c} \equiv \sigma^{ab;c}_{\mathrm{NLD}} + \sigma^{ab;c}_{\mathrm{BCD}} + \sigma^{ab;c}_{\mathrm{QMD}} $, where each piece is given by~\cite{Gao2014,Gao2019,Watanabe2020,Watanabe2021,Sodemann2015,Wang2021,Das2023,Oiwa2022,Kaplan2024,Jia2024}
\begin{align}
\sigma^{ab;c}_{\mathrm{NLD}} & = - \frac{e^3 \tau^2}{\hbar^3} \sum_n \int_{\mathbf{k}} f_n \partial_a \partial_b \partial_c \varepsilon_n \label{eqSmain1} \\
\sigma^{ab;c}_{\mathrm{BCD}} & = \frac{e^3 \tau}{\hbar^2} \sum_n \int_{\mathbf{k}} f_n \frac{1}{2} (\partial_a \Omega_n^{bc} + \partial_b \Omega_n^{ac}) \label{eqSmain2} \\
\sigma^{ab;c}_{\mathrm{QMD}} & = - \frac{e^3}{\hbar} \sum_n \int_{\mathbf{k}} f_n \left[ 2 \partial_c G_n^{ab} - \frac{1}{2}( \partial_a G_n^{bc} + \partial_b G_n^{ac} ) \right]
\label{eqSmain3}
\end{align}

Here, the conductivities are already written exclusively in terms of equilibrium band structure properties, i.~e. all quantities are evaluated in the ground state and carry the superscript $\Box^{(0)}$. Namely, this includes the equilibrium dispersion $\varepsilon_n$ and occupation function $f_n$ for band index $n$, as well as the band-resolved Berry curvature $\Omega^{ab}_n$ and band-resolved, normalized quantum metric $G^{ab}_n$, defined as
\begin{align}
\Omega^{ab}_n &= \sum_{m \neq n} \left( r^a_{nm} r^b_{mn} - r^b_{nm} r^a_{mn} \right)
\\
G^{ab}_n &= \sum_{m \neq n} \frac{ r^a_{nm} r^b_{mn} + r^b_{nm} r^a_{mn} }{\varepsilon_{nm}}.
\end{align}
Both $\Omega^{ab}_n$ and $G^{ab}_n$ are gauge invariant and contain only one band index $n$, which means they are proper low energy quantities. Of course, their definitions entail a sum over all other bands, clearly identifying them as virtual interband contributions. Additionally, we emphasize that $G^{ab}_n$ depends explicitly on the dispersion, which means that it cannot be purely geometric - it is necessarily a \emph{mixed} property of the system. This has to be seen in distinction to  the Berry curvature $\Omega^{ab}_n$, which only depends on the wavefunctions.

We reiterate that NLD, BCD and QMD are each measurable independently, and can for example be isolated by their lifetime dependence~\cite{Gao2023,Wang2023}. We also mention that Kubo formalism with two lifetime has been shown to reproduce the NLD, BCD, and QMD terms~\cite{Kaplan2023}, for further details cf. App.~\ref{appendix:diagrammatics_transport}.

\subsection{Geometric interpretation and quantum anomaly}\label{sec5B}

In the same way that the linear conductivity measures the semiclassical mobility (drift velocity) of a wavepacket, the second-order conductivity couples to the average acceleration that the wavepacket is exposed to~\cite{Holder2021}. In particular, the anomalous components created by the BCD and QMD encode an anomalous acceleration which is created due to the intra-unit cell variation of the electron density that exerts forces on the extended wavepacket which are not locally compensated all the time, thus deforming the propagating state. This view is also completely in line with the observation that the BCD and QMD derive from virtual interband transitions, i.~e. they originate from (virtual) changes of the total wavefunction, thus deforming the real-space charge distribution.

This intuition can be made precise upon turning the nonlinear dc-conductivity according to Eqs.~(\ref{eqSmain1}-\ref{eqSmain3}) into an integral over the Fermi surface by partial integration. Then, the NLD term reads as a Fermi surface average over the k-local effective mass of the quasiparticle, and thus measures the skewness of the drift motion (cf.~Fig.~\ref{fig5}d). Similarly, the BCD weighs the Berry curvature on the Fermi surface along the electric field directions, which results in an angular momentum mismatch (Fig.~\ref{fig5}e). Finally, the QMD accounts for the imbalance in the shift of the dispersion, which will deform the local Fermi surface geometry, and thus the wavepacket (Fig.~\ref{fig5}f).

The fact that anomalous components of the motion derive from virtual interband excitations also constitutes the physical mechanism for semiclassical manifestations of quantum anomalies, for example for the chiral anomaly in Weyl semimetals~\cite{Son2013,Yan2017,Ilan2020}.
A quantum anomaly denotes the non-conservation of a charge even though it is conserved by the classical action. The reason is that the measure in the quantum statistical partition function is not invariant under the corresponding symmetry. One could say more sloppily that the quantum fluctuations enable additional transitions which are classically forbidden and which break the conservation law.
Quantum anomalies are frequently induced by triangle diagrams, so it is no surprise that the triangle diagram of the nonlinear conductivity gives rise to the
mixed axial-gravitational anomaly~\cite{Holder2021a}, which is normally only accessible in the presence of a temperature gradient~\cite{Landsteiner2016,Gooth2017,Chernodub2022}.
The mixed axial-gravitational anomaly results from a shift in the absolute value of the free energy, and thus of the chemical potential. Such a shift is supplied by the renormalization of the dispersion in the QMD response, leading to a non-conservation of the momentum-local neighborhood of a given point on the Fermi surface. For illustration, consider a very simple bandstructure  with two Fermi pockets. If the quantum metric dipole is nonzero, it can for example move one Fermi pocket up in energy, and one down, which conserves total charge in accordance with the global electrostatic requirements. However, the semiclassical continuity equations for each of these (shifted, displaced and resized) Fermi pockets will then be anomalous, allowing for a small leakage current between them. Completely analogous to the more well-known chiral anomaly, this leakage current cannot be removed entirely unless one assumes that the scattering between both Fermi pockets is very strong~\cite{Son2013,Holder2021a}, which is the origin of the longitudinal QMD current.

\subsection{Beyond semiclassics}\label{sec5E}
In a band insulator, the Fermi surface contributions to the conductivity [Eqs.~(\ref{eqSmain1}-\ref{eqSmain3})] vanish. It is a question of fundamental interest whether it is nonetheless possible for a nonlinear quantum anomalous Hall effect (NQAHE) to exist.
This holds true in particular because it is widely appreciated that at linear order the quantum Hall effect and the quantum anomalous Hall effect are highly analogous in terms of the transport properties and quantized edge channels~\cite{Xiao2010,Yu2010}.

This question has recently been answered in the positive~\cite{Kaplan2023a}; a NQAHE can indeed arise both in time-reversal symmetric systems as well as in magnetic insulators, provided that the band structure does not possess any spatial symmetries. The latter property may be due to an intrinsically low internal symmetry, or could be induced by strain.

Given these circumstances, the NQAHE current is then $\bm{j}=\bm{E} \times \bm{\sigma}_{\mathrm{NQAHE}}$, where $\bm{\sigma}_{\mathrm{NQA}}$ is linear in electric field and defined as
\begin{align}
    \sigma^a_{\mathrm{NQAHE}} 
    &= \frac{e^2}{\hbar} \sum_{n} 
    \int_{\mathbf{k}}
    \epsilon^{abc} f_n \Bigl[\frac{\mathcal{A}^b}{\varepsilon} , \mathsf{v}^c + \mathsf{S}^c + \mathsf{\Omega}^c \Bigr]_{nn}.
    \label{eq:semiclass}
\end{align}
The effect arises due to three operator renormalizations which are all linear in the electric field and which are related to, respectively, a velocity shift $\mathsf{v}$, a positional shift $\mathsf{S}$ and finally a Berry curvature shift $\mathsf{\Omega}$ (cf. App.~\ref{appendix:diagrammatics_transport}).

Even though the NQAHE effect typically leads to a rather small correction to the quantized response, it is nonetheless noteworthy for the fact that it can even appear in a time-reversal-breaking insulator with zero Chern number at linear order. Furthermore, the appearance of $\bm{\sigma}_{\mathrm{NQAHE}}$ shows that the close agreement between Quantum Hall and Quantum Anomalous Hall phenomenology does not extend to all orders in the applied field. This does make sense based on the fact that the Quantum Hall effect is exactly linear in electric field, while such a property cannot be expected from a dispersive band structure.

\subsection{Symmetry separation}

\begin{table}[htbp]
    \centering
    \renewcommand{\arraystretch}{1.2} % Adjust row spacing
    \begin{tabular}{cccccc}
    \hline\\[-1.2em]
    \hline
        Symmetry & $\mathcal{P}$ & $\mathcal{T}$ & $\mathcal{PT}$ & $\mathcal{C}_{3z}$ & $(a \leftrightarrow b \leftrightarrow c)$ \\ \hline
        Nonlinear Drude term & \cross & \cross & \checkmark & \checkmark & inherent \\ 
        Berry curvature dipole & \cross & \checkmark & \cross & \cross$^*$ & absent \\ 
        Quantum metric dipole & \cross & \cross & \checkmark & \checkmark & absent \\ 
        \hline\\[-1.2em]
        \hline
    \end{tabular}
    \caption{Symmetry properties of nonlinear transport, listed for the basic symmetry operations as explained in Table.~\ref{table1}. Checkmarks denote allowed response types. A special case is the BCD for $\mathcal{C}_{3z}$ symmetry (denoted by the star), which is symmetry forbidden for all in-plane ($x$-$y$ plane) components, but allowed for out-of-plane directions. Cyclic permutations of the indices are an inherent symmetry only for the NLD term.}
    \label{nonlinear transport symmetry}
\end{table}

Similar to nonlinear optics, it is also possible to utilize the permutation symmetry of the dc conductivity tensor and the symmetry of the material to separate the NLD, BCD, and QMD contributions as detailed in Table~\ref{nonlinear transport symmetry}.

Regarding the role of inversion ($\mathcal{P}$), time-reversal ($\mathcal{T}$), and parity-time reversal ($\mathcal{PT}$) symmetries, let us reiterate that $\mathcal{P}$-symmetry disables all 2nd order charge current responses (cf. Sec.~\ref{optical symmetry}). Regarding $\mathcal{T}$- and $\mathcal{PT}$-symmetry, as further elucidated in Appendix~\ref{transport symmetry}, both NLD and QMD are odd under time reversal transformation, while BCD is even. Therefore, NLD and QMD can only be present in magnetic systems where $\mathcal{T}$ is broken, while the BCD will be excluded in $\mathcal{PT}$-symmetric materials such as antiferromagnets. 

%We then dive into more complicated (magnetic) point group symmetries. 
Regarding pure rotational symmetries, although all 2nd-order conductivity tensors share the same rotational transformation properties, it is still possible to distinguish BCD from NLD and QMD by $\mathcal{C}_{3z}$-symmetry. THe reason being that $\mathcal{C}_{3z}$-symmetry connects a pure Hall (transverse) conductivity to a longitudinal component while BCD is a pure transverse response, i.e. 
\begin{equation}
    \sigma^{yy;x}_{\mathrm{BCD}} \stackrel{\mathcal{C}_{3z}}{=} -\sigma^{xx;x}_{\mathrm{BCD}} = 0.
\end{equation}
For magnetic point group symmetries such as $\mathcal{C}_{n}\mathcal{T}$ and $\mathcal{MT}$, we expect a separation of responses based on spatial directions, similar to the behavior in the optical regime. For instance, a $\mathcal{M}_{z}\mathcal{T}$ system is effectively $\mathcal{T}$-symmetric in $x$-$y$ plane while being $\mathcal{PT}$-symmetric in $z$-direction. Then in $x$-$y$ plane only BCD is allowed while in $z$-direction, only NLD and QMD is allowed. Such separation of responses can also be found for other (magnetic) point group symmetries. Their effective symmetries in different sub-directions are detailed in Table.~\ref{effective symmetry}.

Regarding the permutation symmetry of the conductivity tensor, unlike in optical regime where nonlinear conductivity can generally have both symmetric ($\sigma^{(a,b);c}$) and antisymmetric ($\sigma^{[a,b];c}$) part, the transport conductivity must be symmetric when imposing $(a \leftrightarrow b)$ permutation since $\omega \to 0$ eliminates the difference between $\omega$ and $-\omega$ upon exchange of the spatial indices of the electric field (cf. App.~\ref{app02E}). Additionally, NLD has a fully permutation symmetry $(a \leftrightarrow b \leftrightarrow c)$ since $\sigma^{ab;c}_{\mathrm{NLD}} \propto \int_{\mathbf{k}} \partial_{a} \partial_{b} \partial_{c} \varepsilon_{\mathbf{k}}$, which is notably absent in both BCD and QMD. %which is actually a manifestation of \emph{mixed axial-gravitational anomaly} as mentioned in Sec.~\ref{sec5B}. 
Therefore, even though NLD and QMD are both present in a $\mathcal{PT}$-symmetric system and contribute to both the nonlinear Hall (transverse) and the longitudinal (nonreciprocal) current, it is still possible to isolate certain pieces by mixing and matching spatial components, for example to distill the quantum metric effect~\cite{Holder2021a}. This can be achieved by measuring both transverse conductivity (e.g. $\sigma^{yy;x}$) and off-axis conductivity (e.g. $\sigma^{xy;y}$) with the transport setup shown in Fig.~\ref{symmetry separation} (b). Specifically, NLD contributes the same to both two conductivity components, while QMD generally does not, i.e.
\begin{equation}
    \sigma^{yy;x}_{\mathrm{NLD}} = \sigma^{xy;y}_{\mathrm{NLD}}, \quad \sigma^{yy;x}_{\mathrm{QMD}}  \neq \sigma^{xy;y}_{\mathrm{QMD}}
\end{equation}

\subsection{Extrinsic effects}\label{sec5F}

Finally, we emphasize that the dc-conductivity as given by Eqs.~(\ref{eqSmain1}-\ref{eqSmain3}) only contains anomalous pieces due to virtual interband processes. Additionally, there exist extrinsic contributions due to quasiparticle scattering from impurities, phonons, and collective modes. 
Qualitatively speaking, the anomalous intrinsic parts of the motion are the result of virtual (i.~e. vertical) transitions involving different energies, while extrinsic scattering leads to on-shell (horizontal) transitions involving more than one momentum.

In close analogy to processes found for the first-order extrinsic mechanisms~\cite{Nagaosa2010,Du2021}, nonlinear skew-scattering and side-jump have been identified as major sources of the extrinsic conductivity~\cite{Du2019,Atencia2022}. Additionally, quantum kinetic approaches have discovered genuine second-order relaxation paths~\cite{Nandy2019,Xiao2019,Gao2020a,Du2021a,Koenig2021,Mehraeen2024}.
The study of the role of quantum geometry in such extrinsic processes is still in its infancy, partially because the response function will invariably depend on the microscopic details of the impurities or collective modes.
However, some results are known. For example, in case of a simple potential scatterer, skew and side-jump at linear order can be connected to the Pancharatnam phase~\cite{Sinitsyn2006a}.

Extrinsic processes often lead to currents comparable in size to intrinsic ones and can by no means be neglected unless some symmetry arguments exist which suppress them or render them zero (cf. for example~\cite{Watanabe2021,Du2021}). 
Nevertheless, similar to Hall responses at linear order, recent works have successfully disentangled intrinsic and extrinsic contributions~\cite{Wang2023,Gao2023} by investigating the temperature scaling of the nonlinear Hall conductivity versus the linear longitudinal conductivity squared, which removes the leading order lifetime dependence and thus helps to highlight the anomalous components of the conductivity.

\section{Summary}\label{sec7}
In this review, we elucidated the rich  structure of nonlinear charge responses using several, mutually complementing concepts. The development of these concepts have been ignited and fueled by rapid experimental progress, fertilizing a profound reevaluation of the established intuition regarding the semiclassical electron motion in the bulk solid and culminating in the prediction of a range of novel nonlinear effects.
At the same time, the multi-faceted nature of the electron motion beyond linear order has proven indispensable in the exploration and characterization of modern layered and engineered quantum materials, in particular in moving the observational techniques towards measuring features of quantum geometry in nonlinear quantum materials. 

Looking forward, we firmly believe that further progress can be expected from a close collaboration between experimental and theoretical efforts.  
Currently, most theory estimates are made \emph{a posteriori} because it remains a major challenge to render ab-initio models as well as effective low-energy models  quantitative and predictive.
Only a joint theoretical/experimental effort can reduce these uncertainties and establish protocols to isolate certain materials properties in tractable fashion.

On the other hand, quantum geometry is rapidly developing into a valuable tool for a range of condensed matter questions~\cite{Toermae2023,Yu2025}, for which  multicomponent/tensorial response functions are the primary observational probe.
The versatility of the concepts reviewed here have already sparked several adjacent developments. 
For example, using similar methods and language, it is possible to study nonlinear features with added magnetic field~\cite{Gao2019,Dai2023b,Xiang2024}.
Furthermore, extensions exist for charge transport at third order in the field~\cite{Fregoso2019,Liu2022a}, which is relevant in monolayer
graphene~\cite{Cheng2014,Cheng2015} and in altermagnets~\cite{Fang2024}.

Zooming out, as this review powerfully demonstrates it is worthwhile and insightful to explore nonlinear response functions because they give rise to genuine new effects that can be connected not only with material-specific features, but also with fundamental concepts at the very core of condensed matter physics.

\begin{acknowledgements}

We thank Daniel Kaplan for fruitful discussions.
T.H.\ acknowledges financial support by the 
European Research Council (ERC) under grant QuantumCUSP
(Grant Agreement No. 101077020). B.Y. acknowledges the financial support by the European Research Council (ERC Consolidator Grant ``NonlinearTopo'', No. 815869) and the Israel Science Foundation (ISF: 2932/21, 2974/23)

\end{acknowledgements}

\appendix

\section{Notations}\label{app00}

Let us briefly introduce the rationale and language behind nonlinear conductivities.

\paragraph{Conventions for Fourier transforms}
\begin{equation}
\partial_a \equiv \tfrac{\partial}{\partial k_{a}},\quad 
\int[d\omega] \equiv \int \tfrac{d\omega}{2\pi},\quad
\int[d^D\mathbf{k}] \equiv \int \tfrac{d^D\mathbf{k}}{(2\pi)^D}
\end{equation}
For a time-dependent physical quantity $\mathcal{O}(t)$, we define its Fourier component $\mathcal{O}(\omega)$ as,
\begin{equation}
\mathcal{O}(t) = \int[d \omega] e^{-i \omega t} \mathcal{O}(\omega)
,\quad
\mathcal{O}(\omega) = \int d t e^{i \omega t} \mathcal{O}(t)
\end{equation}

For a spatial uniform, time-dependent electric field $\mathbf{E}(t)$ and its corresponding vector potential $\mathbf{A}(t)$, we have,
\begin{equation}
\begin{aligned}
E^a (t) & 
% = \int [d \omega] e^{-i \omega t} E^a (\omega) \\
=  - \partial_t A^a (t) 
% & = \int [d \omega] (i \omega ) e^{-i \omega t} A^a (\omega)
\end{aligned}
\end{equation}
Therefore,
\begin{equation}
E^a (\omega) = i \omega A^a (\omega) \ \Leftrightarrow \  A^a (\omega) = \frac{E^a(\omega)}{i \omega}
\end{equation}

\paragraph{Definition of current and conductivity}

We define different orders of conductivity in the time-domain as,
\begin{widetext}
\begin{equation}
\begin{aligned}
j^a (t) \equiv & \int d t_1 \ \sigma^{b_1;a}(t;t_1) E^{b_1}(t_1) +\frac{1}{2!} \int d t_1 \int d t_2  \ \sigma^{b_1b_2;a}(t; t_1, t_2) E^{b_1}(t_1) E^{b_2}(t_2) + \dots 
\end{aligned}
\end{equation}
Therefore, the conductivity can be formally expressed in functional derivative form,
\begin{equation}
\sigma^{b_1 \dots b_n;a}(t;t_1,\dots,t_n) = \frac{\delta^n j^a(t)}{\delta E^{b_1}(t_1) \dots \delta E^{b_n}(t_n)}.
\end{equation}
The expression of current in frequency domain can be written as,
\begin{equation}
\begin{aligned}
j^a (\omega) & = \int d t e^{i \omega t} j^a(t) \\
&= \int d t e^{i \omega t} \int d t_1 \ \sigma^{b_1;a}(t;t_1) E^{b_1}(t_1) +\frac{1}{2!} \int d t e^{i \omega t} \int d t_1 \int d t_2 \ \sigma^{b_1b_2;a}(t; t_1, t_2) E^{b_1}(t_1) E^{b_2}(t_2) +\dots \\
&= \int [d \omega_1] \sigma^{b_1;a}(\omega;\omega_1)  E^{b_1}(\omega_1) +\frac{1}{2!} \int [d \omega_1] \int [d \omega_2] \sigma^{b_1b_2;a}(\omega; \omega_1, \omega_2) E^{b_1}(\omega_1) E^{b_2}(\omega_2) +\dots \\
\end{aligned}
\end{equation}
where we have defined the $n$-th order conductivity in frequency domain as,
\begin{equation}
\sigma^{b_1\dots b_n;a}(\omega;\omega_1,\dots,\omega_n) \equiv  \int \dots \int d t d t_1 \dots d t_n e^{i \omega t} \sigma^{b_1\dots b_n;a}(t;t_1,\dots,t_n)  e^{-i \omega_1 t_1 } \dots  e^{-i \omega_n t_n }.
\end{equation}

\end{widetext}

\paragraph{Continuous and discrete form of response theory}\label{app01}
In standard 2nd order current response, the response can be written as,
\begin{equation}
j^c (\bar{\omega}) = \int [d\omega_1] \int [d \omega_2]\sigma^{ab;c} (\bar{\omega}; \omega_1,\omega_2) E_a(\omega_1) E_b(\omega_2).
\end{equation}
To further simplify our notation, we consider a monochromatic electric field.
\begin{equation}
\begin{aligned}
E^a(t) & = \mathcal{E}^a(\omega) e^{-i\omega t} + {\mathcal{E}^a}^*(\omega) e^{i\omega t} \\
& =  \left[ \operatorname{Re}\mathcal{E}^a(\omega) + i \operatorname{Im}\mathcal{E}^a(\omega) \right] \left( \cos\omega t - i \sin \omega t \right) + c.c.  \\
& = 2 \left[\operatorname{Re}\mathcal{E}^a(\omega) \cos \omega t + \operatorname{Im} \mathcal{E}^a(\omega) \sin \omega t \right].
\end{aligned}
\end{equation}
Here $\vec{\mathcal{E}}$ can be regarded as the amplitude vector of the electric field, then for linear polarized light in x direction, we can use $\vec{\mathcal{E}}(\omega) = |\vec{\mathcal{E}}| (1,0,0)$, for circular polarized light in x-y plane propagating in z direction, we can use $\vec{\mathcal{E}}_{\circlearrowleft/\circlearrowright}(\omega) = \frac{|\vec{\mathcal{E}}|}{\sqrt{2}} (1,\pm i,0)$, where $|\vec{\mathcal{E}}| \equiv \sqrt{\vec{\mathcal{E}}^* \cdot \vec{\mathcal{E}}}$ is half the amplitude of the electric field.

Then the electric field in frequency domain can be written as,
\begin{equation}
\begin{aligned}
E^a (\omega') & = \int dt e^{i \omega' t} E^a (t) \\
& = \int dt e^{i \omega' t}  (\mathcal{E}^a(\omega) e^{-i\omega t} + {\mathcal{E}^a}^*(\omega) e^{i\omega t}) \\
& = 2\pi \left[ \mathcal{E}^a(\omega) \delta(\omega'-\omega) + \mathcal{E}^a(-\omega) \delta(\omega'+\omega) \right].
\end{aligned}
\end{equation}
Let us generalize this notation based on $\vec{\mathcal{E}}$ into multi-frequency $\{\omega_i\}$ electric fields by writing,
\begin{equation}
E^a(t) = \sum_i \mathcal{E}^a(\omega_i) e^{-i\omega_i t},
\end{equation}
where we define $\mathcal{E}^a(-\omega_i) = {\mathcal{E}^a}^*(\omega_i)$ to keep the electric field real-valued. Then $E^a(\omega)$ has a discrete spectrum,
\begin{equation}
E^a (\omega) = 2\pi \sum_i \mathcal{E}^a(\omega_i) \delta(\omega - \omega_i).
\end{equation}
The response formula is therefore in terms of the amplitude $\vec{\mathcal{E}}$ given by,
\begin{widetext}
\begin{equation}
\begin{aligned}
j^c (\bar{\omega}) & = \int [d\omega_1] \int [d \omega_2]\sigma^{ab;c} (\bar{\omega}; \omega_1,\omega_2) E^a(\omega_1) E^b(\omega_2) \\
& = \int d\omega_1 \int d \omega_2 \ \sigma^{ab;c} (\bar{\omega}; \omega_1,\omega_2) \sum_{i,j} \mathcal{E}^a(\omega_i) \mathcal{E}^b(\omega_j) \delta(\omega_1 - \omega_i) \delta(\omega_2 - \omega_j) \\
& = \sum_{i,j} \sigma^{ab;c} (\bar{\omega}; \omega_i,\omega_j) \mathcal{E}^a(\omega_i) \mathcal{E}^b(\omega_j)
\end{aligned}
\end{equation}
By time translational symmetry the conductivity is proportional to a delta function as well. This delta function can be resolved by defining a similar discretized frequency spectrum description of current and conductivity,
\begin{equation}
j^c(t) = \sum_{i} \mathcal{J}^c(\bar{\omega}_i) e^{-i \omega_i t},\quad j^c (\bar{\omega}) = 2\pi \sum_i \mathcal{J}^c(\omega_i) \delta(\bar{\omega} - \bar{\omega}_i),\quad \sigma^{ab;c} (\bar{\omega}; \omega_1,\omega_2) = 2\pi \varsigma^{ab;c} (\bar{\omega}; \omega_1,\omega_2) \delta (\bar{\omega} - \omega_1 - \omega_2).
\end{equation}

This leads to the well-known discretized frequency spectrum form of the second order current generation,
\begin{equation}
\begin{aligned}
& j^c (\bar{\omega}) = 2\pi \sum_i \mathcal{J}^c(\omega_i) \delta(\bar{\omega} - \bar{\omega}_i) = 2\pi \sum_{j,k} \varsigma^{ab;c} (\bar{\omega}; \omega_j,\omega_k) \delta (\bar{\omega} - \omega_j - \omega_k) \mathcal{E}^a(\omega_j) \mathcal{E}^b(\omega_k),
\end{aligned}
\end{equation}

which is equivalent to,
\begin{equation}
\mathcal{J}^c(\bar{\omega}_i=\omega_j + \omega_k) = \varsigma^{ab;c} (\bar{\omega}; \omega_j,\omega_k)  \mathcal{E}^a(\omega_j) \mathcal{E}^b(\omega_k).
\end{equation}

\end{widetext}
This is just a transformation from the integral version of original continuous spectrum description into the summation form of discrete spectrum description. To avoid adding complexity in the use of letters in the formulas, in the main text we treat \(E\) and \(\mathcal{E}\) interchangeably, and uniformly write them as \(E\), understanding the meaning based on the specific context. Different symbols will be used for distinction only when necessary.

\section{Symmetry of 2nd order optical conductivity}\label{app02}

Symmetries and symmetry operations are an important tool in studying nonlinear conductivities. In the following we introduce the main ingredients towards such an analysis. 

\subsection{Intrinsic symmetry}\label{intrinsic symmetry}

\paragraph{Intrinsic permutation symmetry}
Consider the 2nd order current generation,
\begin{equation}
j^c (\bar{\omega}) = \sigma^{ab;c} (\bar{\omega}; \omega_1,\omega_2) E^a(\omega_1) E^b(\omega_2).
\end{equation}
The intrinsic permutation symmetry of this response is $(a,\omega_1) \leftrightarrow (b,\omega_2)$, because changing the order of multiplication of $E^a(\omega_1)$ and $E^b(\omega_2)$ leaves the current invariant. Therefore, the conductivity is symmetric with respect to this exchange of indices,
\begin{equation}
j^c (\bar{\omega}) = \sigma^{ba;c} (\bar{\omega}; \omega_2,\omega_1) E^b(\omega_2) E^a(\omega_1).
\end{equation}
It follows for the conductivity that 
\begin{equation}
\sigma^{ab;c} (\bar{\omega}; \omega_1,\omega_2) = \sigma^{ba;c} (\bar{\omega}; \omega_2,\omega_1),
\end{equation}
i.e. it is symmetric under the exchange $(a,\omega_1) \leftrightarrow (b,\omega_2)$.
In the following we consider 2nd order dc current generation, where $\omega_1 = -\omega_2 = \omega$ and $\bar{\omega} = \omega_1 + \omega_2 = 0$.
In the language of adiabatic switching, a finite quasiparticle lifetime enters through the interband and intraband relaxation rates $\Gamma$ and $\gamma$,
\begin{equation}
\omega_{1,2} \rightarrow \omega_{1,2} + i \Gamma, \quad \bar{\omega} \rightarrow \bar{\omega} + i \gamma,
\end{equation}
This yields,
\begin{equation}
\sigma^{ab;c}(\bar{\omega}; \omega_1,\omega_2;\{\gamma, \Gamma\})  = \sigma^{ab;c}(\bar{\omega}+i\gamma;\omega_1+i\Gamma,\omega_2+i\Gamma).
\end{equation}
The 2nd order dc conductivity tensor thus becomes
\begin{equation}
\begin{aligned}
\sigma^{ab;c}_{\mathrm{dc}}(\omega;\{\gamma, \Gamma\}) & \equiv \sigma^{ab;c}(0; \omega,-\omega;\{\gamma, \Gamma\}) \\
& = \sigma^{ab;c} (i\gamma; \omega + i \Gamma, -\omega + i\Gamma).
\end{aligned}
\end{equation}
In that case the intrinsic permutation symmetry assumes the form,
\begin{equation}
\sigma^{ab;c}_{\mathrm{dc}}(\omega;\{\gamma, \Gamma\}) = \sigma^{ba;c}_{\mathrm{dc}}(-\omega;\{\gamma, \Gamma\}).
\end{equation}

\paragraph{Conjugation symmetry for dc-response}\label{app02B}

Since physical observables are always real-valued
$\mathcal{O}^*(t) = \mathcal{O}(t)$,
in the frequency domain it holds that 
$\mathcal{O}^*(\omega)=\mathcal{O}(-\omega)$.
For dc current generation, such symmetry leads to $j^c(\bar{\omega}=0) = {j^c}^*(-\bar{\omega} = 0)$, specifically,
\begin{equation}
\begin{aligned}
j^c(\bar{\omega}=0) & = \sigma^{ab;c} (\bar{\omega};-\omega,\omega;\{\gamma, \Gamma\})E^a(-\omega)  E^b(\omega)\\
& = \sigma^{ab;c}_{\mathrm{dc}}(-\omega;\{\gamma, \Gamma\}) E^a(-\omega)  E^b(\omega) \\
& = \sigma^{ba;c}_{\mathrm{dc}}(\omega;\{\gamma, \Gamma\}) E^a(-\omega) E^b(\omega) \\
= {j^c}^*(-\bar{\omega} = 0) & = \left[\sigma_{ba;c}^{dc}(\omega;\{\gamma, \Gamma\}) \right]^* {E^a}^*(\omega) {E^b}^*(-\omega) \\
& = \left[\sigma_{ab;c}^{dc}(\omega;\{\gamma, \Gamma\}) \right]^* E^a(-\omega) E^b(\omega).
\end{aligned}
\end{equation}
On the third line, we have implemented the intrinsic permutation symmetry $\sigma^{ab;c}_{\mathrm{dc}}(\omega;\{\gamma, \Gamma\}) = \sigma^{ba;c}_{\mathrm{dc}}(-\omega;\{\gamma, \Gamma\})$ as discussed previously. Therefore the 2nd order dc conductivity tensor respects the symmetry,
\begin{equation}
\sigma^{ba;c}_{\mathrm{dc}}(\omega;\{\gamma, \Gamma\}) = \left[\sigma^{ab;c}_{\mathrm{dc}}(\omega;\{\gamma, \Gamma\}) \right]^*.
\end{equation}
One can further separate the 2nd order dc conductivity into its real and imaginary parts,
\begin{equation}
\sigma^{ab;c}_{\mathrm{dc}}(\omega;\{\gamma, \Gamma\}) \equiv \sigma^{ab;c}_{\mathrm{Re}}(\omega;\{\gamma, \Gamma\}) + i \sigma^{ab;c}_{\mathrm{Im}}(\omega;\{\gamma, \Gamma\}).
\end{equation}
This yields a real symmetric and an imaginary antisymmetric part of $\sigma^{ab;c}_{\mathrm{dc}}$ upon exchanging electric field spatial index $(a \leftrightarrow b)$, i.e.
\begin{align}
\sigma^{ab;c}_{\mathrm{Re}}(\omega;\{\gamma, \Gamma\}) &= \sigma^{ba;c}_{\mathrm{Re}}(\omega;\{\gamma, \Gamma\})
\\
\sigma^{ab;c}_{\mathrm{Im}}(\omega;\{\gamma, \Gamma\}) &= -\sigma^{ba;c}_{\mathrm{Im}}(\omega;\{\gamma, \Gamma\})
\end{align}

\paragraph{Transport Limit}\label{app02E}

The nonlinear transport limit is reached upon taking $\omega \rightarrow 0$.
According to the intrinsic permutation symmetry,
\begin{equation}
\left.\sigma_{ab;c}^{dc}(\omega;\{\gamma, \Gamma\})\right|_{\omega \rightarrow 0} = \left.\sigma_{ba;c}^{dc}(-\omega;\{\gamma, \Gamma\})\right|_{\omega \rightarrow 0}.
\end{equation}

Therefore,
\begin{equation}
\sigma^{ab;c}_{\text{tr}} \equiv \sigma^{ab;c}_{\mathrm{dc}}(0;\{\gamma, \Gamma\}) = \sigma^{ba;c}_{\mathrm{dc}}(-\omega;\{\gamma, \Gamma\}) = \sigma^{ba;c}_{\text{tr}},
\end{equation}

or equivalently,
\begin{equation}
\sigma^{ab;c}_{\text{tr}} = \sigma^{(ab);c}_{\text{tr}} 
\end{equation}

\subsection{Linear and circular polarized light}\label{LPLCPL}
Consider incident light with the electric field,
\begin{equation}
\begin{aligned}
E^a(t) & = \mathcal{E}^a(\omega) e^{-i\omega t} + {\mathcal{E}^a}^*(\omega) e^{i\omega t} \\
& = 2 \left[\operatorname{Re}\mathcal{E}^a(\omega) \cos \omega t + \operatorname{Im} \mathcal{E}^a(\omega) \sin \omega t \right].
\end{aligned}
\end{equation}
Without loss of generality, let the incident light propagate along the $+z$ direction and choose linear polarized light to be $\boldsymbol{\mathcal{E}}_{\updownarrow}(\omega) = E_0(\cos\phi,\sin\phi,0)$, while circular polarization is given by $\boldsymbol{\mathcal{E}}_{\circlearrowleft}(\omega) = \frac{E_0}{\sqrt{2}} (1,i,0)$ and $\boldsymbol{\mathcal{E}}_{\circlearrowright}(\omega) = \frac{E_0}{\sqrt{2}} (1,-i,0)$.
Then according to the definition of 2nd-order BPVE conductivity tensor $\sigma^{ab;c}(\bar{\omega};\omega,-\omega)$, the 2nd-order dc current response can be written as,
\begin{align}
\mathcal{J}^c(\bar{\omega}=0) & = \sigma^{ab;c} (\bar{\omega}=0; \omega,-\omega)  \mathcal{E}^a(\omega) {\mathcal{E}^b}^*(\omega) \\
& = \begin{bmatrix}
\mathcal{E}^x & \mathcal{E}^y
\end{bmatrix}\begin{bmatrix}
\sigma^{xx;c} & \sigma^{xy;c} \\
\sigma^{yx;c} & \sigma^{yy;c}
\end{bmatrix} \begin{bmatrix}
{\mathcal{E}^x}^* \\ {\mathcal{E}^y}^*
\end{bmatrix}.
\end{align}
Here, the frequency arguments have been omitted for simplicity. One can further simplify the above expression using the symmetry of the conductivity tensor imposed by the dc current generation, 
\begin{equation}
\sigma^{ba;c}_{\mathrm{dc}}(\omega;\{\gamma, \Gamma\}) = \left[\sigma^{ab;c}_{\mathrm{dc}}(\omega;\{\gamma, \Gamma\}) \right]^*.
\end{equation}
The conductivity matrix is therefore Hermitian and can be rewritten as,
\begin{equation}
\begin{aligned}
\mathcal{J}^c & = \begin{bmatrix}
\mathcal{E}^x & \mathcal{E}^y
\end{bmatrix}\begin{bmatrix}
\sigma^{xx;c}_{\text{Re}} & \sigma^{xy;c}_{\text{Re}} + i\sigma^{xy;c}_{\text{Im}} \\
\sigma^{xy;c}_{\text{Re}} - i\sigma^{xy;c}_{\text{Im}} & \sigma^{yy;c}_{\text{Re}}
\end{bmatrix} \begin{bmatrix}
{\mathcal{E}^x}^* \\ {\mathcal{E}^y}^*
\end{bmatrix}.
\end{aligned}
\end{equation}
For linear polarized light, this yields
\begin{equation}
\begin{aligned}
& \mathcal{J}^c_{\updownarrow} = \sigma^{ab;c} \mathcal{E}^a_{\updownarrow}(\omega) {\mathcal{E}^b_{\updownarrow}}^*(\omega) \\
= & E_0^2\begin{bmatrix}
\cos \phi & \sin \phi
\end{bmatrix}\begin{bmatrix}
\sigma^{xx;c}_{\text{Re}} & \sigma^{xy;c}_{\text{Re}} + i\sigma^{xy;c}_{\text{Im}} \\
\sigma^{xy;c}_{\text{Re}} - i\sigma^{xy;c}_{\text{Im}} & \sigma^{yy;c}_{\text{Re}}
\end{bmatrix} \begin{bmatrix}
\cos \phi \\ \sin \phi
\end{bmatrix} \\
= & E_0^2 \left[ \sigma^{xx;c}_{\text{Re}} \cos^2 \phi + \sigma^{yy;c}_{\text{Re}} \sin^2 \phi + 2 \sigma^{xy;c}_{\text{Re}} \sin \phi \cos \phi \right],
\end{aligned}
\end{equation}
which only depends on the real part of the conductivity tensor.
For circular polarized light, the result is
\begin{equation}
\begin{aligned}
& \mathcal{J}^c_{\circlearrowleft/\circlearrowright} = \sigma^{ab;c} \mathcal{E}^a_{\circlearrowleft/\circlearrowright}(\omega) {\mathcal{E}^b_{\circlearrowleft/\circlearrowright}}^*(\omega) \\
= & \frac{E_0^2}{2} \begin{bmatrix}
1 & \pm i
\end{bmatrix}\begin{bmatrix}
\sigma^{xx;c}_{\text{Re}} & \sigma^{xy;c}_{\text{Re}} + i\sigma^{xy;c}_{\text{Im}} \\
\sigma^{xy;c}_{\text{Re}} - i\sigma^{xy;c}_{\text{Im}} & \sigma^{yy;c}_{\text{Re}}
\end{bmatrix} \begin{bmatrix}
1 \\ \mp i
\end{bmatrix} \\
= & \frac{E_0^2}{2} \left[ \sigma^{xx;c}_{\text{Re}} + \sigma^{yy;c}_{\text{Re}} \pm 2 \sigma^{xy;c}_{\text{Im}} \right].
\end{aligned}
\end{equation}
Although the BPVE generated by circular polarized light depends both on the real and imaginary parts of the conductivity tensor, the difference in the response between left and right circular polarized light only depends on the imaginary part of the conductivity tensor, which can be defined as,
\begin{equation}
\mathcal{J}^c_{\circlearrowleft} - \mathcal{J}^c_{\circlearrowright} = 2 E_0^2 \sigma^{xy;c}_{\text{Im}}.
\end{equation}

\subsection{Spacetime transformation and symmetry}\label{optical spacetime symmetry}

\paragraph{Spatial transformation and symmetry}

Consider a pure spatial transformation $x_a \rightarrow x_a' = \mathcal{R}_{aa'} x_{a'}$. The transformation of the Hamiltonian is accordingly
\begin{equation}
\hat{H} \rightarrow \hat{H}' = \hat{\mathcal{R}} \hat{H} \hat{\mathcal{R}}^{-1},
\end{equation}
where $\hat{\mathcal{R}}$ is the matrix representation of the spatial transformation $\mathcal{R}$, which acts on a real space local basis as $\hat{\mathcal{R}} | \mathbf{r}, \alpha \rangle =  \sum_{\alpha'}  R^{\alpha \alpha'} | \mathbf{r}', \alpha' \rangle$, where $\mathbf{r}'^a = (\mathcal{R} \mathbf{r})^a \equiv \sum_{a'} \mathcal{R}^{aa'} \mathbf{r}^{a'}$ and $\alpha,\alpha'$ indicate the orbital degree of freedom and $R_{\alpha \alpha'}$ is the matrix representation in the corresponding Hilbert space. 
In momentum space, the transformation acts very similarly, with $\hat{\mathcal{R}} | \mathbf{k}, \alpha \rangle =  \sum_{\alpha'}  R^{\alpha \alpha'} | \mathbf{k}', \alpha' \rangle$, where $\mathbf{k}'^a = (\mathcal{R} \mathbf{k})^a \equiv \sum_{a'} \mathcal{R}^{aa'} \mathbf{k}^{a'}$. 
Notably, one can express the transformation of the position/momentum and the orbital degree of freedom separately as $\hat{\mathcal{R}} = \hat{R}  \mathcal{R}$, where $\hat{R}$ denotes the part of the transformation which only acts in orbital space, whereas the matrix $\mathcal{R}$ without operator hat only acts on spatial coordinates/momenta.

Given eigenvalues and eigenstates for momentum $\mathbf{k}$ and band index $n$ satisfying $\hat{H}|\psi_{n\mathbf{k}}\rangle = \varepsilon_{n \mathbf{k}} | \psi_{n \mathbf{k}} \rangle$, the transformed Hamiltonian $\hat{H}' = \hat{\mathcal{R}} \hat{H} \hat{\mathcal{R}}^{-1}$ should therefore have the following eigenvalues and eigenstates:
\begin{equation}
\begin{aligned}
\hat{H}' \left( \hat{\mathcal{R}} |\psi_{n,\mathcal{R}^{-1} \mathbf{k}} \rangle \right) & = \hat{\mathcal{R}} \hat{H} \hat{\mathcal{R}}^{-1} \hat{\mathcal{R}} |\psi_{n,\mathcal{R}^{-1} \mathbf{k}} \rangle \\
& = \hat{\mathcal{R}} \hat{H} |\psi_{n,\mathcal{R}^{-1} \mathbf{k}} \rangle \\
& = \hat{\mathcal{R}} \varepsilon_{n, \mathcal{R}^{-1} \mathbf{k}} |\psi_{n,\mathcal{R}^{-1} \mathbf{k}} \rangle \\
& = \varepsilon_{n, \mathcal{R}^{-1} \mathbf{k}} \left( \hat{\mathcal{R}} | \psi_{n,\mathcal{R}^{-1} \mathbf{k}} \rangle \right).
\end{aligned}
\end{equation}
Since $\left( \hat{\mathcal{R}} |\psi_{n,\mathcal{R}^{-1} \mathbf{k}} \rangle \right)$ must have the momentum $\mathcal{R} \left( \mathcal{R}^{-1} \mathbf{k} \right) = \mathbf{k}$, the corresponding eigenvalues for a specific momentum $\mathbf{k}$ are $\varepsilon'_{n \mathbf{k}} = \varepsilon_{n \mathcal{R}^{-1} \mathbf{k}}$ and the corresponding eigenstate is $|\psi'_{n \mathbf{k}} \rangle = \hat{\mathcal{R}} |\psi_{n,\mathcal{R}^{-1} \mathbf{k}} \rangle$.

The transformation of the Bloch Hamiltonian $\hat{H}(\mathbf{k}) \equiv e^{i \mathbf{k} \cdot \hat{\mathbf{r}}} \hat{H} e^{-i \mathbf{k} \cdot \hat{\mathbf{r}}}$ with momentum $\mathbf{k}$, reads
\begin{equation}
\begin{aligned}
\hat{H}(\mathbf{k}) \rightarrow \hat{H}'(\mathbf{k}) & = e^{i \mathbf{k} \cdot \hat{\mathbf{r}}} \hat{H}' e^{-i \mathbf{k} \cdot \hat{\mathbf{r}}} \\
& = e^{i \mathbf{k} \cdot \hat{\mathbf{r}}} \underbrace{\mathcal{R} \hat{R}}_{\hat{\mathcal{R}}} \hat{H}  \underbrace{\hat{R}^{-1} \mathcal{R}^{-1}}_{\hat{\mathcal{R}}^{-1}} e^{-i \mathbf{k} \cdot \hat{\mathbf{r}}} \\
& = e^{i \mathbf{k} \cdot (\mathcal{R} \hat{\mathbf{r}})} \hat{R} \hat{H} \hat{R}^{-1} e^{-i \mathbf{k} \cdot (\mathcal{R} \hat{\mathbf{r}})} \\
& = \hat{R} e^{i \mathcal{R}^{-1} \mathbf{k} \cdot \hat{\mathbf{r}}} \hat{H} e^{-i \mathcal{R}^{-1} \mathbf{k} \cdot \hat{\mathbf{r}}} \hat{R}^{-1} \\
& = \hat{R} \hat{H}(\mathcal{R}^{-1}\mathbf{k}) \hat{R}^{-1}.
\end{aligned}
\end{equation}
To obtain line three, note that $e^{i \mathbf{k} \cdot \hat{\mathbf{r}}} \mathcal{R} = e^{i \mathbf{k} \cdot (\mathcal{R} \hat{\mathbf{r}})}$, since $e^{i \mathbf{k} \cdot \hat{\mathbf{r}}} \mathcal{R} | \mathbf{r} \rangle = e^{i \mathbf{k} \cdot \hat{\mathbf{r}}} | \mathcal{R} \mathbf{r} \rangle = e^{i \mathbf{k} \cdot (\mathcal{R} \mathbf{r})} | \mathbf{r} \rangle = e^{i \mathbf{k} \cdot (\mathcal{R} \hat{\mathbf{r}})} | \mathbf{r} \rangle $. Similarly, we have $\mathcal{R}^{-1}e^{-i \mathbf{k} \cdot \hat{\mathbf{r}}} = e^{-i \mathbf{k} \cdot (\mathcal{R} \hat{\mathbf{r}})}$ by Hermitian conjugation. Finally, the eigenvalues and eigenstates of the transformed Bloch Hamiltonian are
\begin{equation}
\begin{aligned}
& \hat{H}'(\mathbf{k}) | u'_{n \mathbf{k}} \rangle  = \varepsilon'_{n \mathbf{k}} | u'_{n \mathbf{k}} \rangle \\
\Leftrightarrow & \hat{R} \hat{H}(\mathcal{R}^{-1}\mathbf{k}) \hat{R}^{-1} \left( \hat{R} | u_{n,  \mathcal{R}^{-1} \mathbf{k}} \rangle \right) \\
& = \hat{R} \hat{H}(\mathcal{R}^{-1}\mathbf{k}) | u_{n, \mathcal{R}^{-1} \mathbf{k}} \rangle \\
& = \varepsilon_{n, \mathcal{R}^{-1} \mathbf{k}} \left( \hat{R} | u_{n, \mathcal{R}^{-1} \mathbf{k}} \rangle \right),
\end{aligned}
\end{equation}
confirming that the transformed eigenvalues and eigenstates of the transformed Bloch Hamiltonian are $\varepsilon'_{n \mathbf{k}} = \varepsilon_{n, \mathcal{R}^{-1} \mathbf{k}}$ and $|u'_{n \mathbf{k}} \rangle = \hat{R} | u_{n, \mathcal{R}^{-1} \mathbf{k}} \rangle$.

Next, we calculate the transformation of different matrix elements. For example, the transformation of the dipole matrix element (Berry connection) is,
{\allowdisplaybreaks
\begin{align}
& {r}_{mn}^{a} (\mathbf{k}) \equiv \langle u_{m \mathbf{k}} | i \partial_{k_a} | u_{n \mathbf{k}} \rangle \notag\\
\rightarrow \ & {r'}_{mn}^{a} (\mathbf{k}) = \langle u'_{m \mathbf{k}} | i \partial_{k_a} | u'_{n \mathbf{k}} \rangle \notag\\
& = \langle \hat{R} u_{m, \mathcal{R}^{-1} \mathbf{k}} | i \partial_{k_a} | \hat{R} u_{n, \mathcal{R}^{-1} \mathbf{k}} \rangle \notag\\
& = \langle u_{m, \mathcal{R}^{-1} \mathbf{k}} | \hat{R}^{-1} i \partial_{k_a} \hat{R} | u_{n, \mathcal{R}^{-1} \mathbf{k}} \rangle \notag\\
& = \langle u_{m, \mathcal{R}^{-1} \mathbf{k}} | i \partial_{k_a} | u_{n, \mathcal{R}^{-1} \mathbf{k}} \rangle \notag\\
& = \sum_{b} \frac{\partial (\mathcal{R}^{-1} \mathbf{k})^b}{\partial k_a} \langle u_{m, \mathcal{R}^{-1} \mathbf{k}} |  i \partial_{(\mathcal{R}^{-1} \mathbf{k})_b} | u_{n, \mathcal{R}^{-1} \mathbf{k}} \rangle \notag\\
& = \sum_{b} (\mathcal{R}^{-1})^{ba} r_{mn}^{b} (\mathcal{R}^{-1} \mathbf{k}) \notag\\
& = \sum_{b} \mathcal{R}^{ab} r_{mn}^{b} (\mathcal{R}^{-1} \mathbf{k}).
\end{align}
}
Here, it was assumed that the transformation on the orbital degree of freedom is $\mathbf{k}$-independent, which is appropriate for a point group symmetry operation $\hat{\mathcal{R}}$ (without translation). The dipole matrix element thus transforms like a vector field.

The transformation property of different orders of velocity matrix elements, i.e. $v_{mn}^{a},\,w_{mn}^{ab},\,u_{mn}^{abc}$, or equivalently, the spatial transformation of $h^{a_1 \dots a_n}_{mn}$ follow in the same fashion,
\begin{widetext}
\begin{equation}
\begin{aligned}
& h_{mn}^{a_1 \dots a_n}(\mathbf{k}) \equiv \langle u_{m \mathbf{k}} | (\partial_{k_{a_1}} \dots \partial_{k_{a_n}} \hat{H}(\mathbf{k})) | u_{n \mathbf{k}} \rangle \\
\rightarrow \ & {h'}_{mn}^{a_1 \dots a_n}(\mathbf{k}) \equiv \langle u'_{m \mathbf{k}} | (\partial_{k_{a_1}} \dots \partial_{k_{a_n}} \hat{H}'(\mathbf{k})) | u'_{n \mathbf{k}} \rangle \\
& = \langle \hat{R} u_{m,\mathcal{R}^{-1} \mathbf{k}} | (\partial_{k_{a_1}} \dots \partial_{k_{a_n}} \hat{R} \hat{H}(\mathcal{R}^{-1}\mathbf{k}) \hat{R}^{-1}) | \hat{R} u_{n,\mathcal{R}^{-1} \mathbf{k}} \rangle \\
& = \sum_{b_1,\dots, b_n} \mathcal{R}^{a_1 b_1} \dots \mathcal{R}^{a_n b_n} \langle u_{m,\mathcal{R}^{-1} \mathbf{k}} | (\partial_{(\mathcal{R}^{-1} \mathbf{k})_{b_1}} \dots \partial_{(\mathcal{R}^{-1} \mathbf{k})_{b_n}} \hat{H}(\mathcal{R}^{-1}\mathbf{k})) | u_{n,\mathcal{R}^{-1} \mathbf{k}} \rangle \\
& = \sum_{b_1,\dots, b_n} \mathcal{R}^{a_1 b_1} \dots \mathcal{R}^{a_n b_n} h^{b_1 \dots b_n}_{mn}(\mathcal{R}^{-1} \mathbf{k}).
\end{aligned}
\end{equation}
\end{widetext}
In other words, the $n$-th order velocity matrix element transforms as an $n$-th order tensor field.
The spatial transformation property of nonlinear conductivity is therefore,
\begin{equation}
\sigma^{ab;c} \rightarrow \sigma'^{ab;c} = \sum_{a',b',c'} \mathcal{R}^{a a'} \mathcal{R}^{b b'} \mathcal{R}^{c c'} \sigma^{a' b'; c'},
\end{equation}
in line with the phenomenological expectation.

\paragraph{Time reversal transformation and symmetry}

Similarly, we can also analyze the time reversal transformation property of the Hamiltonian, eigenvalues, eigenstates, and different matrix elements. We denote the time reversal operator as $\hat{T}$, which is an anti-unitary operator.
The time reversal transformed Hamiltonian is,
\begin{equation}
\hat{H} \rightarrow \hat{H}' = \hat{T} \hat{H} \hat{T}^{-1},
\end{equation}
while the time reversal transformed Bloch Hamiltonian is,
\begin{equation}
\begin{aligned}
\hat{H}(\mathbf{k}) \rightarrow \hat{H}'(\mathbf{k}) & = e^{i \mathbf{k} \cdot \hat{\mathbf{r}}} \hat{H}' e^{-i \mathbf{k} \cdot \hat{\mathbf{r}}} \\
& = e^{i \mathbf{k} \cdot \hat{\mathbf{r}}} \hat{T} \hat{H} \hat{T}^{-1} e^{-i \mathbf{k} \cdot \hat{\mathbf{r}}} \\
& = \hat{T} e^{-i \mathbf{k} \cdot \hat{\mathbf{r}}} \hat{H} e^{i \mathbf{k} \cdot \hat{\mathbf{r}}} \hat{T}^{-1} \\
& = \hat{T} \hat{H}(- \mathbf{k}) \hat{T}^{-1}.
\end{aligned}
\end{equation}
Here, note that the time reversal operator $\hat{T}$ is anti-unitary, which adds a complex conjugation when interchanging the order of $e^{i \mathbf{k} \cdot \hat{\mathbf{r}}}$ and $\hat{T}$. The band dispersion and unit cell periodic Bloch states of the time reversal transformed Hamiltonian are $\varepsilon'_{n \mathbf{k}} = \varepsilon_{n, - \mathbf{k}}$ and $|u'_{n \mathbf{k}} \rangle = \hat{T} |u_{n,-\mathbf{k}} \rangle$ respectively. %With this knowledge, we can also analyze the time reversal transformation of different matrix elements.

The time reversal transformation property of the dipole matrix element (Berry connection) is then,
\begin{align}
 {r}_{mn}^{a} (\mathbf{k}) &\equiv \langle u_{m \mathbf{k}} | i \partial_{k_a} | u_{n \mathbf{k}} \rangle \notag\\
\rightarrow \  {r'}_{mn}^{a} (\mathbf{k}) &= \langle u'_{m \mathbf{k}} | i \partial_{k_a} | u'_{n \mathbf{k}} \rangle \notag\\
& = i \langle \hat{T} u_{m, - \mathbf{k}} | \partial_{k_a} | \hat{T} u_{n, - \mathbf{k}} \rangle \notag\\
& = i \langle \hat{T} u_{m, - \mathbf{k}} | \hat{T} \partial_{k_a} u_{n, - \mathbf{k}} \rangle \notag\\
& = i \langle  \partial_{k_a} u_{n, - \mathbf{k}} | u_{m, - \mathbf{k}} \rangle \notag\\
& = - i \langle u_{n, - \mathbf{k}} | \partial_{k_a} u_{m, - \mathbf{k}} \rangle \notag\\
& = \langle u_{n, - \mathbf{k}} | i \partial_{(-\mathbf{k})_a} u_{m, - \mathbf{k}} \rangle \notag\\
& = r_{nm}^{a} (-\mathbf{k}).
\end{align}
Here again, the assumption being that the transformation on the orbital degree of freedom is $\mathbf{k}$ independent, which is appropriate for a point group symmetry operation $\hat{\mathcal{R}}$ (without translation). 

For the derivatives of the Bloch Hamiltonian $h^{a_1 \dots a_n}_{mn}$, one obtains similarly,
\begin{align}
& h_{mn}^{a_1 \dots a_n}(\mathbf{k}) \equiv \langle u_{m \mathbf{k}} | (\partial_{k_{a_1}} \dots \partial_{k_{a_n}} \hat{H}(\mathbf{k})) | u_{n \mathbf{k}} \rangle \notag\\
\rightarrow \ & {h'}_{mn}^{a_1 \dots a_n}(\mathbf{k}) \equiv \langle u'_{m \mathbf{k}} | (\partial_{k_{a_1}} \dots \partial_{k_{a_n}} \hat{H}'(\mathbf{k})) | u'_{n \mathbf{k}} \rangle \notag\\
& = \langle \hat{T} u_{m, - \mathbf{k}} | \hat{T} (\partial_{k_{a_1}} \dots \partial_{k_{a_n}} \hat{H}(- \mathbf{k}) ) \hat{T}^{-1} | \hat{T} u_{n, - \mathbf{k}} \rangle \notag\\
& = \langle \hat{T} u_{m, - \mathbf{k}} | \hat{T} (\partial_{k_{a_1}} \dots \partial_{k_{a_n}} \hat{H}(- \mathbf{k}) )  u_{n, - \mathbf{k}} \rangle \notag\\
& = \langle (\partial_{k_{a_1}} \dots \partial_{k_{a_n}} \hat{H}(- \mathbf{k}) )  u_{n, - \mathbf{k}} | u_{m, - \mathbf{k}} \rangle \notag\\
& = \langle u_{n, - \mathbf{k}} | (\partial_{( - \mathbf{k})_{a_1}} \dots \partial_{( - \mathbf{k})_{a_n}} \hat{H}(- \mathbf{k}) ) | u_{m, - \mathbf{k}} \rangle \notag\\
& = (-1)^{n} h_{nm}^{a_1 \dots a_n}(-\mathbf{k})
\end{align}
As we see, velocity matrix elements with odd numbered order are odd under time reversal transformation, while velocity matrix elements with even numbered order are even under time reversal transformation.

For the derivation of time reversal transformation of nonlinear conductivity, we also need the transformation properties of $f^{FD}_{nm}(\mathbf{k})$, $v^{a}_{nm}(\mathbf{k})$, $\Delta^{c}_{nm}(\mathbf{k})$, $R^c_{mn,a}(\mathbf{k})$, and ${\varepsilon}_{nm}(\mathbf{k})$. They are, respectively,
\begin{align}
{f}_{\mathcal{T},nm}^{FD} (\mathbf{k}) & = f^{FD}({\varepsilon}_{\mathcal{T},n}(\mathbf{k})) - f^{FD}({\varepsilon}_{\mathcal{T},m}(\mathbf{k})) \notag\\
& = f^{FD}({\varepsilon}_{n}(-\mathbf{k})) - f^{FD}({\varepsilon}_{m}(-\mathbf{k})) \notag\\
& = f_{nm}^{FD}(-\mathbf{k}) \notag\\
& = - f_{mn}^{FD}(-\mathbf{k})
\end{align}
\begin{align}
{v}_{\mathcal{T},nm}^a (\mathbf{k}) & = - v_{mn}^a (-\mathbf{k}) 
\end{align}
\begin{align}
{\Delta}_{\mathcal{T},mn}^c (\mathbf{k}) & = {v}_{\mathcal{T},mm}^c (\mathbf{k})  - {v}_{\mathcal{T},nn}^c (\mathbf{k}) \notag\\
& = \left[- v_{mm}^c (-\mathbf{k})\right] - \left[- v_{nn}^c (-\mathbf{k})\right] \notag\\
& = \Delta_{nm}^c (-\mathbf{k})
\end{align}
\begin{align}
& R^c_{\mathcal{T},mn,a}(\mathbf{k}) \notag\\
= & r^c_{\mathcal{T},mm}(\mathbf{k}) - r^c_{\mathcal{T},nn}(\mathbf{k}) + i \partial_{k_c} \operatorname{log} r^a_{\mathcal{T},mn}(\mathbf{k}) \notag\\
= & r^c_{mm}(-\mathbf{k}) - r^c_{nn}(-\mathbf{k}) + i \partial_{k_c} \operatorname{log} r^a_{nm}(\mathbf{k}) \notag\\
= & - \left[ r^c_{nn}(-\mathbf{k}) - r^c_{mm}(-\mathbf{k}) + i \partial_{-\mathbf{k}_c} \operatorname{log} r^a_{nm}(-\mathbf{k})\right] \notag\\
= & - R^c_{nm,a} (-\mathbf{k})
\end{align}
\begin{align}
{\varepsilon}_{\mathcal{T},nm}(\mathbf{k}) & = \varepsilon_{nm}(-\mathbf{k})
\end{align}

\section{Feynman diagrammatics}\label{app1}

\begin{figure*}[htbp]
\includegraphics[width=0.92\linewidth]{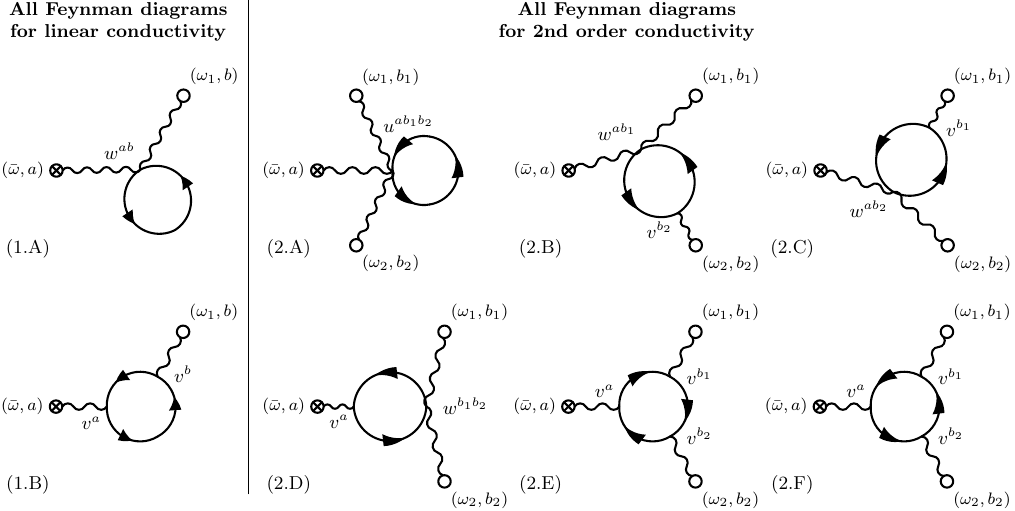}
\caption{ Structrure of the Feynman diagrams for (1) linear and  (2) second order conductivity.
In the two (six) contributions at first (second) the wavy lines connected to the ``$\circ$'' and ``$\otimes$'' represent the incoming and outgoing photons respectively, while the solid lines with arrows represents the fermion propagators.
}\label{fig:appendix-feynman}
\end{figure*}

% \subsection{Perturbation theory}

Here, we introduce the quantum perturbative analysis. Further details can be found in Refs.~\cite{Parker2019,Holder2020}.
An applied electric field enters via minimal coupling to the electric gauge potential,
\begin{equation}
\hat{H}_0(\mathbf{k}) \rightarrow \hat{H}'(\mathbf{k},t) = \hat{H}_0(\mathbf{k} - \frac{q}{\hbar} \mathbf{A}(t)).
\end{equation}
where $q = -e$ is the charge of the electron, here we keep the notation of $q$ for smooth generalization of our theoretical derivation to any kinds of charge carriers.
The perturbation expansion yields,
\begin{equation}
\hat{H}'(\mathbf{k},t) = \hat{H}_0(\mathbf{k}) + \sum_{k=1}^{\infty} \frac{(-q/\hbar)^k}{k!} A^{b_1}(t)\dots A^{b_k}(t)\hat{h}^{b_1\dots b_k}
\end{equation}
where $\hat{h}^{b_1\dots b_k}\equiv\partial_{b_1}\dots\partial_{b_k}\hat{H}_0(\mathbf{k})$ and the Einstein summation convention is used. The $\hat{h}^{b_1\dots b_k}$'s constitute the electron-photon interaction vertices. It is common to introduce new notation for commonly used low order vertices and suppress the momentum dependence, such as,
\begin{equation}
\hat{v}^a \equiv \hat{h}^a,\,\hat{w}^{ab} \equiv \hat{h}^{ab},\,\hat{u}^{abc} \equiv \hat{h}^{abc}.
\end{equation}
The current operator can be obtained by,
\begin{align}
\hat{J}^a (t) 
&\equiv \frac{1}{\mathcal{V}}\frac{\delta \hat{H}'}{\delta A^{a}(t)} \notag\\
&= -\frac{q}{\mathcal{V}\hbar} \sum_{k=1}^{\infty} \frac{(-q/\hbar)^k}{k!} A^{b_1}(t)\dots A^{b_k}(t)\hat{h}^{a b_1\dots b_k}
\end{align}
Therefore the current generation vertices are the same as the electron-photon interaction vertices.

In imaginary time perturbation theory, the partition function is given by,
\begin{equation}
\begin{aligned}
\mathcal{Z} & = \int \mathcal{D} \bar{\psi} \mathcal{D} \psi \exp \left\{ - \int_0^{\beta} d \tau \left[ \bar{\psi} (\partial_\tau + \hat{H}') \psi  \right] \right\} \\
\end{aligned}
\end{equation}
The current expectation value is therefore,
\begin{equation}
\begin{aligned}
& \langle \hat{J}^a (\tau) \rangle \\
= & \frac{1}{\mathcal{Z}} \int \mathcal{D} \bar{\psi} \mathcal{D} \psi \hat{J}^a (t) \exp \left\{ - \int_0^{\beta} d \tau \left[ \bar{\psi} (\partial_\tau + \hat{H}') \psi  \right] \right\} \\
= & \frac{1}{\mathcal{Z}} \int \mathcal{D} \bar{\psi} \mathcal{D} \psi \frac{\delta \hat{H}'}{\delta A^a(\tau)} \exp \left\{ - \int_0^{\beta} d \tau \left[ \bar{\psi} (\partial_\tau + \hat{H}') \psi  \right] \right\} \\
= & -\frac{1}{\mathcal{Z}} \frac{\delta \mathcal{Z}}{\delta A^a(\tau)}.
\end{aligned}
\end{equation}
Transforming to the Matsubara frequency domain yields
\begin{equation}
\begin{aligned}
\langle \hat{J}^a (i \nu_n) \rangle & = \int_0^{\beta} d \tau e^{i \nu_n \tau} \langle \hat{J}^a (\tau) \rangle = -\frac{\delta \operatorname{ln} \mathcal{Z}}{\delta A^a(-i \nu_n)}.
\end{aligned}
\end{equation}
The susceptibility with respect to the vector potential in Matsubara frequency domain is defined by,
\begin{equation}
\begin{aligned}
& \langle \hat{J}^a (i \bar{\nu}) \rangle \\
= & \sum_{i\nu_1} \chi^{b;a}(i \bar{\nu}; i \nu_1) A^b(i \nu_1) \\
+ & \frac{1}{2!} \sum_{i\nu_1} \sum_{i\nu_2} \chi^{b_1b_2;a}(i \bar{\nu}; i \nu_1, i \nu_2) A^{b_1}(i \nu_1) A^{b_2}(i \nu_2) \\
+ & \dots
\end{aligned}
\end{equation}
which is synonymous with the definition in terms of functional derivatives,
\begin{equation}
\begin{aligned}
\chi^{b;a}(i \bar{\nu}; i \nu_1) & = -\frac{\delta^2 \operatorname{ln} \mathcal{Z}}{\delta A^b(i \nu_1) \delta A^a(-i \bar{\nu})} \\
\chi^{b_1b_2;a}(i \bar{\nu}; i \nu_1, i \nu_2) & = -\frac{\delta^3 \operatorname{ln} \mathcal{Z}}{\delta A^{b_1}(i \nu_1) \delta A^{b_2}(i \nu_2) \delta A^a(-i \bar{\nu})}.
\end{aligned}
\end{equation}
Here we can see the $n$-th order susceptibility consists of connected Feynman diagrams (as imposed by the $\operatorname{ln} \mathcal{Z}$ function) with $n$ incoming external photon lines with frequency $\nu_1,\nu_2,\dots$ (as imposed by $\frac{\delta}{\delta A^b(i \nu_1)} \frac{\delta}{\delta A^b(i \nu_2)} \dots$), and one outgoing external photon line with frequency $\bar{\nu}$ (as imposed by $\frac{\delta}{\delta A^a(-i \bar{\nu})}$).

% \subsection{Feynman diagrammatics}
\begin{widetext}

Let us calculate the susceptibility. The first order susceptibility follows from two diagrams,
\begin{equation}
\chi^{b;a} = \chi^{b;a}_{\text{(1.A)}} + \chi^{b;a}_{\text{(1.B)}},
\end{equation}
where the susceptibilities are defined using the fermionic Matsubara green's function $\mathcal{G}(i\omega)=\frac{1}{i\hbar\omega - H_0}$,
\begin{equation}
\chi^{b;a}_{\text{(1.A)}}(i \bar{\nu}; i \nu_1) = - \frac{1}{\beta} \sum_{i\omega_l}  \int_{\mathbf{k}} \operatorname{Tr} \left[ \hat{w}^{ab} \mathcal{G}(i \omega_l) \right] \delta_{i\bar{\nu},i\nu_1} 
\end{equation}
\begin{equation}
\chi^{b;a}_{\text{(1.B)}}(i \bar{\nu}; i \nu_1) = - \frac{1}{\beta} \sum_{i\omega_l}  \int_{\mathbf{k}} \operatorname{Tr} \left[ \hat{v}^a \mathcal{G}(i \nu_1 + i \omega_l) \hat{v}^b \mathcal{G}(i \omega_l) \right] \delta_{i\bar{\nu},i\nu_1},
\end{equation}

Similarly, the second order susceptibility is contributed by six diagrams that correspond to,
\begin{equation}
\chi^{b_1b_2;a} = \chi^{b_1b_2;a}_{\text{(2.A)}} + \chi^{b_1b_2;a}_{\text{(2.B)}} + \chi^{b_1b_2;a}_{\text{(2.C)}} + \chi^{b_1b_2;a}_{\text{(2.D)}} + \chi^{b_1b_2;a}_{\text{(2.E)}} + \chi^{b_1b_2;a}_{\text{(2.F)}},
\end{equation}
where
\begin{equation}
\chi^{b_1b_2;a}_{\text{(2.A)}}(i \bar{\nu}; i \nu_1, i \nu_2) = -\frac{1}{\beta} \sum_{i\omega_m}  \int_{\mathbf{k}} \operatorname{Tr} \left[ \hat{u}^{ab_1b_2} \mathcal{G}(i \omega_m) \right] \delta_{i\bar{\nu},i\nu_1 + i\nu_2} 
\end{equation}
\begin{equation}
\chi^{b_1b_2;a}_{\text{(2.B)}}(i \bar{\nu}; i \nu_1, i \nu_2) = - \frac{1}{\beta} \sum_{i\omega_m}  \int_{\mathbf{k}} \operatorname{Tr} \left[ \hat{w}^{ab_1} \mathcal{G}(i \nu_2 + i \omega_m) \hat{v}^{b_2} \mathcal{G}(i \omega_m) \right] \delta_{i\bar{\nu},i\nu_1 + i\nu_2}
\end{equation}
\begin{equation}
\chi^{b_1b_2;a}_{\text{(2.C)}}(i \bar{\nu}; i \nu_1, i \nu_2) = - \frac{1}{\beta} \sum_{i\omega_m}  \int_{\mathbf{k}} \operatorname{Tr} \left[ \hat{w}^{ab_2} \mathcal{G}(i \nu_1 + i \omega_m) \hat{v}^{b_1} \mathcal{G}(i \omega_m) \right] \delta_{i\bar{\nu},i\nu_1 + i\nu_2}
\end{equation}
\begin{equation}
\chi^{b_1b_2;a}_{\text{(2.D)}}(i \bar{\nu}; i \nu_1, i \nu_2) = - \frac{1}{\beta} \sum_{i\omega_m}  \int_{\mathbf{k}} \operatorname{Tr} \left[ \hat{v}^{a} \mathcal{G}(i \bar{\nu} + i \omega_m) \hat{w}^{b_1b_2} \mathcal{G}(i \omega_m) \right] \delta_{i\bar{\nu},i\nu_1 + i\nu_2}
\end{equation}
\begin{equation}
\chi^{b_1b_2;a}_{\text{(2.E)}}(i \bar{\nu}; i \nu_1, i \nu_2) = - \frac{1}{\beta} \sum_{i\omega_m}  \int_{\mathbf{k}} \operatorname{Tr} \left[ \hat{v}^{a} \mathcal{G}(i \bar{\nu} + i \omega_m) \hat{v}^{b_2} \mathcal{G}(i \nu_1 + i \omega_m) \hat{v}^{b_1} \mathcal{G}(i \omega_m) \right] \delta_{i\bar{\nu},i\nu_1 + i\nu_2}
\end{equation}
\begin{equation}
\chi^{b_1b_2;a}_{\text{(2.F)}}(i \bar{\nu}; i \nu_1, i \nu_2) = - \frac{1}{\beta} \sum_{i\omega_m}  \int_{\mathbf{k}} \operatorname{Tr} \left[ \hat{v}^{a} \mathcal{G}(i \bar{\nu} + i \omega_m) \hat{v}^{b_1} \mathcal{G}(i \nu_2 + i \omega_m) \hat{v}^{b_2} \mathcal{G}(i \omega_m) \right] \delta_{i\bar{\nu},i\nu_1 + i\nu_2}
\end{equation}
We note that the topology of the diagrams up to second order do not lead to nontrivial multiplicities~\cite{Parker2019}.

After analytical continuation $i\nu \rightarrow \omega + i 0^+$, this yields the susceptibility for real frequencies. In order to obtain the conductivity, note that $E^a (\omega) = i \omega A^a (\omega)$, which leads to the conductivity,
\begin{equation}
\sigma^{b_1 \dots b_n;a} (\bar{\omega}; \{\omega_i\}) = \frac{\chi^{b_1b_2;a} (i\bar{\nu} \rightarrow \bar{\omega} + i 0^+; \{i \nu_i \rightarrow \omega_i + i 0^+\}) }{\prod_{i=1}^n (i \omega_i)}.
\end{equation}
We introduce the Fermi-DIrac distribution $f_m \equiv \frac{1}{e^{\beta\hbar(\varepsilon_m-\mu)}+1}$ of band $m$ where $\mu$ is the chemical potential and $\beta^{-1}=k_BT$ for temperature $T$.
Restoring the electric charge and Planck constant, 
the 2nd order conductivity then becomes,
\begin{equation}
\sigma^{b_1b_2;a}_{\text{(2.A)}} (\bar{\omega}; \omega_1, \omega_2) = -\frac{e^3}{\hbar^2 \omega_1 \omega_2} \sum_{n} \int_{\mathbf{k}} f_n u^{b_1b_2a}_{nn}
\end{equation}
\begin{equation}
\sigma^{b_1b_2;a}_{\text{(2.B)}} (\bar{\omega}; \omega_1, \omega_2) = -\frac{e^3}{\hbar^2 \omega_1 \omega_2} \sum_{nm} \int_{\mathbf{k}} \frac{f_{nm} v_{nm}^{b_2} w_{mn}^{a b_1}}{\omega_2 + \varepsilon_{nm} + i 0^+}
\end{equation}
\begin{equation}
\sigma^{b_1b_2;a}_{\text{(2.C)}} (\bar{\omega}; \omega_1, \omega_2) = -\frac{e^3}{\hbar^2 \omega_1 \omega_2} \sum_{nm} \int_{\mathbf{k}} \frac{f_{nm} v_{nm}^{b_1} w_{mn}^{a b_2}}{\omega_1 + \varepsilon_{nm} + i 0^+}
\end{equation}
\begin{equation}
\sigma^{b_1b_2;a}_{\text{(2.D)}} (\bar{\omega}; \omega_1, \omega_2) = -\frac{e^3}{\hbar^2 \omega_1 \omega_2} \sum_{nm} \int_{\mathbf{k}} \frac{f_{nm} w_{nm}^{b_1 b_2} v_{mn}^a}{\bar{\omega} - \varepsilon_{mn} + i 0^+}
\end{equation}
\begin{equation}
\sigma^{b_1b_2;a}_{\text{(2.E)}} (\bar{\omega}; \omega_1, \omega_2) = -\frac{e^3}{\hbar^2 \omega_1 \omega_2} \sum_{nml} \int_{\mathbf{k}} \left( \frac{f_{nm} v_{nm}^{b_2} v_{ml}^{b_1} v_{ln}^{a}}{(\omega_2 + \varepsilon_{nm} + i 0^+)(\bar{\omega} - \varepsilon_{ln} + i 0^+)} - 
\frac{f_{nm} v_{nm}^{b_1} v_{ml}^{a} v_{ln}^{b_2}}{(\omega_1 + \varepsilon_{nm} + i 0^+)(\bar{\omega} - \varepsilon_{ml} + i 0^+)}\right)
\end{equation}
\begin{equation}
\sigma^{b_1b_2;a}_{\text{(2.F)}} (\bar{\omega}; \omega_1, \omega_2) = -\frac{e^3}{\hbar^2 \omega_1 \omega_2} \sum_{nml} \int_{\mathbf{k}} \left( \frac{f_{nm} v_{nm}^{b_1} v_{ml}^{b_2} v_{ln}^{a}}{(\omega_1 + \varepsilon_{nm} + i 0^+)(\bar{\omega} - \varepsilon_{ln} + i 0^+)} - \frac{f_{nm} v_{nm}^{b_2} v_{ml}^{a} v_{ln}^{b_1}}{(\omega_2 + \varepsilon_{nm} + i 0^+)(\bar{\omega} - \varepsilon_{ml} + i 0^+)} \right)
\end{equation}
A more compact and symmetric notation reads
\begin{equation}
\sigma^{b_1b_2;a}_{\text{(2.A)}} (\bar{\omega}; \omega_1, \omega_2) = -\frac{e^3}{\hbar^2 \omega_1 \omega_2} \sum_{n} \int_{\mathbf{k}} f_n \frac{u^{b_1b_2a}_{nn}}{2} + (b_1,\omega_1 \leftrightarrow b_2,\omega_2)
\end{equation}
\begin{equation}
\sigma^{b_1b_2;a}_{\text{(2.B)}+\text{(2.C)}} (\bar{\omega}; \omega_1, \omega_2) = -\frac{e^3}{\hbar^2 \omega_1 \omega_2} \sum_{n} \int_{\mathbf{k}} f_n \left[ \frac{v^{b_1}}{\omega_1 +\varepsilon + i 0^+} , w^{ab_2} \right]_{nn} + (b_1,\omega_1 \leftrightarrow b_2,\omega_2)
\end{equation}
\begin{equation}
\sigma^{b_1b_2;a}_{\text{(2.D)}} (\bar{\omega}; \omega_1, \omega_2) = -\frac{e^3}{\hbar^2 \omega_1 \omega_2} \sum_{n} \int_{\mathbf{k}} \frac{f_n}{2} \left[ w^{b_1 b_2} , \frac{v^{a}}{\bar{\omega} - \varepsilon + i 0^+} \right]_{nn} + (b_1,\omega_1 \leftrightarrow b_2,\omega_2)
\end{equation}
\begin{equation}
\sigma^{b_1b_2;a}_{\text{(2.E)}+\text{(2.F)}} (\bar{\omega}; \omega_1, \omega_2) = -\frac{e^3}{\hbar^2 \omega_1 \omega_2} \sum_{n} \int_{\mathbf{k}} f_n \left[ \frac{v^{b_1}}{\omega_1 + \varepsilon + i 0^+} , \left[ v^{b_2} , \frac{v^{a}}{\bar{\omega} - \varepsilon + i 0^+} \right] \right]_{nn} + (b_1,\omega_1 \leftrightarrow b_2,\omega_2)
\end{equation}

Here the infinitesimal imaginary part in the denominator is the inverse interband/intraband lifetime, which is detailed in Sec.~\ref{appendix:diagrammatics_transport}.

\section{Comments on nonlinear transport conductivity}
\label{appF}
The literature on nonlinear transport is extensive~\cite{Gao2019,Bhalla2023,
Sodemann2015,Watanabe2020,
Sodemann2015,Zhang2018a, Matsyshyn2019,Nandy2019,
Gao2014,Liu2021,Kaplan2024,Das2023,
Kaplan2023a,Holder2021a}, with various different approaches leading to similar but not identical results~\cite{Gao2014,Gao2019,Watanabe2020,Watanabe2021,Sodemann2015,Wang2021,Das2023,Oiwa2022,Kaplan2024,Jia2024,jiang2024electrical}. For the sake of clarity, in the following we elucidate the  main steps towards the semiclassical and quantum perturbative calculations, emphasizing the underlying ingredients rather than the technical aspects. The argumentation mostly follows Refs.~\cite{Kaplan2023a,Kaplan2024}.

\end{widetext}

\subsection{Semiclassical Formalism}\label{appF1}

% \subsection{Perturbation Theory and Schrieffer–Wolff transformation}

\paragraph{Schrieffer–Wolff transformation}
As mentioned above, second order response requires eigenvalues up to the 3rd order perturbative correction, i.e., \(\mathcal{O}(\mathbf{E}^3)\), and wavefunction up to the 2nd order, i.e., \(\mathcal{O}(\mathbf{E}^2)\).

One way to avoid a painstaking derivation of higher-order perturbation theory, is to implement a Schrieffer-Wolff (SW) transformation, which changes the wave function basis, transforming the lowest order perturbation to the same order of \(\mathcal{O}(\mathbf{E}^2)\), i.e.,
\begin{equation}
H \rightarrow H' \equiv e^{S} H e^{-S},\,|n\rangle \rightarrow |n'\rangle \equiv e^S |n\rangle 
\end{equation}
\begin{equation}
H' = H_0' + H_{1}'
\end{equation}
\begin{equation}
H_{1}' = \mathcal{O}(\mathbf{E}^2)
\end{equation}
In the new basis, the perturbative expansions of the eigenvalue and eigenstate only require the 2nd order eigenvalue and 1st order eigenstate perturbation theory for \(H'\) to cover \(\mathcal{O}(\mathbf{E}^3)\) for the former and \(\mathcal{O}(\mathbf{E}^2)\) for the latter. Throughout the discussion below, the prime \('\) represents the eigenstates and operators after the SW transformation.

Let us first separate the Hamiltonian \(H\) into diagonal and off-diagonal parts,
\begin{equation}
H = H^{(0)} + e \mathbf{E} \cdot \mathbf{r} \equiv H_0 + H_1
\end{equation}
where \(H^{(0)}\) is the original lattice Hamiltonian, and \(H_0\), \(H_1\) correspond to the diagonal and off-diagonal parts of the full Hamiltonian after applying the external field, which read explicitly,
\begin{align}
H_0 & = H^{(0)} + \sum_{n,\mathbf{k}} \left( e \mathbf{E} \cdot \langle \psi_{n\mathbf{k}}^{(0)} | \mathbf{r} | \psi_{n\mathbf{k}}^{(0)} \rangle \right) | \psi_{n\mathbf{k}}^{(0)} \rangle \langle \psi_{n\mathbf{k}}^{(0)} | \\
& = \sum_{n,\mathbf{k}} \left( \varepsilon_n^{(0)} (\mathbf{k}) + e \mathbf{E} \cdot \mathbf{A}_{nn}^{(0)}(\mathbf{k}) \right) | \psi_{n\mathbf{k}}^{(0)} \rangle \langle \psi_{n\mathbf{k}}^{(0)} |
\end{align}
\begin{align}
H_1 & = \sum_{\mathbf{k} \neq \mathbf{k}'} \sum_{n \neq m} \left( e \mathbf{E} \cdot \langle \psi_{n\mathbf{k}}^{(0)} | \mathbf{r} | \psi_{m\mathbf{k}'}^{(0)} \rangle \right) | \psi_{n\mathbf{k}}^{(0)} \rangle \langle \psi_{m\mathbf{k}'}^{(0)} | \\
& =  \sum_{\mathbf{k}} \sum_{n \neq m} \left( e \mathbf{E} \cdot \langle \psi_{n\mathbf{k}}^{(0)}  | \mathbf{r} | \psi_{m\mathbf{k}}^{(0)}  \rangle \right) | \psi_{n\mathbf{k}}^{(0)}  \rangle \langle \psi_{m\mathbf{k}}^{(0)}  | \\
& \equiv \sum_{\mathbf{k}} \sum_{n \neq m} \left( e \mathbf{E} \cdot \mathbf{A}_{nm}^{(0)} (\mathbf{k}) \right) | \psi_{n\mathbf{k}}^{(0)}  \rangle \langle \psi_{m\mathbf{k}}^{(0)} |
\end{align}
Here, \(\varepsilon_n^{(n)}(\mathbf{k})\), \(|\psi_{n\mathbf{k}}^{(0)}\rangle\) represent the eigenvalues and eigenstates (Bloch wavefunction) of the \(n\)-th band at crystal momentum \(\mathbf{k}\) of the original Hamiltonian \(H^{(0)}\) before applying the electric field, and \(\mathbf{A}_{nm}^{(0)}(\mathbf{k})\) denotes the (0th order) Berry connection.

We emphasize that the upper index \((\dots)^{(n)}\) indicates considering \(e \mathbf{E} \cdot \mathbf{r}\) as the perturbation, since we only know about the dispersion \(\varepsilon_n^{(0)}(\mathbf{k})\) and the eigenstate (Bloch wavefunction) \(|\psi_{n\mathbf{k}}^{(0)} \rangle\), so we have to transform all the quantities we want to calculate into forms of \((\dots)^{(n)}\).

To this end, we write the SW transformation as,
\begin{equation}
H' = e^{\epsilon S} H e^{-\epsilon S},
\end{equation}
and we regard the perturbative parameter \(\epsilon\) as the marker for the order of the original perturbation theory (i.~e. $
\epsilon \propto |\mathbf{E}|$).

The separation of the diagonal and off-diagonal parts of \(H\) now becomes,
\begin{equation}
H = H_0 + \epsilon H_1.
\end{equation}
One can check the correspondence of the order of the perturbation theory both before and after the SW transformation by expanding the SW transformation order by order with respect to the parameter \(\epsilon\),
\begin{align}
H' & = \sum_{n = 0}^{\infty} \frac{\epsilon^n}{n!} \underbrace{[S,[S,\dots,[S,H]\dots]]}_{n\text{ nested commutators}} \notag\\
& = H + \epsilon [S, H] + \frac{\epsilon^2}{2!}[S,[S,H]] + \frac{\epsilon^3}{3!} [S,[S,[S,H]]] + \dots \notag\\
& = H_0 + \epsilon \left( H_1 + [S,H_0] \right) + \epsilon^2 \left( [S,H_1] + \frac{1}{2} [S,[S,H_0]] \right) \notag\\
& + \dots
\end{align}
This imposes consistency conditions on the operator \(S\), which should also be anti-Hermitian to guarantee the SW transformation is unitary.
To enforce that the SW-transformed Hamiltonian has a perturbation term only at order \(\mathcal{O}(\mathbf{E}^2) \sim \mathcal{O}(\epsilon^2)\), it should hold that
\begin{equation}
H_1 + [S,H_0] = 0.
\end{equation}

As long as \(S\) has the same discrete translational symmetry as \(H^{(0)}\), we can write the operator \(S\) in the Bloch basis, which is block diagonal in momentum space,
\begin{align}
S 
& = \sum_{\mathbf{k}} \sum_{n,m} S_{nm} | \psi_{n\mathbf{k}}^{(0)} \rangle \langle \psi_{m\mathbf{k}}^{(0)} |,
\end{align}
where the matrix element is \(S_{nm}\equiv\langle \psi_{n\mathbf{k}}^{(0)} | S | \psi_{m\mathbf{k}}^{(0)} \rangle\). We can also write this constraint in the basis of \(| \psi_{n \mathbf{k}}^{(0)} \rangle\) as:
\begin{align}
\sum_{\mathbf{k}} & \sum_{n \neq m} \left( e \mathbf{E} \cdot \mathbf{A}_{nm}^{(0)} \right) | \psi_{n\mathbf{k}}^{(0)}  \rangle \langle \psi_{m\mathbf{k}}^{(0)} | \notag\\
+\sum_{\mathbf{k}} & \sum_{n,m} \left[ S_{nm} \left( \varepsilon_m^{(0)} (\mathbf{k}) + e \mathbf{E} \cdot \mathbf{A}_{mm}^{(0)}(\mathbf{k}) \right) \right. \notag\\
& \left. -\left( \varepsilon_n^{(0)} (\mathbf{k}) + e \mathbf{E} \cdot \mathbf{A}_{nn}^{(0)}(\mathbf{k}) \right) S_{nm} \right] | \psi_{n\mathbf{k}}^{(0)} \rangle \langle \psi_{m\mathbf{k}}^{(0)} | \notag\\
= 0 & 
\end{align}
The diagonal part is identically fulfilled by choosing $S_{nn} = 0$, while the off-diagonal part reads,
\begin{align}
& e \mathbf{E} \cdot \mathbf{A}_{nm}^{(0)} + S_{nm} \left( \varepsilon_m^{(0)} (\mathbf{k}) + e \mathbf{E} \cdot \mathbf{A}_{mm}^{(0)}(\mathbf{k}) \right) \notag\\
& - \left( \varepsilon_n^{(0)} (\mathbf{k}) + e \mathbf{E} \cdot \mathbf{A}_{nn}^{(0)}(\mathbf{k}) \right) S_{nm} = 0,
\end{align}
which has the solution,
\begin{align}
S_{nm} & = \frac{- e \mathbf{E} \cdot \mathbf{A}_{nm}^{(0)}}{\varepsilon_{nm}^{(0)} - e \mathbf{E} \cdot (\mathbf{A}_n - \mathbf{A}_m) } \notag\\
& \approx \underbrace{ \frac{- e \mathbf{E} \cdot \mathbf{A}_{nm}^{(0)}}{\varepsilon_{nm}^{(0)}} }_{\mathcal{O}(\mathbf{E})} - \underbrace{\frac{e^2 E^a E^b A_{nm}^a (A_n^b - A_m^b)}{(\varepsilon_{nm}^{(0)})^2}}_{\mathcal{O}(\mathbf{E}^2)} + \mathcal{O}(\mathbf{E}^3) \notag\\
& \equiv \epsilon S^{(1)}_{nm} + \epsilon^2 S_{nm}^{(2)} + \mathcal{O}(\mathbf{E}^3)
\end{align}
where
$\varepsilon^{(0)}_{nm} \equiv \varepsilon^{(0)}_n - \varepsilon^{(0)}_m$.

With this choice of  \(S\), the expression for \(H'\) is simplified to,
\begin{align}
H' & = H_0 + \frac{1}{2} \epsilon [S, H_1] + \frac{1}{3} \epsilon [S,[S,H_1]] + \dots \notag\\
& = H_0 + \frac{1}{2} \epsilon^2 [S^{(1)}, H_1] \notag\\
& + \epsilon^3 \left[ \frac{1}{2}  [S^{(2)}, H_1] + \frac{1}{3} [S^{(1)},[S^{(1)}, H_1]] \right] + \mathcal{O}(\mathbf{E}^4) \notag\\
& =H_0 + \underbrace{H_1'}_{\mathcal{O}(\mathbf{E}^2)}
\end{align}
Now, the unperturbed part of the SW-transformed Hamiltonian can be regarded as the diagonal part of the original Hamiltonian \(H_0\), whose unperturbed eigenstates are just \(|\psi_{n\mathbf{k}}^{(0)}\rangle\), and the corresponding eigenvalues are \((\varepsilon_n^{(0)} - e \mathbf{E} \cdot \mathbf{A}_{nn}^{(0)})\). We have transformed the perturbation Hamiltonian into a \(\mathcal{O}(\mathbf{E}^2)\) term.

Thus, the 1st order perturbation theory for this SW-transformed Hamiltonian is enough to tell us the correction of dispersion up to 3rd order, i.e., \(\varepsilon_n^{(3)}\)~\cite{Fang2024}. Since the perturbation is \(\mathcal{O}(\mathbf{E}^2)\) and contains \(\mathcal{O}(\mathbf{E}^4)\), the correction of the dispersion up to \(\varepsilon_n^{(4)}\) will contain contributions from both the 1st order perturbation theory’s \(\mathcal{O}(\mathbf{E}^4)\) term and the 2nd order perturbation theory’s \((H_1')^2 \sim \mathcal{O}(\mathbf{E}^4)\) term. However, this discussion is complex and will not be covered here. Additionally, since the SW transformation is unitary, the calculation yields the same eigenvalue for the original Hamiltonian \(H\).

\paragraph{Correction of band dispersion}

We calculate the correction to eigenvalues order by order. The SW-transformed Hamiltonian's 0th order eigenvalue already contains \(\mathcal{O}(\mathbf{E}^0)\) and \(\mathcal{O}(\mathbf{E}^1)\) corrections for the original Hamiltonian eigenvalues:
\begin{equation}
\varepsilon_n^{(1)} = - e E^a A_{nn}^a
\end{equation}

The SW-transformed Hamiltonian's 1st order eigenvalue contains the \(\mathcal{O}(\mathbf{E}^2)\) correction, where \(\varepsilon_n^{(2)}\) is:
{\allowdisplaybreaks
\begin{align}
\varepsilon_n^{(2)} & = \langle \psi_{n \mathbf{k}}^{(0)} | \frac{1}{2} \epsilon^2 [S^{(1)}, H_1] | \psi_{n \mathbf{k}}^{(0)} \rangle \notag\\
& = \frac{1}{2} \sum_{m \neq n} \epsilon S_{nm}^{(1)} (- e \mathbf{E} \cdot \mathbf{A}_{mn}^{(0)} ) - (- e \mathbf{E} \cdot \mathbf{A}_{nm}^{(0)} ) \epsilon S_{mn}^{(1)} \notag\\
& = \frac{1}{2} e^2 E^a E^b \left( \sum_{m \neq n} \frac{A_{nm}^a A_{mn}^b + A_{nm}^b A_{mn}^a}{\varepsilon_{nm}^{(0)}} \right).
\end{align}}
Here, the \(a\)-th component of unperturbed Berry connection has been abbreviated as \(A_{nm}^a = (\mathbf{A}_{nm}^{(0)})^a\).

\paragraph{Correction of wave function and Berry connection}

The correction to the wave function corresponds to
\begin{align}
|n\rangle & \equiv e^{-S} |n'\rangle \notag\\
& = |n'\rangle - S |n'\rangle + \frac{1}{2!} S^2 |n'\rangle + \dots \notag\\
& = (|n^{(0)}\rangle + \mathcal{O}(\mathbf{E}^2) ) - S (|n^{(0)}\rangle + \mathcal{O}(\mathbf{E}^2) ) + \mathcal{O}(\mathbf{E}^2) \notag\\
& = |n^{(0)}\rangle - S |n^{(0)}\rangle + \mathcal{O}(\mathbf{E}^2)
\end{align}
Here, \(|n \rangle\) is the eigenstate of the original perturbed system (before the SW transformation), and \(|n' \rangle\) is the eigenstate of the Hamiltonian after the SW transformation. We've used the fact that \(|n'\rangle\) is \(|n^{(0)}\rangle\) up to \(\mathcal{O}(\mathbf{E}^2)\) and \(S \sim \mathcal{O}(\mathbf{E})\), where \(|n^{(0)}\rangle\) represents the eigenstate of the original unperturbed Hamiltonian.

The corrected Berry connection is,
\begin{align}
& \mathbf{A}_{nm} = \langle \psi_{n\mathbf{k}} | \mathbf{r} | \psi_{m \mathbf{k}} \rangle \notag\\
& = \left( \langle \psi_{n\mathbf{k}}^{(0)} | - \langle \psi_{n\mathbf{k}}^{(0)} | S^\dagger + \mathcal{O}(\mathbf{E}^2) \right)  \mathbf{r} \left( | \psi_{n\mathbf{k}}^{(0)} \rangle - S | \psi_{n\mathbf{k}}^{(0)} \rangle + \mathcal{O}(\mathbf{E}^2) \right) \notag\\
& = \underbrace{\mathbf{A}_{nm}^{(0)}}_{\mathcal{O}(\mathbf{E}^0)} + \underbrace{\langle \psi_{n\mathbf{k}}^{(0)} | [S, \mathbf{r}] | \psi_{m \mathbf{k}}^{(0)} \rangle}_{\mathcal{O}(\mathbf{E}^1)} + \mathcal{O}(\mathbf{E}^2).
\end{align}
Therefore, the 1st order correction to the Berry connection reads explicitly,
\begin{align}
& (\mathbf{A}_{nm}^{(1)})^b \notag\\
& = \langle \psi_{n\mathbf{k}}^{(0)} | [S^{(1)}, \mathbf{r}^b] | \psi_{m \mathbf{k}}^{(0)} \rangle \notag\\
& = \sum_{l} S_{nl}^{(1)} (\mathbf{A}_{lm}^{(0)})^b - (\mathbf{A}_{nl}^{(0)})^b S_{lm}^{(1)} \notag\\
& = -e E^a \sum_{l} \frac{ (\mathbf{A}_{nl}^{(0)})^a (\mathbf{A}_{lm}^{(0)})^b}{\varepsilon_{nl}^{(0)}} - \frac{ (\mathbf{A}_{nl}^{(0)})^b (\mathbf{A}_{lm}^{(0)})^a }{\varepsilon_{lm}^{(0)}} \notag\\
& = -e E^a G^{ab}_{nm}.
\end{align}
The quantity
\(G^{ab}_{nm} \equiv \sum_{l} 
(\mathbf{A}_{nl}^{(0)})^a (\mathbf{A}_{lm}^{(0)})^b/\varepsilon_{nl}^{(0)} - (\mathbf{A}_{nl}^{(0)})^b (\mathbf{A}_{lm}^{(0)})^a/\varepsilon_{lm}^{(0)}\) is defined as the (non-Abelian) band-normalized quantum metric. For response up to 2nd order, we only need its diagonal (Abelian) part, i.e., the Band-normalized Quantum metric of the \(n\)-th band,
\begin{equation}
G_n^{ab} \equiv \sum_{m \neq n} \frac{ (\mathbf{A}_{nm}^{(0)})^a (\mathbf{A}_{mn}^{(0)})^b + (\mathbf{A}_{nm}^{(0)})^b (\mathbf{A}_{mn}^{(0)})^a }{\varepsilon_{nm}^{(0)}}.
\end{equation}

The corresponding 1st order correction of Berry curvature is,
\begin{equation}
(\mathbf{\Omega}^{(1)}_n)^c = \epsilon^{abc} \partial_a (\mathbf{A}^{(1)}_{nn}) = - e \mathbf{E}^d \epsilon^{abc} \partial_a \Omega_n^{db},
\end{equation}
where $\epsilon^{abc}$ is the Levi-Civita tensor.

This completes the semiclassical perspective, which after some algebra results in Eqs.~(\ref{eqSmain1}-\ref{eqSmain3}) quoted in the main text.

\subsection{Spacetime transformation and symmetry}\label{transport symmetry}

With the knowledge of the spatial transformation property of dipole and velocity matrix elements, one can straightforwardly deduce the symmetry of Drude, Berry Curvature Dipole and Quantum Metric Dipole.
%
%\paragraph{Spatial transformation property of Berry curvature and band-normalized quantum metric}
To this end, consider the symmetry of Berry curvature and band-normalized quantum metric,
\begin{align}
\Omega^{ab}_n = \sum_{m \neq n} \left( r^a_{nm} r^b_{mn} - r^b_{nm} r^a_{mn} \right)
\\
G^{ab}_n = \sum_{m \neq n} \frac{ r^a_{nm} r^b_{mn} + r^b_{nm} r^a_{mn} }{\epsilon_{nm}}
\end{align}
As explained in Sec.~\ref{app02}, the spatial transformation of dipole matrix element behave like a vector field,
\begin{equation}
r^a_{nm} (\mathbf{k}) \rightarrow r'^a_{nm} (\mathbf{k}) = \sum_{b} \mathcal{R}^{a b} r^b_{nm} (\mathcal{R}^{-1} \mathbf{k}).
\end{equation}
The spatial transformation of energy (difference) behaves like a scalar field,
\begin{equation}
\epsilon_{nm} (\mathbf{k}) \rightarrow \epsilon'_{nm} (\mathbf{k}) = \epsilon_{nm} (\mathcal{R}^{-1} \mathbf{k}).
\end{equation}

Thus, the Berry curvature and band-normalized quantum metric both transform as rank two tensors,
\begin{equation}
\Omega^{ab}_n (\mathbf{k}) \rightarrow {\Omega'}^{ab}_n (\mathbf{k}) = \sum_{a'b'} \mathcal{R}^{a a'} \mathcal{R}^{b b'} \Omega^{a'b'}_n (\mathcal{R}^{-1} \mathbf{k}),
\end{equation}
\begin{equation}
G^{ab}_n (\mathbf{k}) \rightarrow {G'}^{ab}_n (\mathbf{k}) = \sum_{a'b'} \mathcal{R}^{a a'} \mathcal{R}^{b b'} G^{a'b'}_n (\mathcal{R}^{-1} \mathbf{k}).
\end{equation}
Regarding the pure spatial transformation of nonlinear Drude, Berry Curvature Dipole and Quantum Metric Dipole, we conclude that they are all cubic in the dipole matrix element $r_{nm}$ and momentum derivative $\partial_{k}$. Therefore, the pure spatial transformation property of nonlinear Drude, Berry Curvature Dipole and Quantum Metric Dipole behave like a 3rd rank order tensor field.

Turning to the time reversal transformation, the dipole matrix element reads
\begin{equation}
r^a_{nm} (\mathbf{k}) \rightarrow r'^a_{nm} (\mathbf{k}) = r^a_{mn} (-\mathbf{k})
\end{equation}
while the energy (difference) is
\begin{equation}
\epsilon_{nm} (\mathbf{k}) \rightarrow \epsilon'_{nm} (\mathbf{k}) = \epsilon_{nm} ( - \mathbf{k}).
\end{equation}
The time reversal transformation of Berry curvature and band-normalized quantum metric then follow as,
\begin{align}
\Omega^{ab}_n (\mathbf{k}) 
&\rightarrow {\Omega'}^{ab}_n (\mathbf{k}) 
\notag\\& = \sum_{m \neq n} \left[  r'^a_{nm} (\mathbf{k}) r'^b_{mn} (\mathbf{k}) - r'^b_{nm} (\mathbf{k}) r'^a_{mn} (\mathbf{k}) \right] \notag\\
& = \sum_{m \neq n} \left[  r^a_{mn} (-\mathbf{k}) r^b_{nm} (-\mathbf{k}) - r^b_{nm} (-\mathbf{k}) r^a_{mn} (-\mathbf{k}) \right] \notag\\
& = \sum_{m \neq n} \left[  r^b_{nm} (-\mathbf{k}) r^a_{mn} (-\mathbf{k}) - r^a_{mn} (-\mathbf{k}) r^b_{nm} (-\mathbf{k})\right] \notag\\
& = \Omega^{ba}_n(-\mathbf{k}) = - \Omega^{ab}_n (-\mathbf{k})
\end{align}

\begin{align}
G^{ab}_n (\mathbf{k}) &\rightarrow {G'}^{ab}_n (\mathbf{k}) 
\notag\\& = \sum_{m \neq n} \frac{ r'^a_{nm} (\mathbf{k}) r'^b_{mn} (\mathbf{k}) + r'^b_{nm} (\mathbf{k}) r'^a_{nm} (\mathbf{k}) }{\epsilon'_{nm} (\mathbf{k})} \notag\\
& = \sum_{m \neq n} \frac{ r^a_{mn} (-\mathbf{k}) r^b_{nm} (-\mathbf{k}) + r^b_{nm} (-\mathbf{k}) r^a_{mn} (-\mathbf{k}) }{\epsilon_{nm} (-\mathbf{k})} \notag\\
& = \sum_{m \neq n} \frac{ r^b_{nm} (-\mathbf{k}) r^a_{mn} (-\mathbf{k}) + r^a_{mn} (-\mathbf{k}) r^b_{nm} (-\mathbf{k}) }{\epsilon_{nm} (-\mathbf{k})} \notag\\
& = G^{ba}_n(-\mathbf{k}) = G^{ab}_n (-\mathbf{k})
\end{align}
Combining both transformations, we note that the Berry curvature is odd under $\mathcal{PT}$ transformation, i.e.
\begin{equation}
\Omega^{ab}_n (\mathbf{k}) \xrightarrow{\mathcal{T}} -{\Omega}^{ab}_n (-\mathbf{k}) \xrightarrow{\mathcal{P}} -{\Omega}^{ba}_n (\mathbf{k}).
\end{equation}
Therefore if the system has $\mathcal{PT}$ symmetry, the Berry curvature should be zero, which directly rules out the Berry Curvature Dipole term in the 2nd order dc conductivity.
As an instructive example, we analyze the QMD term,
\begin{widetext}
\begin{align}
\sigma^{ab;c}_{\mathrm{QMD}} &\equiv - \frac{e^3 }{\hbar} \sum_n \int_k f_n(\mathbf{k}) \left[ 2 \partial_c G_n^{ab}(\mathbf{k}) - \frac{1}{2}( \partial_a G_n^{bc}(\mathbf{k}) + \partial_b G_n^{ac}(\mathbf{k}) ) \right]\\
\rightarrow\quad  {\sigma'}^{ab;c}_{\mathrm{QMD}} 
& = - (-1) \frac{e^2}{\hbar^3} \sum_n \int_k f_n(-\mathbf{k}) \left[ 2 \partial_{(-\mathbf{k})_c} G_n^{ab}(-\mathbf{k}) - \frac{1}{2}( \partial_{(-\mathbf{k})_a} G_n^{bc}(-\mathbf{k}) + \partial_{(-\mathbf{k})_b} G_n^{ac}(-\mathbf{k}) ) \right] = -\sigma^{ab;c}_{\mathrm{QMD}}.
\end{align}
In conclusion, the nonlinear Drude and quantum metric dipole terms are odd under time reversal transformation, while the Berry curvature dipole term is even under time reversal transformation, which means nonlinear Drude and quantum metric dipole terms can only appear when time reversal symmetry is broken.

\subsection{Quantum pertubative approach}\label{appendix:diagrammatics_transport}
We elucidate the two key steps how the dc-limit can be taken in a Green-Kubo approach, leading back to the same result of Eqs.~(\ref{eqSmain1}-\ref{eqSmain3}) obtained from the semiclassical approach.

\paragraph{Expanding velocity vertices}

In the diagrammatic approach, all geometric information is contained in different orders of velocity vertices, e.g. $\hat{v}^a$, $\hat{w}^{ab}$, and $\hat{u}^{abc}$. However, they the quantum geometric quantities are far from explicit in these vertices. Namely, the Berry connection $r^a$, the quantum metric $g^{ab}$, the Berry curvature $\Omega^{ab}$, the quantum connection $Q^{a;bc}$, the Berry curvature dipole $\partial_{k_c} \Omega^{ab}$, and the quantum metric dipole $\partial_{k_c} g^{ab}$ have to be first extracted by sum rules.

To this end, note that the geometric quantities are essentially momentum derivatives of the wavefunction, e.g. $\partial_{k} | m \rangle$ rather than derivatives of the Hamiltonian $\partial_k\dots\partial_k \hat{H}(\mathbf{k})$, the latter of which contains both dispersion $\varepsilon_m$ and wavefunction information $| m \rangle$.

The crucial ingredient is the covariant derivative~\cite{Ventura2017,Parker2019}, which connects the matrix element of momentum derivative to the momentum derivative of matrix elements in Bloch basis, i.e.
\begin{align}
(\partial_a \hat{\mathcal{O}})_{mn} & = \langle m | \partial_a (\sum_{m'n'} | m' \rangle \mathcal{O}_{m'n'} \langle n' | ) | n \rangle \notag\\
& = \sum_{m'n'} \langle m | \partial_a m' \rangle \mathcal{O}_{m'n'} \langle n' | n \rangle + \sum_{m'n'} \langle m | m' \rangle \partial_a \mathcal{O}_{m'n'} \langle n' | n \rangle + \sum_{m'n'} \langle m | m' \rangle \mathcal{O}_{m'n'} \langle \partial_a n' | n \rangle \notag\\
& = \partial_a \mathcal{O}_{mn} + \sum_{m'} \langle m | \partial_a m' \rangle \mathcal{O}_{m'n} + \sum_{n'}  \mathcal{O}_{mn'} \langle \partial_a n' | n \rangle \notag\\
& = \partial_a \mathcal{O}_{mn} + i [ \hat{\mathcal{O}}, \hat{r} ]_{mn},
\end{align}

where the Berry connection $\hat{r}$ is, as usual given by,
$$
r^a_{mn} \equiv i \langle m | \partial_a n \rangle = - i \langle \partial_a m | n \rangle,
$$
which follows from $0 = \partial_a (\langle m | n \rangle) = \langle \partial_a m | n \rangle + \langle m | \partial_a n \rangle$. 
The matrix element of commutator should be understood as,
$$
[ \hat{A}, \hat{B} ]_{mn} = \sum_{m'} \left( A_{mm'} B_{m'n} - B_{mm'} A_{m'n} \right).
$$

Then we can start from analyzing the 1st order velocity vertex $\hat{v}^a$,
\begin{align}
v^a_{mn} & = (\partial_{a} \hat{H})_{mn} = \partial_a H_{mn} + i [ \hat{H}, \hat{r}^a ]_{mn} = \partial_a \varepsilon_n \delta_{mn} + i (\hat{\varepsilon} \circ\hat{r})_{mn}.
\end{align}
Here we used the property that the Bloch Hamiltonian is undoubtedly diagonal in the Bloch basis, i.e. $H_{mn} = \varepsilon_n \delta_{mn}$. Hadamard product $\circ$ denotes the element-wise product of two matrices,
$(\hat{A} \circ \hat{B})_{mn} = A_{mn} B_{mn}$.
The matrix element of $\hat{\varepsilon}$ is the energy difference between different bands,
$(\hat{\varepsilon})_{mn} = \varepsilon_m - \varepsilon_n$.

The expected geometric content in a nth-order vertex is thus a collection of at most n number of pure momentum derivatives of Bloch wavefunctions. In order to capture this, we define a (gauge-dependent) higher-order derivative via
$$
\Lambda^{a_1\dots a_n}_{mn} \equiv \frac{i}{2} \left( \langle m | \partial_{a_1} \dots \partial_{a_n}  n \rangle - \langle \partial_{a_1} \dots \partial_{a_n} m | n \rangle \right).
$$
At first order this is just the Berry connection, i.e. $\Lambda^{a}_{mn} = r^a_{mn}$, while the second order gauge-dependent geometric quantity becomes,
$$
\lambda^{a_1a_2}_{mn} \equiv \Lambda_{mn}^{a_1 a_2} \equiv \frac{i}{2} \left( \langle m | \partial_{a_1} \partial_{a_2} n \rangle - \langle \partial_{a_1} \partial_{a_2} m | n \rangle \right).
$$

Now we keep on expanding higher order velocity vertices, the matrix elements of the second order velocity vertex $\hat{w}^{a_1 a_2}$ is,

\begin{align}
w^{ab}_{mn} & = (\partial_{a} \hat{v}^b)_{mn} = \partial_a v^b_{mn} + i [ \hat{v}^b, \hat{r}^a ]_{mn} \notag\\
& = \partial_a \left( \partial_b \varepsilon_n \delta_{mn} + i (\hat{\varepsilon} \circ \hat{r}^{b})_{mn} \right) + i [ \left( \partial_b \varepsilon_n \delta_{mn} + i (\hat{\varepsilon} \circ \hat{r}^{b}) \right), \hat{r}^b ]_{mn} \notag\\
& = \partial_a \partial_b \varepsilon_n \delta_{mn} + i \partial_a (\hat{\varepsilon} \circ \hat{r}^{b})_{mn} + i (\partial_a \varepsilon_{mn}) r^b_{mn} - [ \hat{\varepsilon} \circ \hat{r}^a, \hat{r}^b ]_{mn} \notag\\
& = \partial_a \partial_b \varepsilon_n \delta_{mn} + i  ( \hat{\Delta}^a \circ \hat{r}^{b} + \hat{\Delta}^b \circ \hat{r}^{a})_{mn} + i \varepsilon_{mn} \partial_a r^b_{mn} - [ \hat{\varepsilon} \circ \hat{r}^a, \hat{r}^b ]_{mn} 
\end{align}
where the velocity shift is $\Delta^a_{mn} \equiv \partial_a \varepsilon_{mn} = \partial_a \varepsilon_{m} - \partial_a \varepsilon_{n} = v^a_{mm} - v^a_{nn}$.

Yet, the second order velocity vertex, can also be expanded for $(a\leftrightarrow b)$, i.e. $\hat{w}^{ab} = \partial_b \hat{v}^a$. Both expressions are not manifestly the same because the expansion in terms of $\hat{r}^a$ breaks the explicit permutation symmetry in spatial indices $a$ and $b$. We can recover such explicit permutation symmetry by manually symmetrizing the expansion in terms of $(a \leftrightarrow b)$, which affects both $i \varepsilon_{mn} \partial_a r^b_{mn}$ and $- [ \hat{\varepsilon} \circ \hat{r}^a, \hat{r}^b ]_{mn}$ which are not symmetric under $(a \leftrightarrow b)$. A useful organizing principle is to introduce a higher order analogue to the Berry curvature as,
\begin{align}
\tilde{\Omega}^{ab,N}_{mn} \equiv \left\{ \begin{array}{ll}
i \left[\hat{\varepsilon}^N \circ \hat{r}^a, \hat{r}^b\right]_{mn} & \text{if } N \text{ even} \\
\left[\hat{\varepsilon}^N \circ \hat{r}^b, \hat{r}^a\right]_{mn} & \text{if } N \text{ odd}
\end{array} \right.
\end{align}
it is easy to see that the zeroth order generalized Berry curvature is just the ordinary Berry curvature, i.e. $\tilde{\Omega}^{ab,0}_{mn} = \Omega^{ab}_{mn}$.
The derivative of the position operator is then,
\begin{align}
\partial_b r^a_{mn} & = \frac{i}{2} \partial_b \left( \langle m | \partial_a n \rangle - \langle \partial_a m | n \rangle \right) \notag\\
& = \frac{i}{2} \left(  \langle m | \partial_a \partial_b n \rangle - \langle \partial_a \partial_b m | n \rangle \right) + \frac{i}{2} \left( \langle \partial_b m | \partial_a n \rangle - \langle \partial_a m | \partial_b n \rangle \right) \notag\\
& = \lambda^{ab}_{mn} + \frac{i}{2} \sum_l \left( \langle \partial_b m | l \rangle \langle l | \partial_a n \rangle - \langle \partial_a m | l \rangle \langle l | \partial_b n \rangle \right) \notag\\
& = \lambda^{ab}_{mn} + \frac{i}{2} \sum_l \left( r^b_{ml} r^a_{ln} - r^a_{ml} r^b_{ln} \right) \notag \\
& = \lambda^{ab}_{mn} - \frac{1}{2} \Omega^{ab}_{mn}
\end{align}
In this notation, the first term is symmetric while the second term is antisymmetric under $(a \leftrightarrow b)$.
Symmetrizing the second order vertex thus yields,
\begin{align}
w^{ab}_{mn} & = \frac{1}{2} \left[ (\partial_b \hat{r}^a)_{mn} + (a \leftrightarrow b) \right] \notag\\
& = \partial_a \partial_b \varepsilon_n \delta_{mn} + i  ( \hat{\Delta}^a \circ \hat{r}^{b} + \hat{\Delta}^b \circ \hat{r}^{a})_{mn} + i (\hat{\varepsilon} \circ \hat{\lambda}^{ab})_{mn} - \frac{1}{2} \left( \tilde{\Omega}^{ab,1}_{mn} + \tilde{\Omega}^{ba,1}_{mn} \right).
\end{align}

The remaining expansions follow the same principles. For example, the diagonal matrix elements of the third order velocity vertex $\hat{u}^{abc}$ read (note that $\varepsilon_{nn} = 0$, $\Delta^a_{nn} = 0$),
\begin{align}
u^{abc}_{nn} & = \partial_c w^{ab}_{nn} + i [ w^{ab}, \hat{r}^c ]_{nn} \\
& = \underbrace{\partial_a \partial_b \partial_c \varepsilon_n - \frac{1}{2} \partial_c \left( \tilde{\Omega}^{ab,1}_{nn} + \tilde{\Omega}^{ba,1}_{nn} \right)}_{\partial_c w^{ab}_{nn}} \underbrace{- [\hat{\Delta}^a \circ \hat{r}^b + \hat{\Delta}^b \circ \hat{r}^a, \hat{r}^c]_{nn} - \frac{i}{2} [\tilde{\Omega}^{ab,1} + \tilde{\Omega}^{ba,1}, \hat{r}^c]_{nn} - [\hat{\varepsilon} \circ \hat{\lambda}^{ab}, \hat{r}^c]_{nn}}_{i [ w^{ab}, \hat{r}^c ]_{nn}}.
\end{align}
We mention in passing that only the $\partial_c w^{ab}_{nn}$ part is needed when deriving the 2nd order conductivity, since the $i [ w^{ab}, \hat{r}^c ]_{nn}$ part will be canceled by diagram (2.D) that contributes a $i [ \hat{r}^c, w^{ab} ]_{nn}$ term.
Additionally, for a fully filled band (insulator) with $f_n = 1 \text{ or } 0$, it is $\int_{\mathbf{k}} f_n \partial_c w^{ab}_{nn} = 0$.

The last ingredient is the expansion of the momentum derivative of generalized Berry curvature, i.e.
\begin{align}
\partial_c \tilde{\Omega}^{ab,1}_{nn} & = \partial_c [\hat{\varepsilon} \circ \hat{r}^a, \hat{r}^b]_{nn} \\
& = [\hat{\Delta}^c \circ \hat{r}^a, \hat{r}^b]_{nn} + [\hat{\varepsilon} \circ \partial_c \hat{r}^a, \hat{r}^b]_{nn} + [\hat{\varepsilon} \circ \hat{r}^a, \partial_c \hat{r}^b]_{nn} \\
& = [\hat{\Delta}^c \circ \hat{r}^a, \hat{r}^b]_{nn} + [\hat{\varepsilon} \circ (\hat{\lambda}^{ca} + \frac{1}{2} \tilde{\Omega}^{ca,0}), \hat{r}^b]_{nn} + [\hat{\varepsilon} \circ \hat{r}^a, (\hat{\lambda}^{cb} + \frac{1}{2} \tilde{\Omega}^{cb,0})]_{nn}
\end{align}
Using the property $\left[ \hat{\varepsilon} \circ \hat{A}, \hat{B} \right]_{nn} = \left[ \hat{\varepsilon} \circ \hat{B}, \hat{A} \right]_{nn}$, it is,
\begin{align}
\partial_c \tilde{\Omega}^{ab,1}_{nn} & = [ \hat{\Delta}^c \circ \hat{r}^a, \hat{r}^b ]_{nn} + [ \hat{\varepsilon} \circ \hat{\lambda}^{ca} , \hat{r}^b ]_{nn} + [ \hat{\varepsilon} \circ \hat{r}^a, \hat{\lambda}^{cb} ]_{nn} + \frac{i}{2} \left[ \hat{\varepsilon} \circ \left[ \hat{r}^c, \hat{r}^a \right] , \hat{r}^b \right]_{nn} + \frac{i}{2} \left[ \hat{\varepsilon} \circ \left[ \hat{r}^c, \hat{r}^b \right] , \hat{r}^a \right]_{nn} \\
& = [ \hat{\Delta}^c \circ \hat{r}^a, \hat{r}^b ]_{nn} + [ \hat{\varepsilon} \circ \hat{\lambda}^{ca} , \hat{r}^b ]_{nn} + [ \hat{\varepsilon} \circ \hat{r}^a, \hat{\lambda}^{cb} ]_{nn} - \frac{i}{2} \left\{ \left[ \hat{r}^b , \left[ \hat{H} , \left[ \hat{r}^c, \hat{r}^a \right] \right] \right]_{nn} + (a \leftrightarrow b) \right\}.
\end{align}
With the help of the Jacobi identities for nested commutators, 
\begin{align}
0&=\left[ \hat{A}, \left[ \hat{B}, \hat{C} \right] \right] + \left[ \hat{B}, \left[ \hat{C}, \hat{A} \right] \right] + \left[ \hat{C}, \left[ \hat{A}, \hat{B} \right] \right],
\\
0&=\left[ \hat{A}, \left[ \hat{B}, \left[ \hat{C}, \hat{D} \right] \right] \right] + \left[ \hat{B}, \left[ \hat{C}, \left[ \hat{D}, \hat{A} \right] \right] \right] + \left[ \hat{C}, \left[ \hat{D}, \left[ \hat{A}, \hat{B} \right] \right] \right] + \left[ \hat{D}, \left[ \hat{A}, \left[ \hat{B}, \hat{C} \right] \right] \right],
\end{align}
one obtains
\begin{align}
-\left[ \hat{r}^b , \left[ \hat{H} , \left[ \hat{r}^c, \hat{r}^a \right] \right] \right]_{nn} & = \left[ \hat{H}, \left[ \hat{r}^c , \left[ \hat{r}^a, \hat{r}^b \right] \right] \right]_{nn} + \left[ \hat{r}^c , \left[ \hat{r}^a , \left[ \hat{r}^b, \hat{H} \right] \right] \right]_{nn} + \left[ \hat{r}^a , \left[ \hat{r}^b , \left[ \hat{H}, \hat{r}^c \right] \right] \right]_{nn}.
\end{align}
Therefore, we can rewrite the expression of $\partial_c \tilde{\Omega}^{ab,1}_{nn}$ as,
\begin{align}
\partial_c \tilde{\Omega}^{ab,1}_{nn} & = [ \hat{\Delta}^c \circ \hat{r}^a, \hat{r}^b ]_{nn} + [ \hat{\varepsilon} \circ \hat{\lambda}^{ca} , \hat{r}^b ]_{nn} + [ \hat{\varepsilon} \circ \hat{r}^a, \hat{\lambda}^{cb} ]_{nn} \\
& + \frac{i}{2} \{ [ \hat{H}, [ \hat{r}^c , \underbrace{[ \hat{r}^a, \hat{r}^b ]}_{+(a \leftrightarrow b) = 0} ] ]_{nn} + [ \hat{r}^c ,\underbrace{[ \hat{r}^a , [ \hat{r}^b, \hat{H} ] ]}_{\tilde{\Omega}^{ba,1}} ]_{nn} + [ \hat{r}^a , \underbrace{[ \hat{r}^b , [ \hat{H}, \hat{r}^c ] ]}_{-\tilde{\Omega}^{cb,1}} ]_{nn} + (a \leftrightarrow b) \} \\
& = [ \hat{\Delta}^c \circ \hat{r}^a, \hat{r}^b ]_{nn} + [ \hat{\varepsilon} \circ \hat{\lambda}^{ca} , \hat{r}^b ]_{nn} + [ \hat{\varepsilon} \circ \hat{r}^a, \hat{\lambda}^{cb} ]_{nn} + \frac{i}{2} \left\{ - [\tilde{\Omega}^{ba,1}, \hat{r}^c]_{nn} + [\tilde{\Omega}^{cb,1}, \hat{r}^a]_{nn} + (a \leftrightarrow b) \right\}.
\end{align}

The final expression of $u^{abc}_{nn}$ is then, after some lengthy algebra,
\begin{align}
u^{abc}_{nn} 
& = \partial_a \partial_b \partial_c \varepsilon_n + [\hat{r}^c, \hat{\Delta}^a \circ \hat{r}^b]_{nn} + [\hat{r}^a, \hat{\Delta}^b \circ \hat{r}^c]_{nn} + [ \hat{r}^b, \hat{\Delta}^c \circ \hat{r}^a ]_{nn} \notag\\
& - [\hat{\varepsilon} \circ \hat{\lambda}^{ab}, \hat{r}^c]_{nn} - [ \hat{\varepsilon} \circ \hat{\lambda}^{ca} , \hat{r}^b ]_{nn} - [ \hat{\varepsilon} \circ \hat{\lambda}^{bc}, \hat{r}^a ]_{nn} + \frac{i}{2} \left\{ [\hat{r}^a, \tilde{\Omega}^{cb,1}]_{nn} + (a \leftrightarrow b) \right\}
\end{align}

\paragraph{Expansion with respect to $\tau$}

In the transport regime, where \(\omega \to 0\), the quasiparticle lifetime \(\tau\) becomes the only parameter available for expansion. As discussed in the main text, the most general case inevitably involves at least two distinct lifetimes: the interband lifetime \(\tau_{\mathrm{inter}}\) and the intraband lifetime \(\tau_{\mathrm{intra}}\). Let us define two corresponding relaxation rates: \(\Gamma = \tau_{\mathrm{inter}}^{-1}\) and \(\gamma = \tau_{\mathrm{intra}}^{-1}\).
Taking the interband relaxation time as the reference timescale $\tau=\tau_{\mathrm{inter}}$, the intraband relaxation time becomes $\tau_{\mathrm{intra}} = \alpha^{-1} \tau$, with the undetermined ratio $\alpha = \gamma / \Gamma$ distinguishing them.
As discussed in the main text, the ratio $\alpha=2$ implies that the quasiparticle lifetime in the conduction band is much longer than in the valence band. In the transport regime, this ensures charge conservation in the valence band as now the localization effect dominates the relaxation process in transports. In the following, we keep the $\alpha$ as a free parameter to discuss the most general case, and we will show $\alpha = 2$ arises as a natural choice not only in the sense of guaranteeing charge conservation but also in the sense of sharing consistent results (including NLD, BCD, and QMD terms) with the derivation from semiclassical formalism.

Inserting the proper lifetimes into the formula for the second-order conductivity derived from Feynman diagrammatics, yields,

\begin{align}
\sigma^{ab;c}_{\text{(2.A)}} (0; \omega, -\omega) &= -\frac{e^3}{\hbar^2 (\omega + i \tau^{-1}) (-\omega + i \tau^{-1})} \sum_{n} \int_{\mathbf{k}} f_n \frac{u^{abc}_{nn}}{2} + (a,\omega \leftrightarrow b,-\omega)
\\
\sigma^{ab;c}_{\text{(2.B)}+\text{(2.C)}} (0; \omega, -\omega) &= -\frac{e^3}{\hbar^2 (\omega + i \tau^{-1}) (-\omega + i \tau^{-1})} \sum_{nm} \int_{\mathbf{k}} \frac{f_{nm} v_{nm}^{a} w_{mn}^{cb}}{\omega + \varepsilon_{nm} + i \tau^{-1}} + (a,\omega \leftrightarrow b,-\omega)
\\
\sigma^{ab;c}_{\text{(2.D)}} (0; \omega, -\omega) &= -\frac{e^3}{\hbar^2 (\omega + i \tau^{-1}) (-\omega + i \tau^{-1})} \sum_{nm} \int_{\mathbf{k}} \frac{1}{2} \frac{f_{nm} w_{nm}^{ab} v_{mn}^c}{ - \varepsilon_{mn} + i \alpha \tau^{-1}} + (a,\omega \leftrightarrow b,-\omega)
\\
\sigma^{ab;c}_{\text{(2.E)}+\text{(2.F)}} (0; \omega, -\omega) &= -\frac{e^3}{\hbar^2 (\omega + i \tau^{-1}) (-\omega + i \tau^{-1})} \sum_{nml} \int_{\mathbf{k}} \frac{f_{nm} v_{nm}^{a}}{\omega + \varepsilon_{nm} + i \tau^{-1}} \left( \frac{v_{ml}^{b} v_{ln}^{c}}{- \varepsilon_{ln} + i \alpha \tau^{-1}} - \frac{v_{ml}^{c} v_{ln}^{b}}{- \varepsilon_{ml} + i \alpha \tau^{-1}} \right) \notag\\&+ (a,\omega \leftrightarrow b,-\omega)
\end{align}

In these expressions, the interband/intraband lifetime have been introduced according to whether $\pm \omega$ (interband) or $\bar{\omega}$ (intraband) is present in the denominator . Note that the relaxation time is also included in the global coefficient $-\frac{e^3}{\hbar^2 \omega_1 \omega_2}$ which is a consequence of the (quasi-) adiabatic perturbation protocol with $E^a(\omega) = i \omega A^a(\omega)$, which is also subject to the relaxation time. 

We note that compared to the expressions at first order, at second order it is not possible to perform adiabatic switching by simply replacing $\omega\rightarrow i\tau^{-1}$ because mixed terms appear not only in the Green's functions but also in the global frequency denominator.
Instead, the zero frequency limit $\omega \to 0$ should be taken by expanding with respect to $\tau$. Let us demonstrate with a few examples.

Regarding (2.B)+(2.C), recall that $f_{nm}$ always excludes terms where $n=m$, therefore the expansion can be performed while assuming that $|\varepsilon_{nm} \tau| \gg 1$,
\begin{align}
\lim_{\omega \to 0} \frac{1}{(\omega + i \tau^{-1})(- \omega + i \tau^{-1})} \frac{1}{\omega + \varepsilon_{nm} + i \tau^{-1}} = - \frac{\tau^2}{\varepsilon_{nm}} + i \frac{\tau}{\varepsilon_{nm}^2} + \frac{1}{\varepsilon_{nm}^3} - i \frac{\tau^{-1}}{\varepsilon_{nm}^4} + \mathcal{O}(\tau^{-2})
\end{align}
(2.D) becomes similarly,
\begin{align}
\lim_{\omega \to 0} \frac{1}{(\omega + i \tau^{-1})(- \omega + i \tau^{-1})} \frac{1}{- \varepsilon_{mn} + i \alpha \tau^{-1}} = \frac{\tau^2}{\varepsilon_{mn}} + i \alpha \frac{\tau}{\varepsilon_{mn}^2} - \alpha^2 \frac{1}{\varepsilon_{mn}^3} - i \alpha^3 \frac{\tau^{-1}}{\varepsilon_{mn}^4} + \mathcal{O}(\tau^{-2})
\end{align}
However, regarding (2.E)+(2.F)
we point out that for the denominator $\frac{1}{- \varepsilon_{ln} + i \alpha \tau^{-1}}$, it is possible to have both $n \neq l$ and $n=l$ cases, therefore it becomes necessary to discuss both cases separately when doing the expansion. In particular, in the $n=l$ case one obtains terms of size $\mathcal{O}(\tau^3)$ for each diagram, which cancel each other in the final result.

\paragraph{Quantum metric dipole and in-gap nonlinear Hall response}

Using the sum rules for the velocity vertices and with the help of the lifetime expansion, one can recover the semiclassical NLD at order $\mathcal{O}(\tau^2)$ and the BCD at order $\mathcal{O}(\tau)$ by straightforward but tedious algebra. However, at order $\mathcal{O}(\tau^0)$, the quantum perturbative result differs from the semiclassical one by the presence of an additional Fermi sea term. 
For completeness, we state this observation using the language introduced above.

It is possible to construct the band-normalized quantum metric dipole in terms of the sum rule expressions, 
\begin{align}
& \partial_c G^{ab} = \partial_c \left[ \hat{\varepsilon}^{-1} \circ \hat{\mathcal{A}}^a , \hat{\mathcal{A}}^b \right] \notag\\
& = \left[ \hat{\varepsilon}^{-2} \circ \hat{\mathcal{A}}^b , \hat{\Delta}^c \circ \hat{\mathcal{A}}^a \right] + \left[ \hat{\varepsilon}^{-1} \circ \hat{\mathcal{A}}^b , \hat{S}^{ac} \right] + \frac{i}{2} \left[ \hat{\varepsilon}^{-1} \circ \hat{\mathcal{A}}^b , \left[ \hat{\mathcal{A}}^a , \hat{\mathcal{A}}^c \right] \right] + \left[ \hat{\varepsilon}^{-1} \circ \hat{\mathcal{A}}^a , \hat{S}^{bc} \right] + \frac{i}{2} \left[ \hat{\varepsilon}^{-1} \circ \hat{\mathcal{A}}^a , \left[ \hat{\mathcal{A}}^b , \hat{\mathcal{A}}^c \right] \right] 
\end{align}

From the semiclassical derivation, one expects this band-normalized quantum metric dipole term, i.e. the nonlinear conductivity should contain a piece $ \beta \partial_c G^{ab}_{nn} + \gamma \partial_a G^{bc}_{nn} + (a \leftrightarrow b)$, where the coefficients $\beta$ and $\gamma$ are parameters which still need to be determined. 

A lengthy calculation~\cite{Kaplan2023a} then reveals that three Fermi sea terms are left over besides the quantum metric dipole in the conductivity at order $\tau^0$,
\begin{align}
\sigma^{ab;c}_{\tau^0} = \frac{e^3}{\hbar^2} \int_{\mathbf{k}} \sum_{n} f_n \{ \beta \partial_c G^{ab}_{nn} + \gamma \partial_a G^{bc}_{nn} + \mathcal{O}(\varepsilon^{-2} \Delta \mathcal{A}^2) + \mathcal{O}(\varepsilon^{-1} S \mathcal{A}) + \mathcal{O}(\varepsilon^{-1} \mathcal{A}^3) \} + (a \leftrightarrow b)
\end{align}
The three Fermi sea contributions to the intrinsic nonlinear response are purely quantum and do not appear in the semiclassical derivation. Individually, they correspond to the contribution from the velocity shift, the positional shift, and the Berry curvature renormalization during the nonlinear transport process, and read explicitly
\begin{align}
\mathcal{O}(\varepsilon^{-2} \Delta \mathcal{A}^2) 
&= \left( \frac{1}{\alpha} - 1 - \beta \right) \left[ \hat{\varepsilon}^{-2} \circ \hat{\mathcal{A}}^a, \hat{\Delta}^c \circ \hat{\mathcal{A}}^b \right] + (\alpha - \gamma) \left[ \hat{\varepsilon}^{-2} \circ \hat{\mathcal{A}}^c, \hat{\Delta}^a \circ \hat{\mathcal{A}}^b \right]
\\
\mathcal{O}(\varepsilon^{-1} S \mathcal{A}) 
&= ( - 2 \beta - \gamma + 1) \left[ \hat{\varepsilon}^{-1} \circ \hat{\mathcal{A}}^a , \hat{S}^{bc} \right] + \left( - \gamma + \frac{\alpha^2}{2} \right) \left[ \hat{\varepsilon}^{-1} \circ \hat{\mathcal{A}}^c , \hat{S}^{ab} \right]
\\
\mathcal{O}(\varepsilon^{-1} \mathcal{A}^3) 
&= \left( - \beta - \frac{\gamma}{2} + \alpha \right) \left[ \hat{\varepsilon}^{-1} \circ \hat{\mathcal{A}}^a, i \left[ \hat{\mathcal{A}}^b, \hat{\mathcal{A}}^c \right] \right] + i \alpha \left( 1 - \frac{\alpha}{2} \right) \left[ \hat{\mathcal{A}}^a, \left[ \hat{\mathcal{A}}^b, \hat{\varepsilon}^{-1} \circ \hat{\mathcal{A}}^c \right] \right].
\end{align}
One can ask whether it is possible to eliminate all terms other than the quantum metric dipole term by choosing a certain combination of $\alpha$, $\beta$ and $\gamma$ so that all the coefficients becomes zero. However, this is not possible. The second term in $\mathcal{O}(\varepsilon^{-1} \mathcal{A}^3)$ enforces $\alpha = 2$, then the second term in $\mathcal{O}(\varepsilon^{-1} S \mathcal{A})$ enforces $\gamma = 2$, which further leads to $\beta = -\frac{1}{2}$ to cancel the first term in $\mathcal{O}(\varepsilon^{-1} S \mathcal{A})$. Although this cancels all terms in $\mathcal{O}(\varepsilon^{-1} S \mathcal{A})$ and $\mathcal{O}(\varepsilon^{-2} \Delta \mathcal{A}^2)$, yet we are still left with $\mathcal{O}(\varepsilon^{-1} \mathcal{A}^3) = \frac{3}{2} \left[ \hat{\varepsilon}^{-1} \circ \hat{\mathcal{A}}^a, i \left[ \hat{\mathcal{A}}^b, \hat{\mathcal{A}}^c \right] \right] \neq 0$. Thus the diagrammatic derivation contains at least one more Fermi sea contribution that is missed in the semiclassical derivation.

Although different values have been reported for $\beta$ and $\gamma$~\cite{Gao2014,Gao2019,Das2023,Kaplan2024,Jia2024}, there is an emerging agreement that the intrinsic nonlinear transport conductivity should contain both a longitudinal and transverse QMD. 

\end{widetext}

%\bibliography{reference,literature}
%merlin.mbs apsrev4-1.bst 2010-07-25 4.21a (PWD, AO, DPC) hacked
%Control: key (0)
%Control: author (0) dotless jnrlst
%Control: editor formatted (1) identically to author
%Control: production of article title (0) allowed
%Control: page (1) range
%Control: year (0) verbatim
%Control: production of eprint (0) enabled
%

\end{document}